\documentclass[useAMS,usenatbib]{mn2e}
\usepackage{graphicx}
\usepackage{amssymb}
\usepackage{rotating}

\newcommand{\apj}{ApJ}

\newcommand{\mnras}{MNRAS}

\newcommand{\MC}{\multicolumn}
\newcommand{\kms}{km~s$^{-1}$}

\newcommand{\HI}{H{\sc i}}
\newcommand{\HII}{H{\sc ii}}
\newcommand{\sunn}{$_{\odot}$}

\newcounter{qub}
\newcommand{\qq}{\addtocounter{qub}{1}\arabic{qub}}

\title[Nearby Void XMP gas-rich dwarfs: SALT spectroscopy]
{XMP gas-rich dwarfs in Nearby Voids: results of SALT spectroscopy}

\author[S.A.~Pustilnik, A.Y.~Kniazev, Y.A.~Perepelitsyna,  E.S.~Egorova]
{S.A.~Pustilnik,$^{1}$\thanks{E-mail: sap@sao.ru (SAP)} A.Y.~Kniazev,$^{2,3,4}$
  Y.A.~Perepelitsyna,$^{1}$ E.S.~Egorova$^{4}$  \\
$^1$ Special Astrophysical Observatory of RAS, Nizhnij Arkhyz,Karachai-Circassia 369167, Russia \\
$^2$ South African Astronomical Observatory, PO Box 9, 7935 Observatory,  Cape Town, South Africa \\
$^3$ Southern African Large Telescope Foundation, PO Box 9, 7935 Observatory, Cape Town, South Africa \\
$^4$ Sternberg Astronomical Institute, Lomonosov Moscow State University, Universitetskij Pr. 13,  Moscow 119992, Russia
}

\begin{document}

\label{firstpage}

\date{Accepted December 23, 2019, Received August 3, 2019}

\pagerange{\pageref{firstpage}--\pageref{lastpage}} \pubyear{2019}

\maketitle

\begin{abstract} 
In the framework of an ongoing project aimed at searching for and
studying
eXtremely Metal-Poor (XMP) very gas-rich blue dwarfs in nearby voids, we
conducted spectroscopy with the 11-m Southern African Large Telescope (SALT)
of 26 candidates, preselected in the first paper of this series (PEPK19). For
23 of them, we detected Oxygen lines, allowing us to estimate  the gas O/H
ratio. For ten of them, the oxygen abundance is found to be very low,
in the range of 12+$\log$(O/H)=6.95--7.30~dex. Of those, four void
dwarfs have
12+$\log$(O/H) $<$ 7.19, or Z $<$ Z\sunn/30. For the majority of
observed galaxies, the faint line [O{\sc iii}]$\lambda$4363~\AA\ used to
estimate O/H with the direct T$_{\rm e}$ method appeared either too
noisy or was not detected. We therefore use the semi-empirical
method of \citet{IT07}  for these spectra, or, when applicable,
the new 'Strong line' method of Izotov et al. (2019b).
We present and discuss the results for all void dwarfs observed in this
work. We also compare their O/H values with O/H values of
$\sim$140 void galaxies available from our recent papers. We address the
properties of the newly found unusual void XMP dwarfs and compare them with
those for  ten known prototype void XMP objects. The latter small
group is
outstanding based on their very small mass fraction of stars (only 0.01--0.02
of the baryonic mass), the blue colours of stars in the outer body
(indicating a non-cosmological age for the main star-forming episode), and the
low gas metallicity (several times lower than expected for their luminosity).
\end{abstract}

\begin{keywords}
galaxies: dwarf -- galaxies: evolution -- galaxies: photometry --
galaxies: abundances -- cosmology: large-scale structure of Universe
\end{keywords}

\section[]{Introduction}
\label{sec:intro}
\setcounter{figure}{0}

In a recent paper \citep[][hereafter PTM19]{PTM19}
we presented a sample of 1354 Nearby Void Galaxies (the NVG sample),
which reside within the boundaries of 25 Nearby Voids,
in the volume with distances of $<$ 25 Mpc from the Local Group.

The nearby voids were defined over the entire celestial sphere based
on the sample of luminous galaxies defined via the K-band luminosity.
They were adopted as those with absolute K-band magnitude $M_{\rm K} < -22.0$.
This galaxy sample was used to define empty spheres with minimum radii
of 6 Mpc. When the individual empty spheres were close enough, they were
joined in groups of spheres.
Finally, groups of empty spheres close in space and having common
spheres, were also combined into larger entities defined as individual
voids with major sizes from $\sim$14 to $\sim$35 Mpc. Of the initial
sample of $\sim$7000 objects within the considered volume, 1354 galaxies
which appeared inside the found empty spheres, are defined as residing in the
 abovementioned 25 Nearby Voids. See more details on the Nearby Voids and
the galaxies residing in them in PTM19.

One of the goals of selecting this void galaxy sample, was
an opportunity  to substantially increase the number of unusual
void XMP very gas-rich dwarfs from the handful currently known.
Most of the known prototype XMP dwarfs were found as a result of
the systematic study of a hundred galaxies in the nearby Lynx-Cancer void
\citep[][and references therein]{PaperI,PaperVII}.

The efficient search for new void XMP dwarfs is based on the preliminary
selection of good candidates for the subsequent spectroscopy. The mass
selection from the NVG sample is based on the galaxy properties available in
public databases and in the literature. This selection is described in the
first paper of this series by Pustilnik et al. (2019, MNRAS, accepted,
hereafter PEPK19), where a list of 60 selected XMP dwarf void candidates
is presented.
An overview of the XMP dwarf search and the motivation for the current
project are described in detail in the Introduction to that paper.
Therefore, we only briefly outline below the main points of the project
and the criteria used to select XMP dwarf candidates.

The issue of XMP dwarf galaxies has attracted the attention of astrophysicists
since the discovery of the extremely low metallicity
of the blue compact dwarf (BCD) IZw18=MRK~116 \citep{Searle72}, with
 12+$\log$(O/H) = 7.17 dex or $Z \sim Z$\sunn/30\footnote{We
adopt hereafter the solar value of 12+$\log$(O/H)=8.69 after
\citet{Asplund09}.}.

The continuing interest in IZw18, its companion IZw18C, the system
SBS0335-052~E,W and similar
objects with $Z \lesssim Z$\sunn/30, was motivated by the original idea
on their recent first star formation (SF) burst. However, after the discovery
of the red giant branch (RGB) population in IZw18, this idea transformed to
the less exotic. Namely, such unusual galaxies have probably
non-cosmological ages ($\sim$1--2~Gyr) of the main stellar population
\citep{Pustilnik04,Izotov2009,PO12,Annibali2013}.
 Besides, the study of such XMP galaxies is important for
understanding star-formation processes at extremely low metallicities
typical of galaxies at much earlier epochs.

The term 'very low metallicity' galaxy was originally applied to
objects with gas metallicity below $Z$\sunn/10. Due to the known relation
between $Z_{\rm gas}$ and galaxy mass/luminosity \citep[e.g.][]{Kunth2000}
and the many
spectroscopic studies of dwarfs in the nearby Universe, metallicities
of $Z  \lesssim Z$\sunn/10  are now more or less routinely found, so that
their current number reaches several hundred \citep[e.g.][]{Guseva17}.

\setcounter{qub}{0}
\begin{table}
\begin{center}
\caption{Journal of SALT RSS spectral observations}
\label{tab:journal}
\hoffset=-2cm
\begin{tabular}{r|l|l|l|r|c|c} \hline  \hline \\ [-0.2cm]
\MC{1}{r|}{No.} &
\MC{1}{c|}{Name} &
\MC{1}{c|}{Date} &
\MC{1}{c|}{Expos.}&
\MC{1}{c|}{PA} &
\MC{1}{c|}{$\theta$\arcsec}&
\MC{1}{c}{Air}  \\

\MC{1}{r|}{} &
\MC{1}{c|}{ } &
\MC{1}{c|}{ } &
\MC{1}{c|}{time, s}&
\MC{1}{c|}{ } &
\MC{1}{c|}{} &
\MC{1}{c}{mass}\\

\\[-0.2cm] \hline \\[-0.2cm]
\qq&PGC000389          & 2018.10.10  &2$\times$1200&  26.5 & 1.5 & 1.21 \\ 
\qq&PGC736507          & 2018.12.10  &2$\times$1200&  23.5 & 1.7 & 1.29 \\ 
\qq&HIJ0021+08$\dagger$& 2018.11.07  &2$\times$1200& 169.0 & 2.0 & 1.33 \\ 
\qq&AGC104227          & 2017.11.10  &2$\times$1200& 121.5 & 1.5 & 1.32 \\ 
\qq&PGC493444          & 2017.11.18  &2$\times$1200& 355.0 & 1.3 & 1.30 \\ 
\qq&PGC1190331         & 2018.09.16  &6$\times$1200& 341.0 & 1.6 & 1.19 \\ 
\qq&AGC411446          & 2017.12.08  &2$\times$1200& 108.0 & 1.1 & 1.29 \\ 
   &-\#-               & 2017.12.12  &2$\times$1200& 108.0 & 1.8 & 1.27 \\ 
   &-\#-               & 2017.12.14  &2$\times$1200& 108.0 & 1.8 & 1.27 \\ 
   &-\#-               & 2017.12.15  &2$\times$1200& 108.0 & 1.1 & 1.29 \\ 
\qq&AGC114584          & 2018.10.10  &2$\times$1200& 157.0 & 1.5 & 1.14 \\ 
   &-\#-               & 2018.10.31  &2$\times$1200& 157.0 & 1.7 & 1.14 \\ 
\qq&AGC123223          & 2018.11.09  &2$\times$1200&   1.0 & 1.5 & 1.14 \\ 
\qq&AGC124629          & 2018.11.05  &2$\times$1200& 310.5 & 2.0 & 1.24 \\ 
   &-\#-               & 2018.12.29  &2$\times$1200& 310.5 & 2.0 & 1.24 \\ 
   &-\#-               & 2018.12.30  &2$\times$1200& 310.5 & 2.0 & 1.24 \\ 
\qq&AGC132121          & 2017.11.11  &2$\times$1200& 320.5 & 1.3 & 1.34 \\ 
\qq&ESO121-020         & 2018.11.09  &2$\times$1300& 142.0 & 1.3 & 1.26 \\ 
   &-\#-               & 2019.03.03  &2$\times$1300&  43.2 & 1.3 & 1.26 \\ 
\qq&PGC385975          & 2018.10.10  &2$\times$1200& 149.0 & 1.9 & 1.23 \\ 
   &-\#-               & 2018.11.09  &2$\times$1200& 142.0 & 1.3 & 1.17 \\ 
\qq&AGC174605          & 2018.02.22  &2$\times$1200& 138.0 & 1.5 & 1.32 \\ 
\qq&AGC188955          & 2019.02.27  &2$\times$1250&  79.0 & 1.5 & 1.25 \\ 
\qq&AGC198454          & 2018.12.31  &2$\times$1150& 202.5 & 1.1 & 1.32 \\ 
\qq&PGC1314481         & 2018.02.25  &2$\times$1200& 172.5 & 2.5 & 1.33 \\ 
\qq&J1001+0846         & 2017.12.27  &2$\times$1200& 303.0 & 1.3 & 1.33 \\ 
\qq&PGC1230703         & 2017.12.25  &2$\times$1200&  22.0 & 1.6 & 1.28 \\ 
\qq&PGC1178576         & 2018.02.22  &2$\times$1200& 137.0 & 1.5 & 1.27 \\ 
\qq&AGC208397          & 2019.02.09  &2$\times$1300& 348.0 & 1.5 & 1.28 \\ 
   &-\#-               & 2019.02.27  &2$\times$1300& 348.0 & 1.5 & 1.28 \\ 
   &-\#-               & 2019.02.28  &2$\times$1300& 348.0 & 1.5 & 1.28 \\ 
   &-\#-               & 2019.03.03  &2$\times$1300& 348.0 & 1.5 & 1.28 \\ 
\qq&PGC044681          & 2018.07.06  &2$\times$1200& 186.5 & 1.6 & 1.27 \\ 
\qq&PGC135827          & 2018.02.26  &2$\times$1200& 324.5 & 1.5 & 1.22 \\ 
\qq&AGC258574          & 2018.02.27  &2$\times$1200& 132.0 & 1.8 & 1.27 \\ 
   &-\#-               & 2018.07.08  &2$\times$1200& 132.0 & 1.8 & 1.27 \\ 
\qq&KK246              & 2018.07.04  &2$\times$1200&  46.0 & 1.4 & 1.22 \\ 
\qq&AGC335193          & 2017.11.10  &2$\times$1200& 227.0 & 1.8 & 1.28 \\ 
\hline \hline \\[-0.2cm]
\multicolumn{7}{p{8.1cm}}{$\dagger$ HIJ0021+08 means HIPASSJ0021+08} \\
\end{tabular}
\end{center}
\end{table}

Our aim is to search for much rarer objects, which we call XMP
galaxies, with $Z \lesssim Z$\sunn/30. The main motivation is that
in this metallicity range we found several very unusual void dwarfs
\citep[see][and references therein]{PaperIV,PaperVII}. The main advancement
in the finding of XMP galaxies is related to the dedicated search for
such objects in the enormous spectral database of the SDSS \citep{DR7} project
\citep[e.g.][and references therein]{Guseva17,Izotov18,Izotov19,Sanchez16a}.
However, in the whole SDSS DR14 database \citep{DR14} it was possible to
identify only about dozen such objects \citep{Izotov19DR14}.
Together with a handful of XMP dwarfs found by the alternative means
\citep[e.g.][]{Searle72,Izotov1990,Izotov2009,DDO68,J0926,Triplet,Skillman13,
Hirschauer16,U3672,LittleCub,Takashi19}, the
list of such objects found to date, comprises of only $\sim$20.
While a sizable fraction of XMP dwarfs is found outside voids, the
majority of such galaxies known to date resides in voids. Besides, for
several XMP dwarfs (with Z $<$Z\sunn/30) found via SDSS spectra, the type of
environment is not yet published.

Since it was shown that among the least luminous blue dwarfs in voids the
fraction of such XMPs can reach $\sim$30\% \citep{PaperIV,PaperVII},
we use the NVG sample to produce a list of 60 suitable candidates for
further spectral study (PEPK19).  The selection was based on the
similarity of candidate properties known from public databases and the
literature to those of about ten prototype XMP dwarfs (see PEPK19). Those
include the elevated gas content, blue colours and low luminosity
as well as the indicative strong Oxygen line to H$\beta$ flux ratios,
when they were available. See PEPK19
for more detail. In this paper we present the results for 26 of them,
available from observations with SALT.
The complementary part of the candidate list in the Northern sky was
observed at BTA (Big Telescope Alt-azimuth, the SAO 6m telescope) and will
appear in an accompanying paper (in preparation).

The rest of this paper is arranged as follows:
The description of the SALT spectral observations and data processing
is presented in Sec.~\ref{sec:observing}.
In Sec.~\ref{sec:OH} we describe emission line measurements and methods
used for O/H determination.
In Sec.~\ref{sec:results} we show the estimates of O/H for the observed
galaxies.
In Sec.~\ref{sec:discussion} we discuss the obtained results along with other
available information.
Finally, in Sec.~\ref{sec:conclusions} we present our conclusions.
In Appendix~A  we present plots of 1D spectra for each of the observed
galaxies and in Appendix~B, tables with line intensities, derived
physical parameters and O/H ratios.

\section[]{SALT observations and data processing}
\label{sec:observing}

Spectral observations with the Southern African Large Telescope
\citep[SALT;][]{Buck06,Dono06} were conducted in service mode
in the period from November 2017 to March 2019.
Several of the 26 target galaxies were observed from two to four times.
See Table~\ref{tab:journal}.
We used the SALT Robert Stobie Spectrograph  \citep[RSS;][]{Burgh03,Kobul03}
with VPH grating PG0900 with the long slit of 1.5\arcsec\ by 8\arcmin\
to cover the range from 3600~\AA\ to 6700~\AA\  with the resulting spectral
resolution of FWHM$\sim$6.0~\AA.
All spectral data were obtained with a binning factor of four for the
spatial scale and factor of two for the spectral coordinate, to give a final
spatial sampling of 0\farcs51 pixel$^{-1}$ and spectral sampling of 0.97
\AA\,pixel$^{-1}$ .
Since the RSS is equipped with an Atmospheric Dispersion Compensator (ADC),
this allowed us to escape the effect of atmospheric dispersion at arbitrary
long-slit position angles (PA). Spectrophotometric standards were observed
during twilight as part of the SALT standard calibrations program.

SALT is a telescope where the unfilled entrance pupil of the telescope
moves during the observations. This implies that the part of the mirror
collecting light changes continuously during each specific
observation. For that reason absolute flux calibration is not
feasible with SALT. However, since all optical elements and instrumentation
are always the same, relative flux calibration could be used.
Hence, the relative distribution of energy in the spectra
could be obtained with SALT data.

Since the majority of \HII-regions in the program galaxies are too faint and
low-contrast, we used nearby offset stars for the SALT pointing. Position
angles (PAs, in degrees) of the slit (from the North counterclockwise)
were selected in most cases to include a nearby offset star and to cover
a faint \HII-knot in a program galaxy. See the journal of observation in
Table~\ref{tab:journal} for the main information on each observation
including the dates of observations, exposure times, seeing $\theta$
in arc seconds and air mass.

\begin{figure*}
\includegraphics[width=4.0cm,angle=0,clip=]{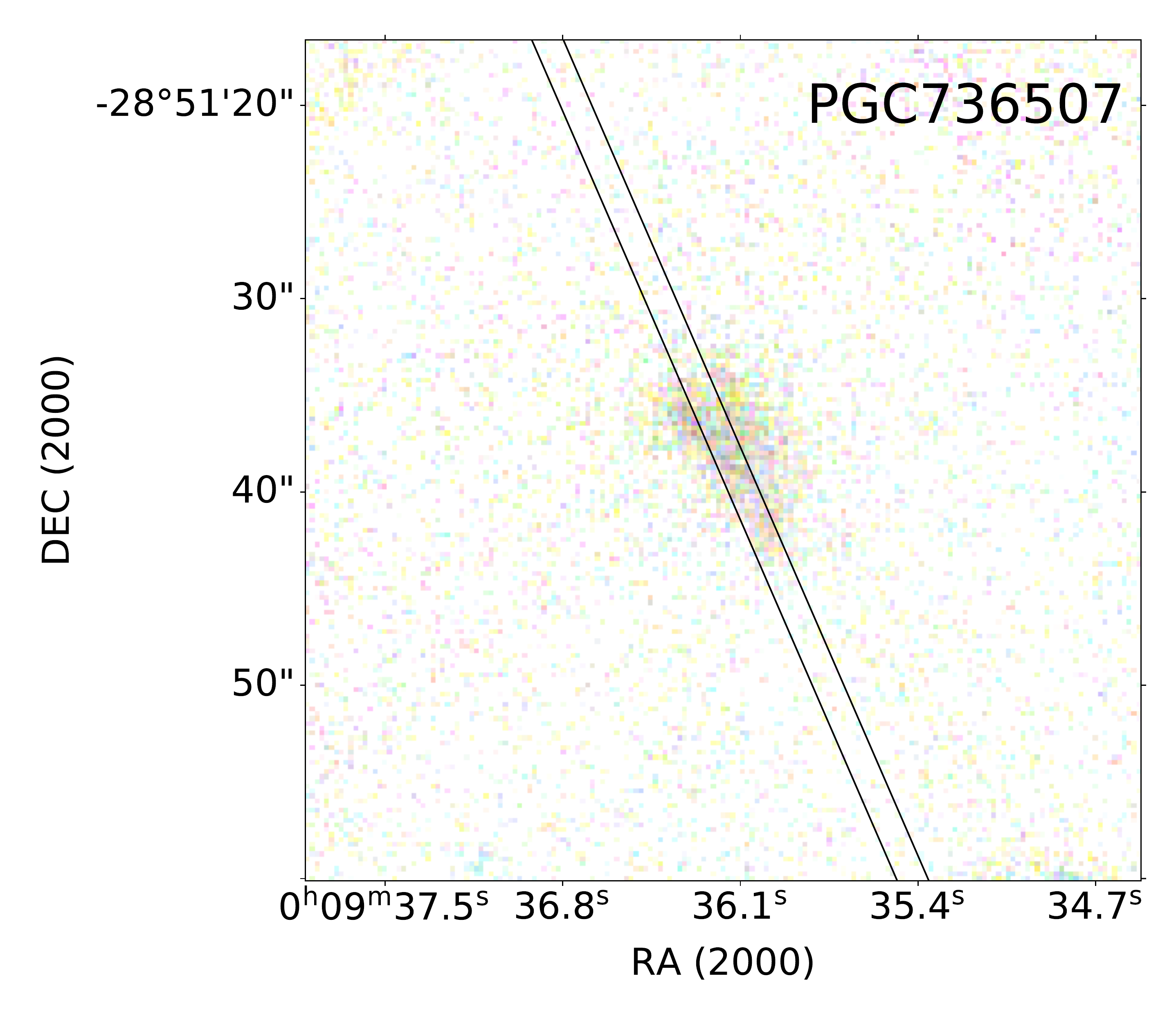}
\includegraphics[width=4.0cm,angle=0,clip=]{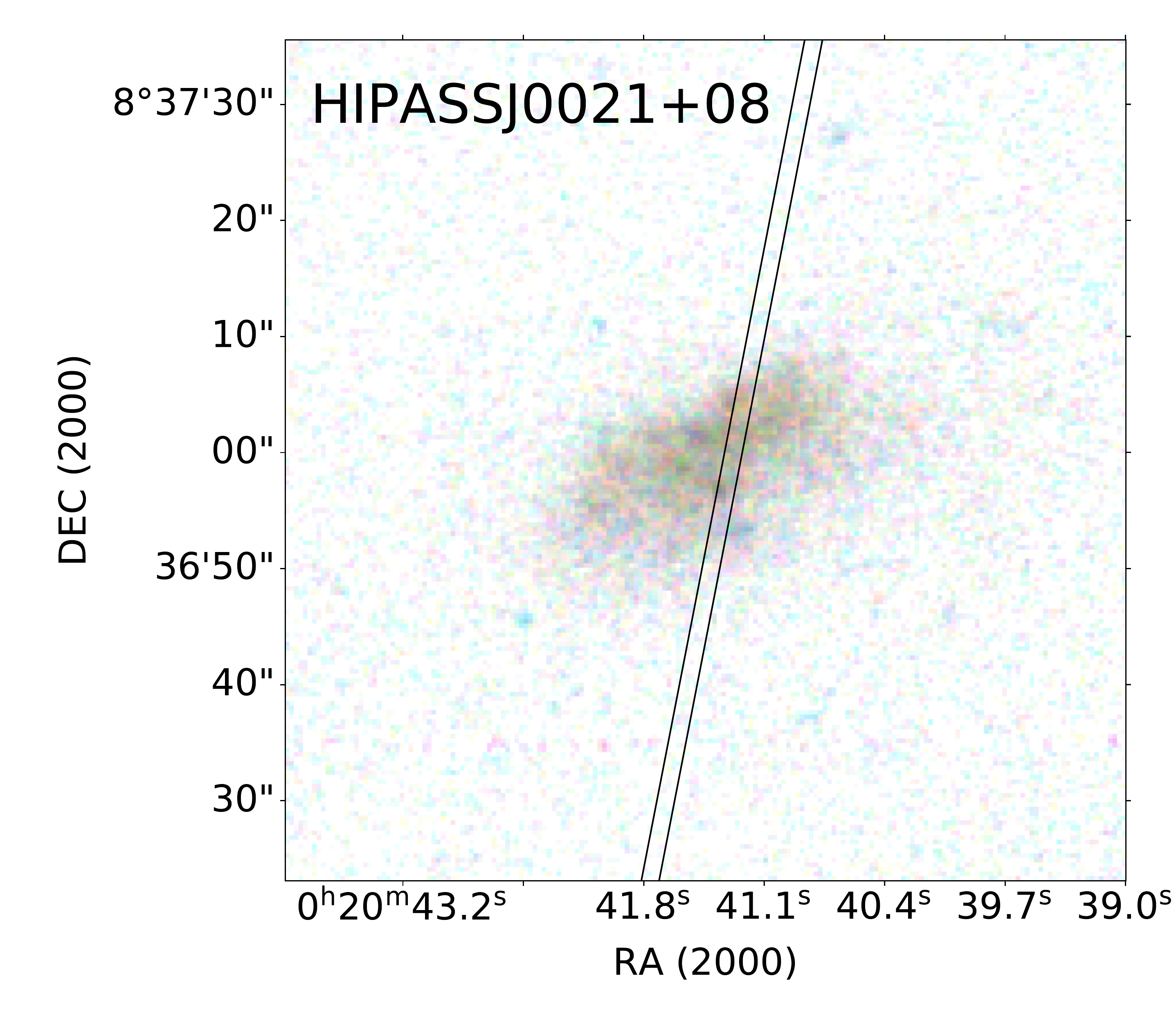}
\includegraphics[width=4.0cm,angle=0,clip=]{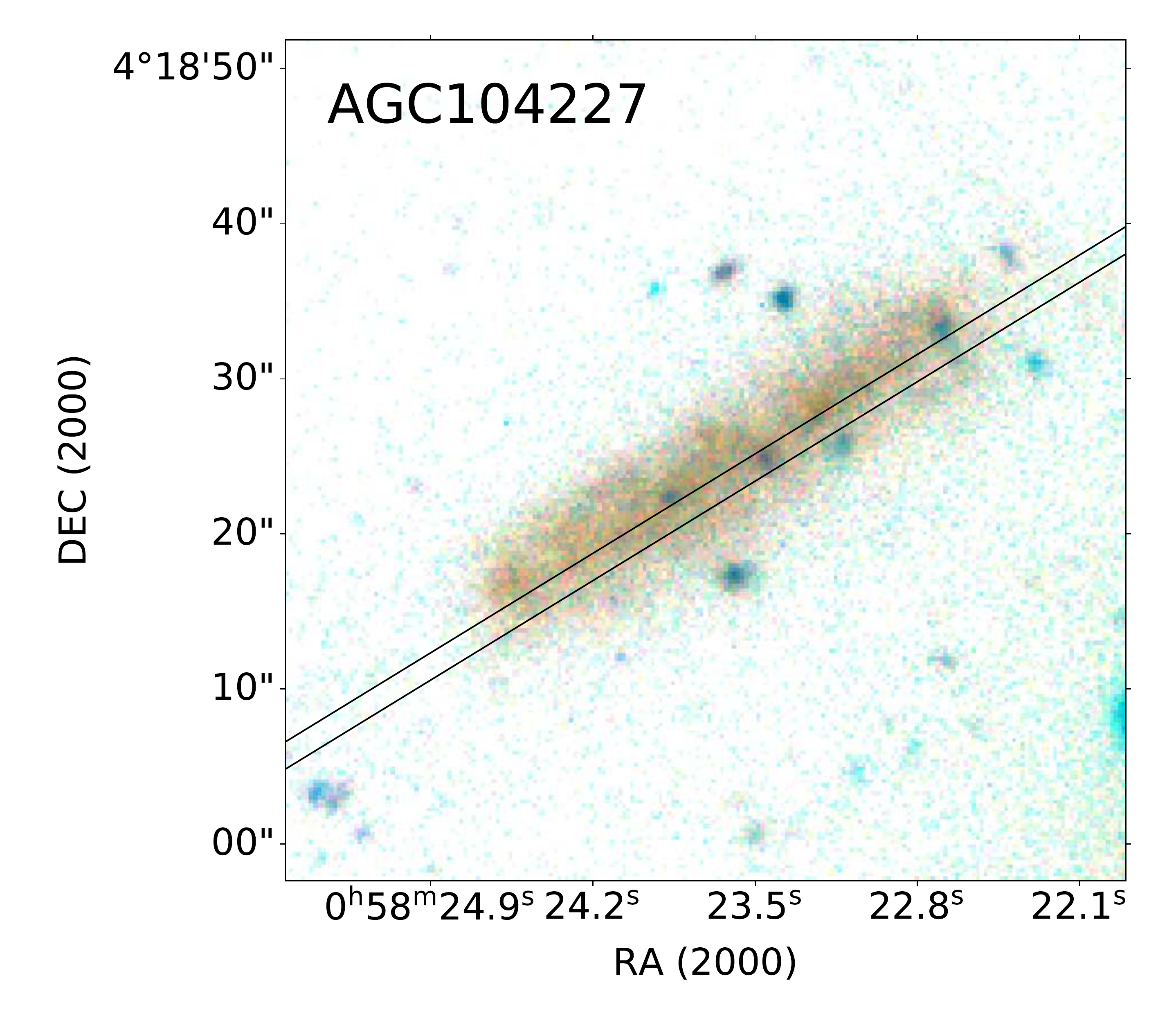}
\includegraphics[width=4.0cm,angle=0,clip=]{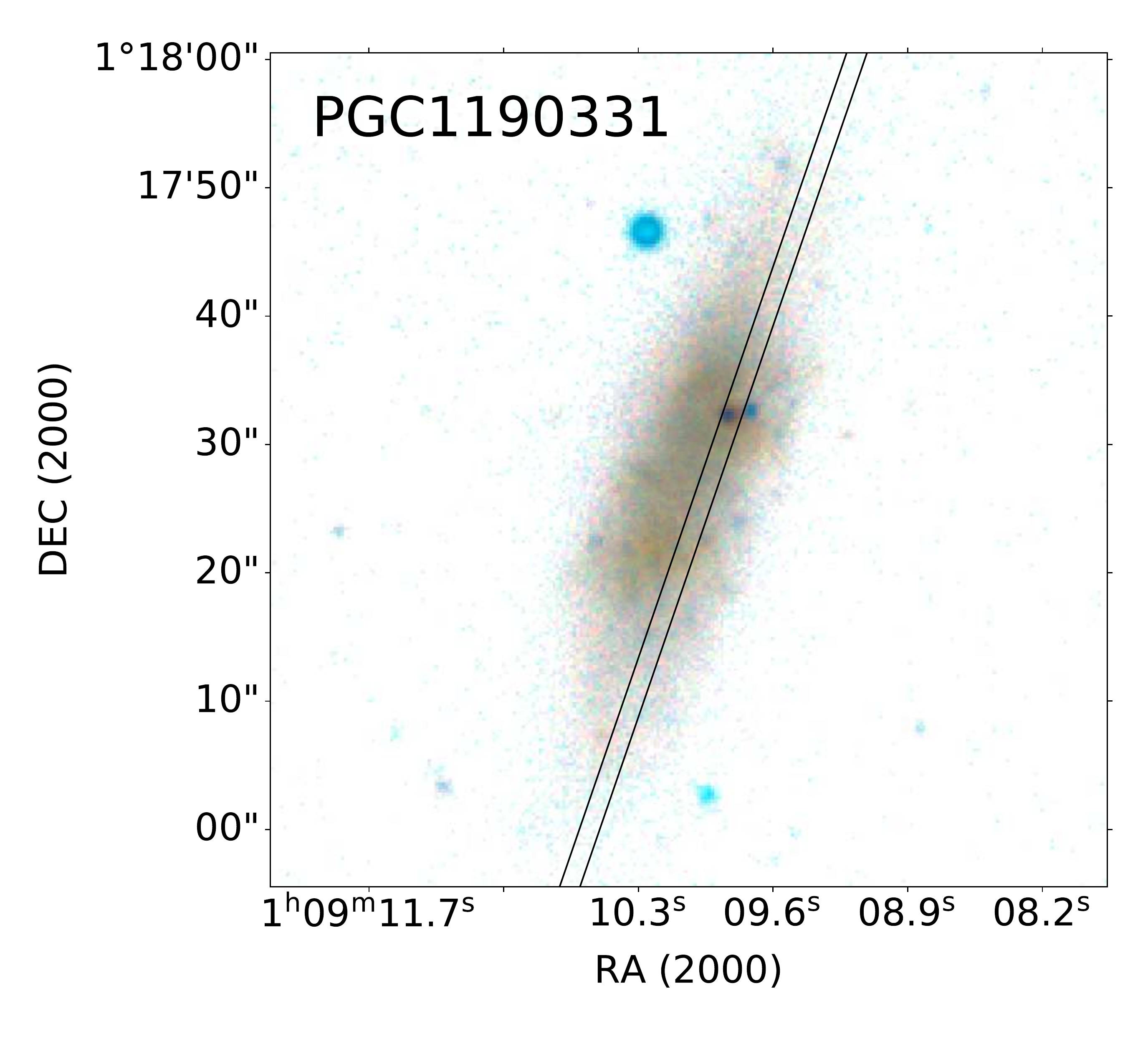}
\includegraphics[width=4.0cm,angle=0,clip=]{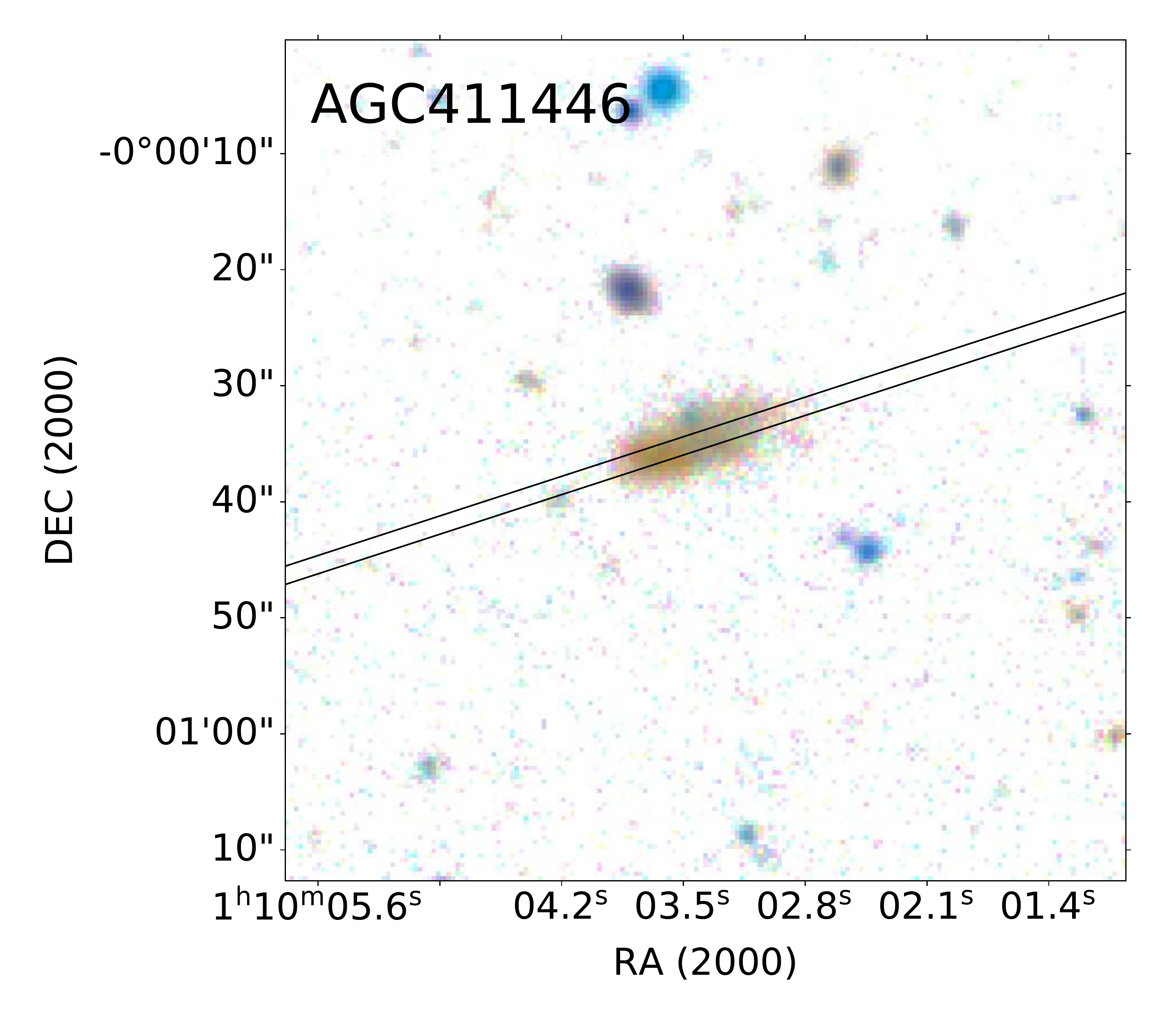}
\includegraphics[width=4.0cm,angle=0,clip=]{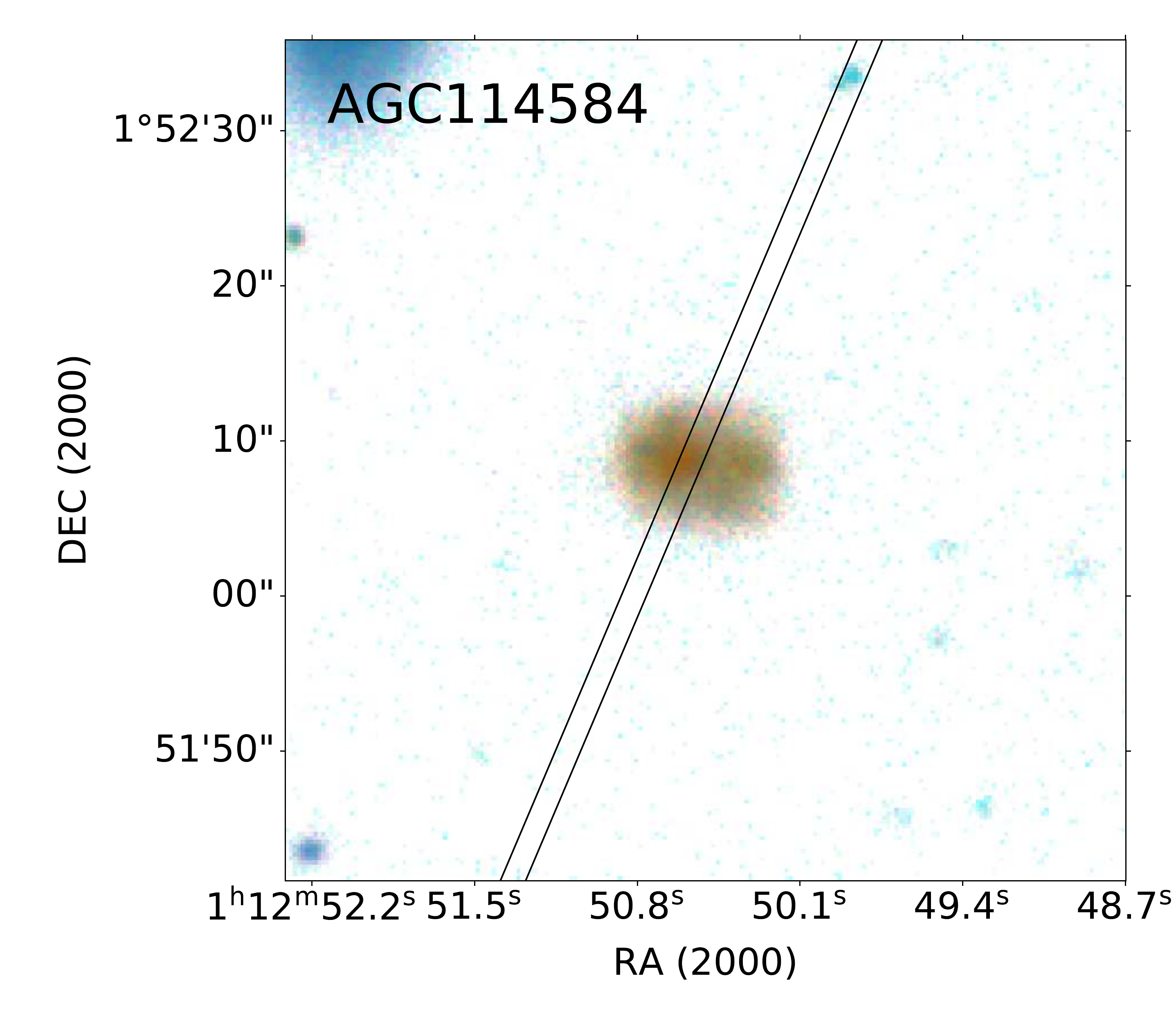}
\includegraphics[width=4.0cm,angle=0,clip=]{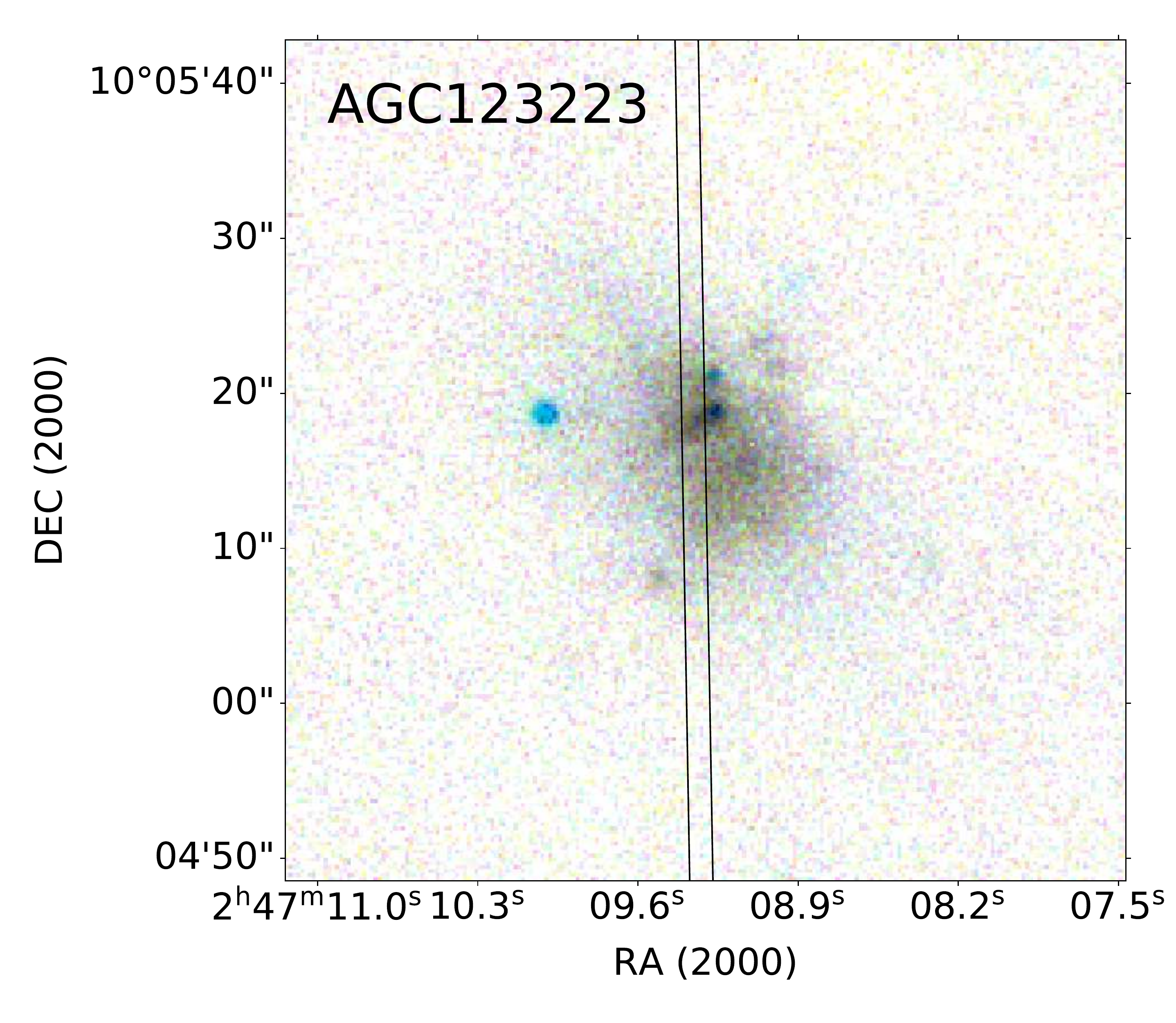}
\includegraphics[width=4.0cm,angle=0,clip=]{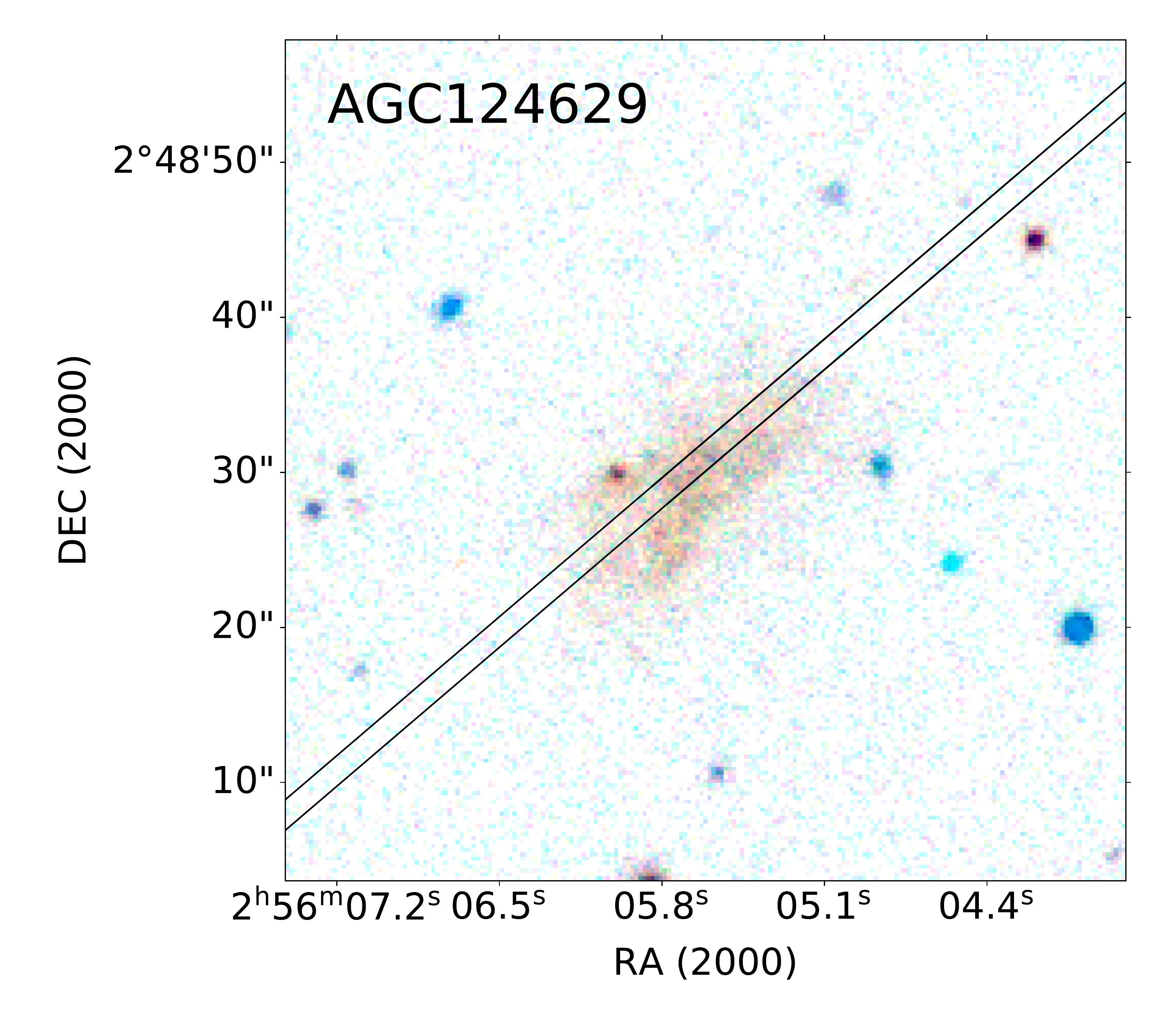}
\includegraphics[width=4.0cm,angle=0,clip=]{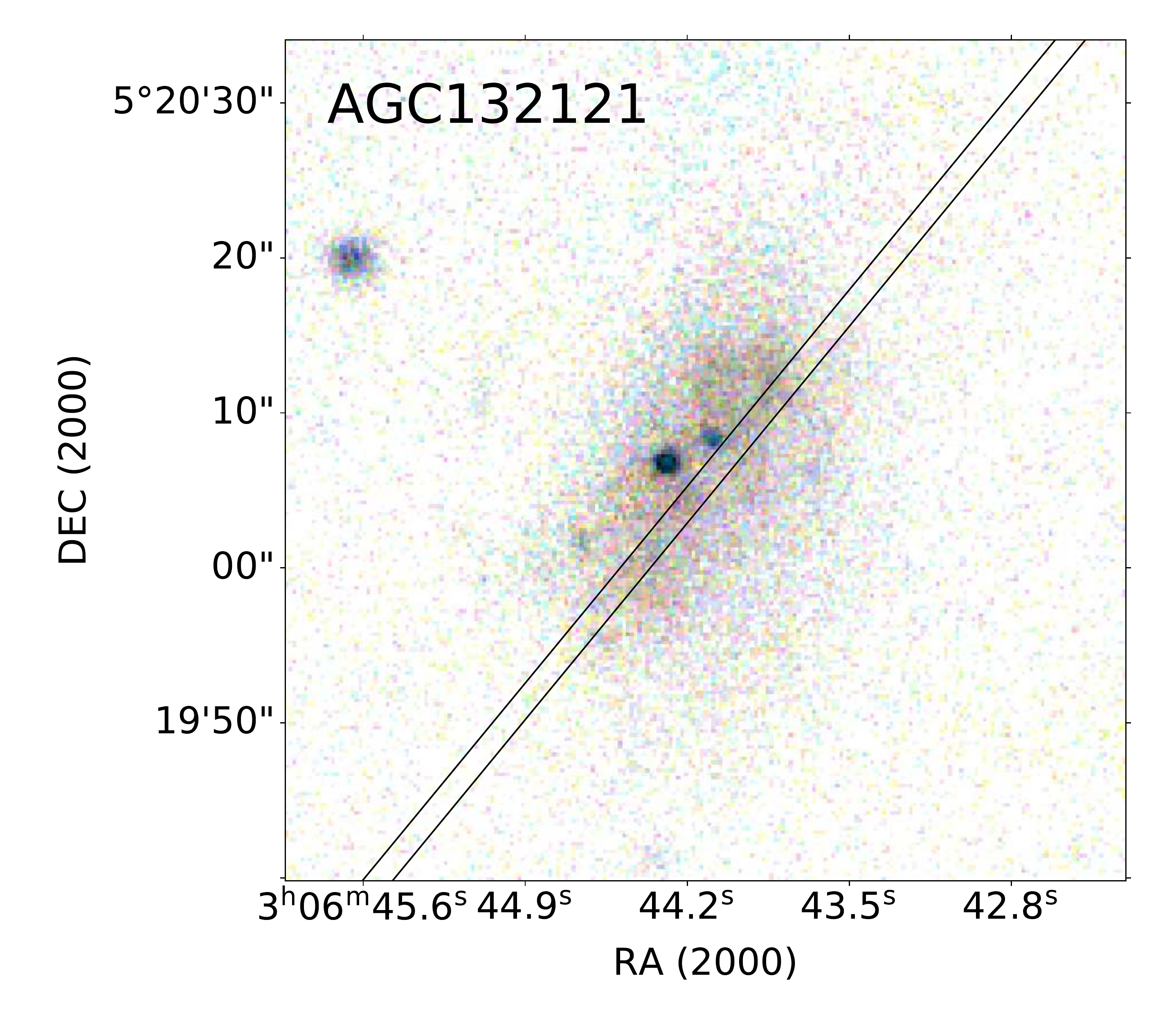}
\includegraphics[width=4.0cm,angle=0,clip=]{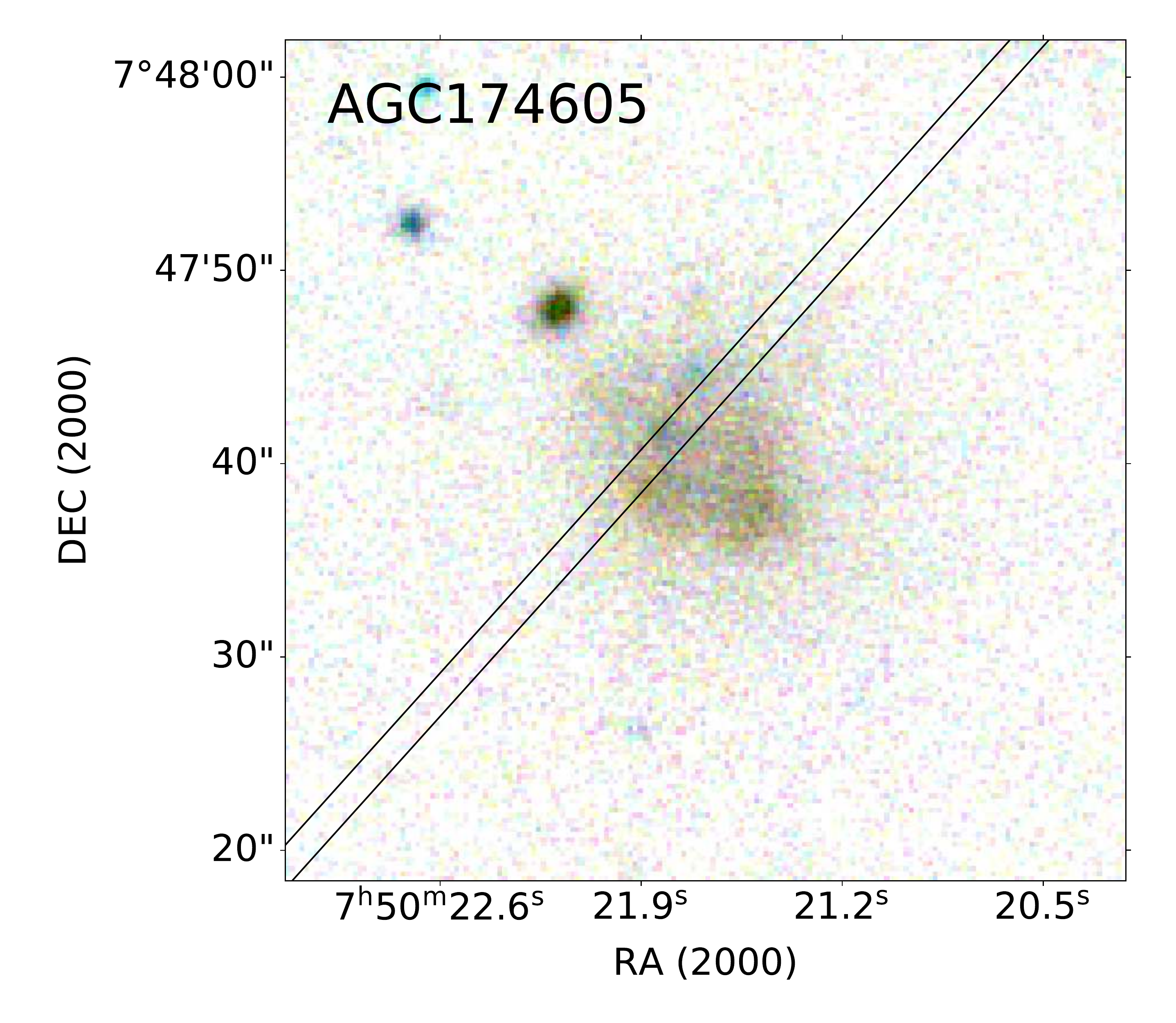}
\includegraphics[width=4.0cm,angle=0,clip=]{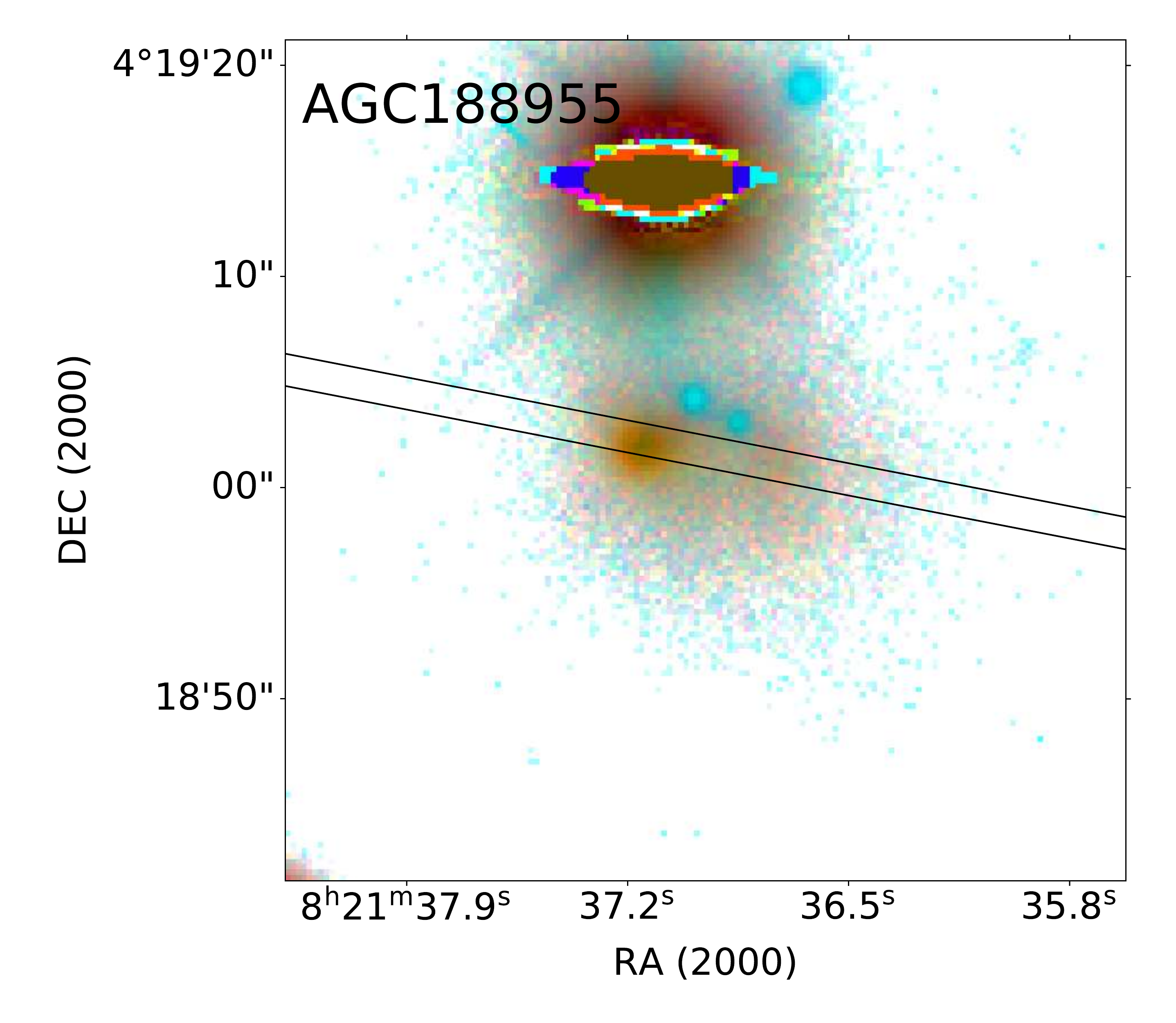}
\includegraphics[width=4.0cm,angle=0,clip=]{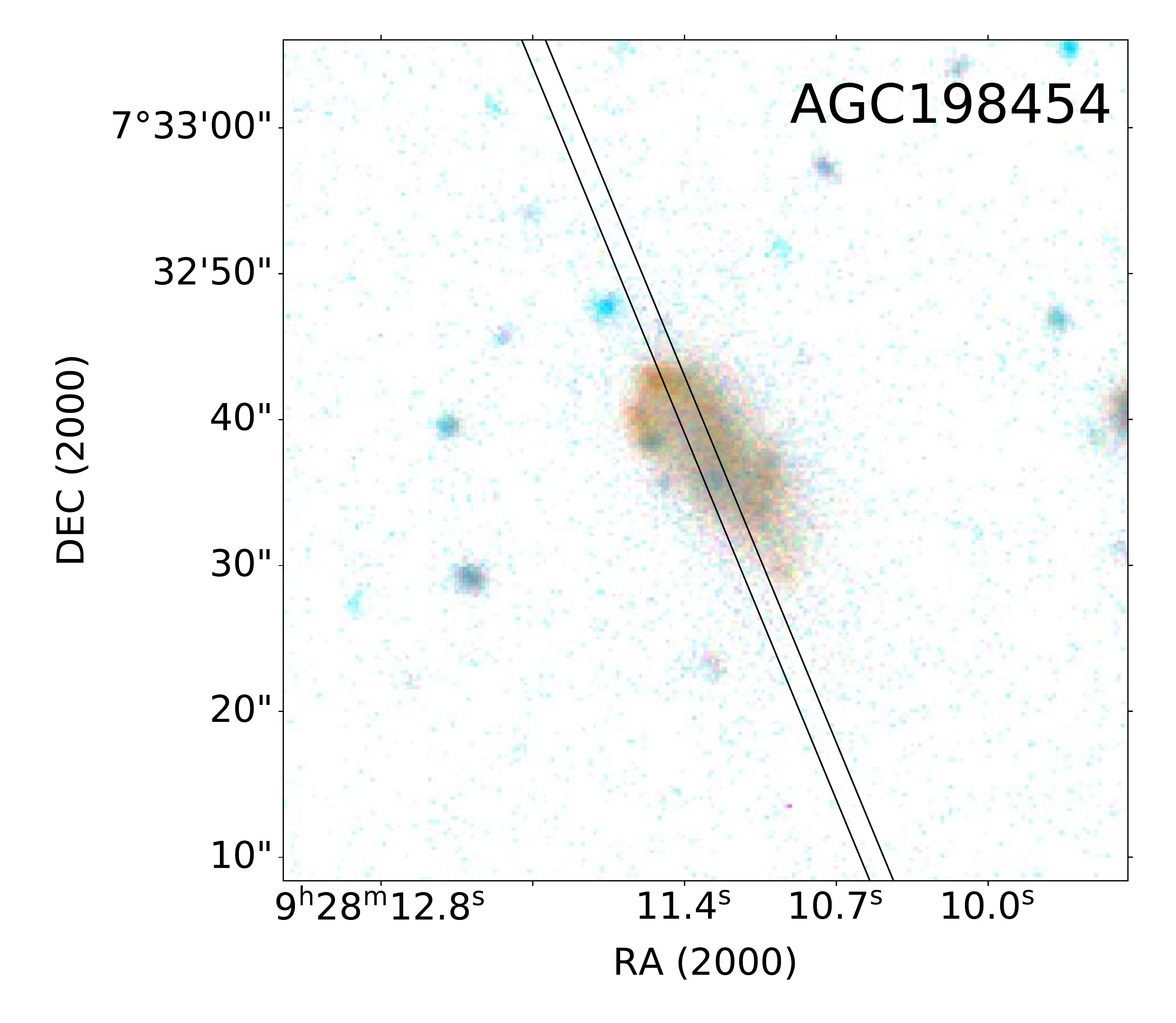}
\includegraphics[width=4.0cm,angle=0,clip=]{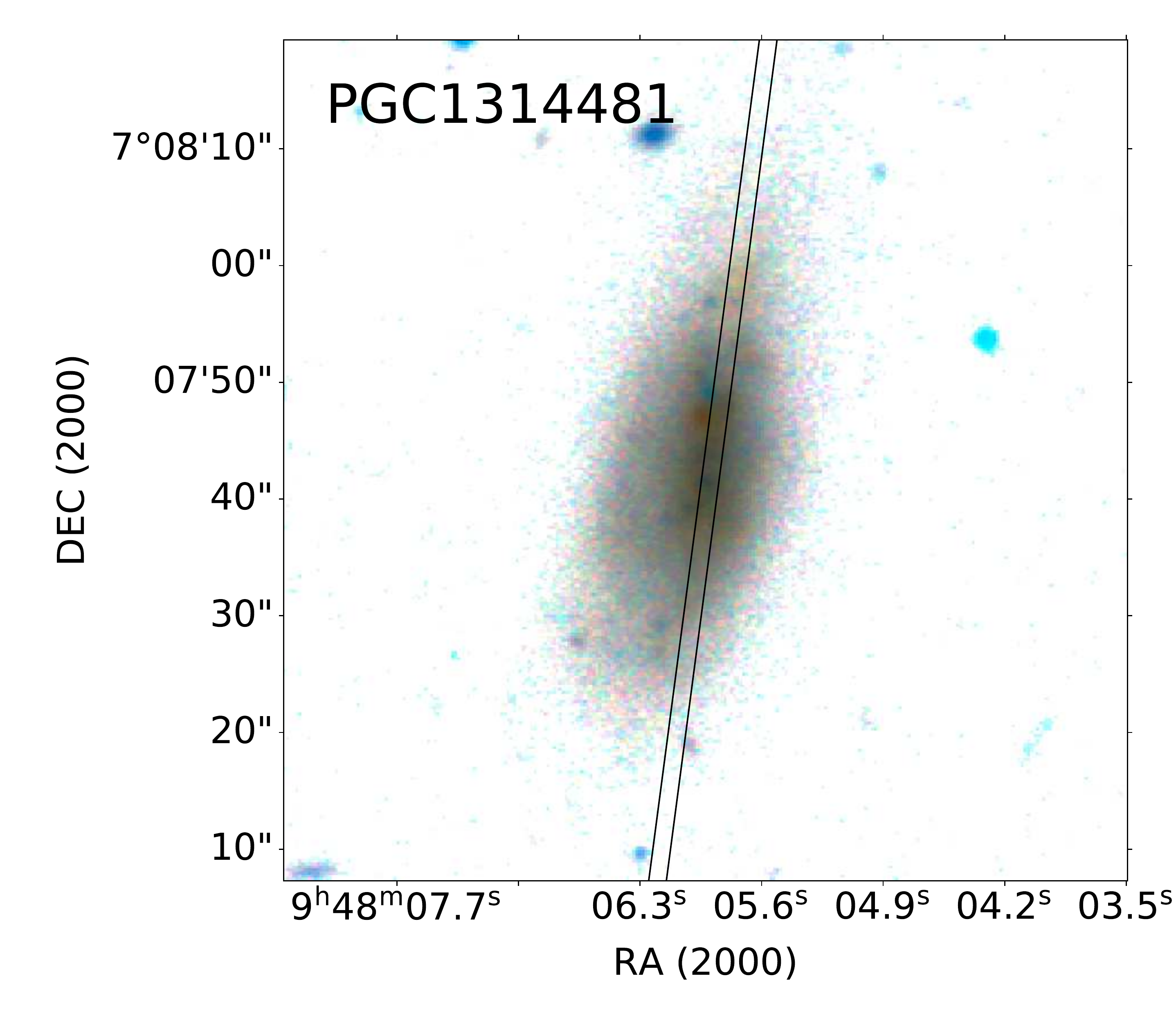}
\includegraphics[width=4.0cm,angle=0,clip=]{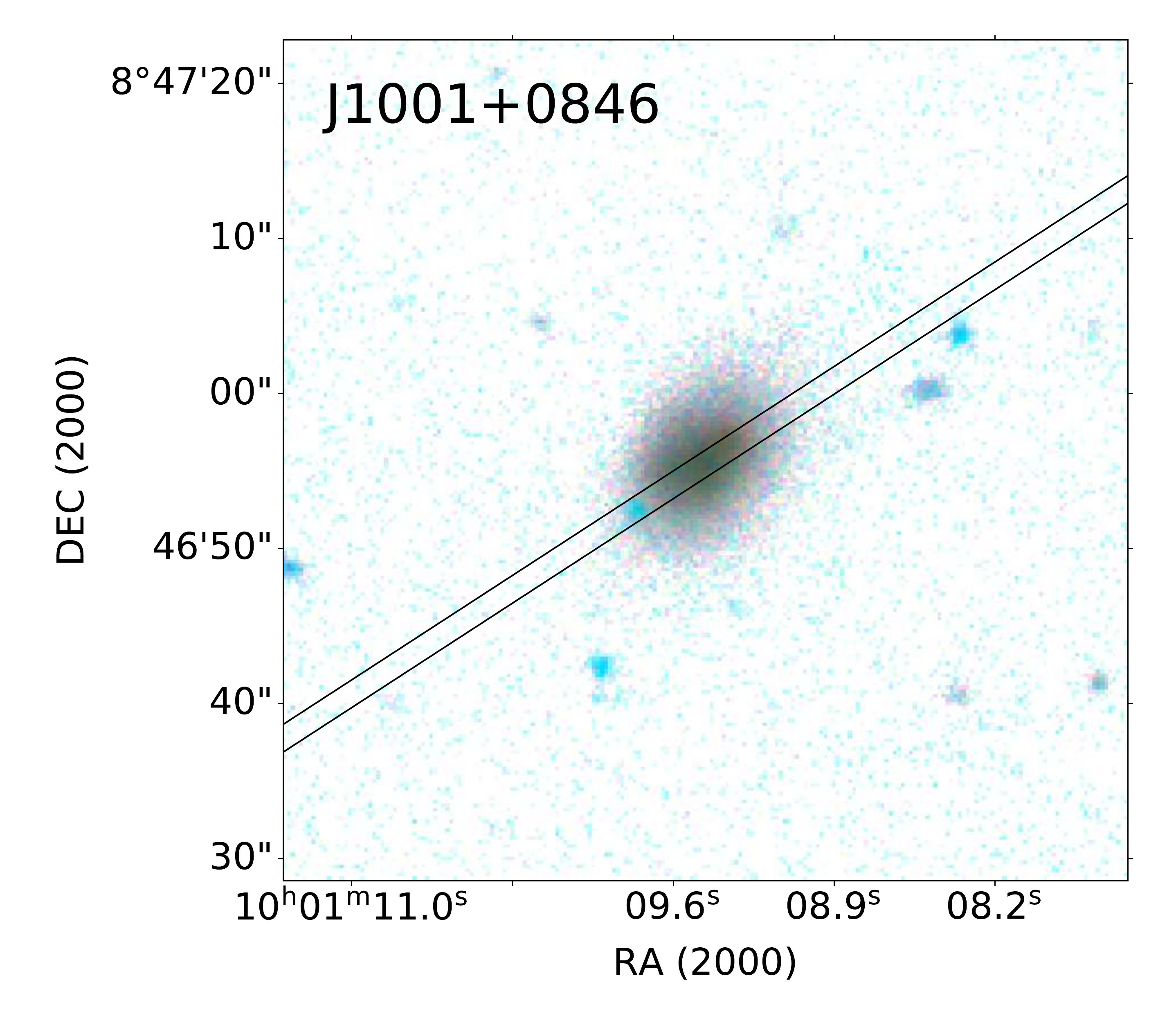}
\includegraphics[width=4.0cm,angle=0,clip=]{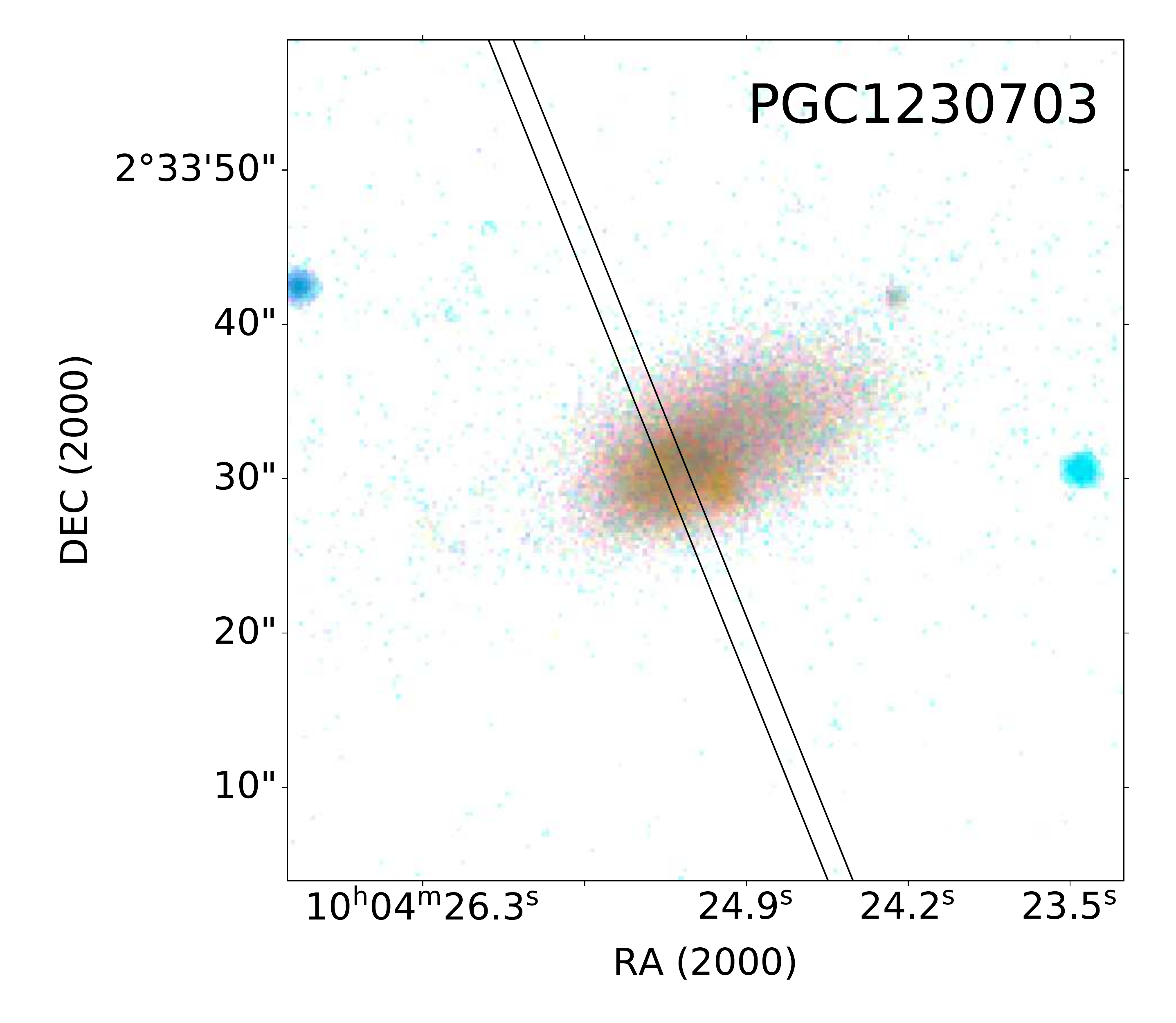}
\includegraphics[width=4.0cm,angle=0,clip=]{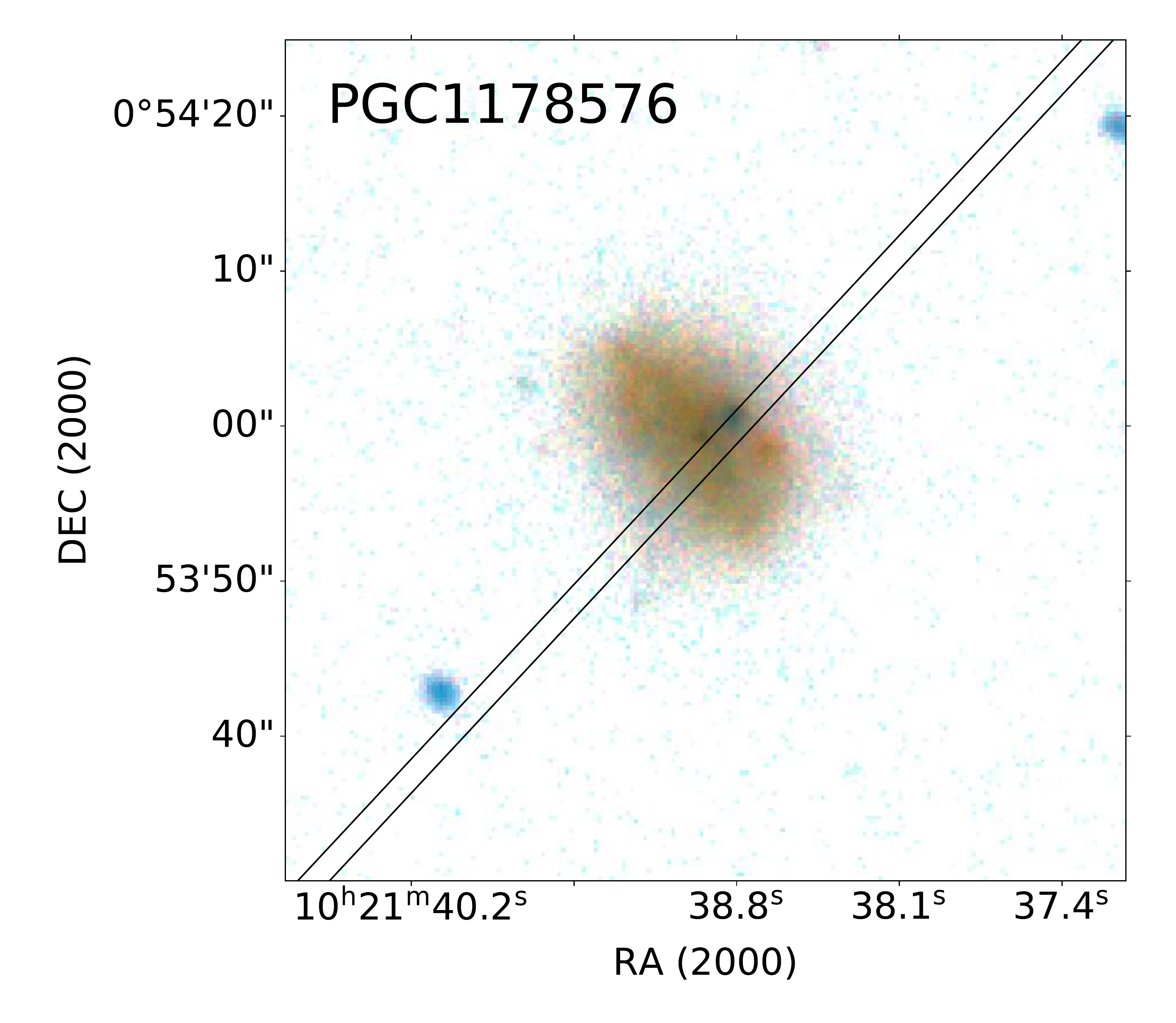}
\includegraphics[width=4.0cm,angle=0,clip=]{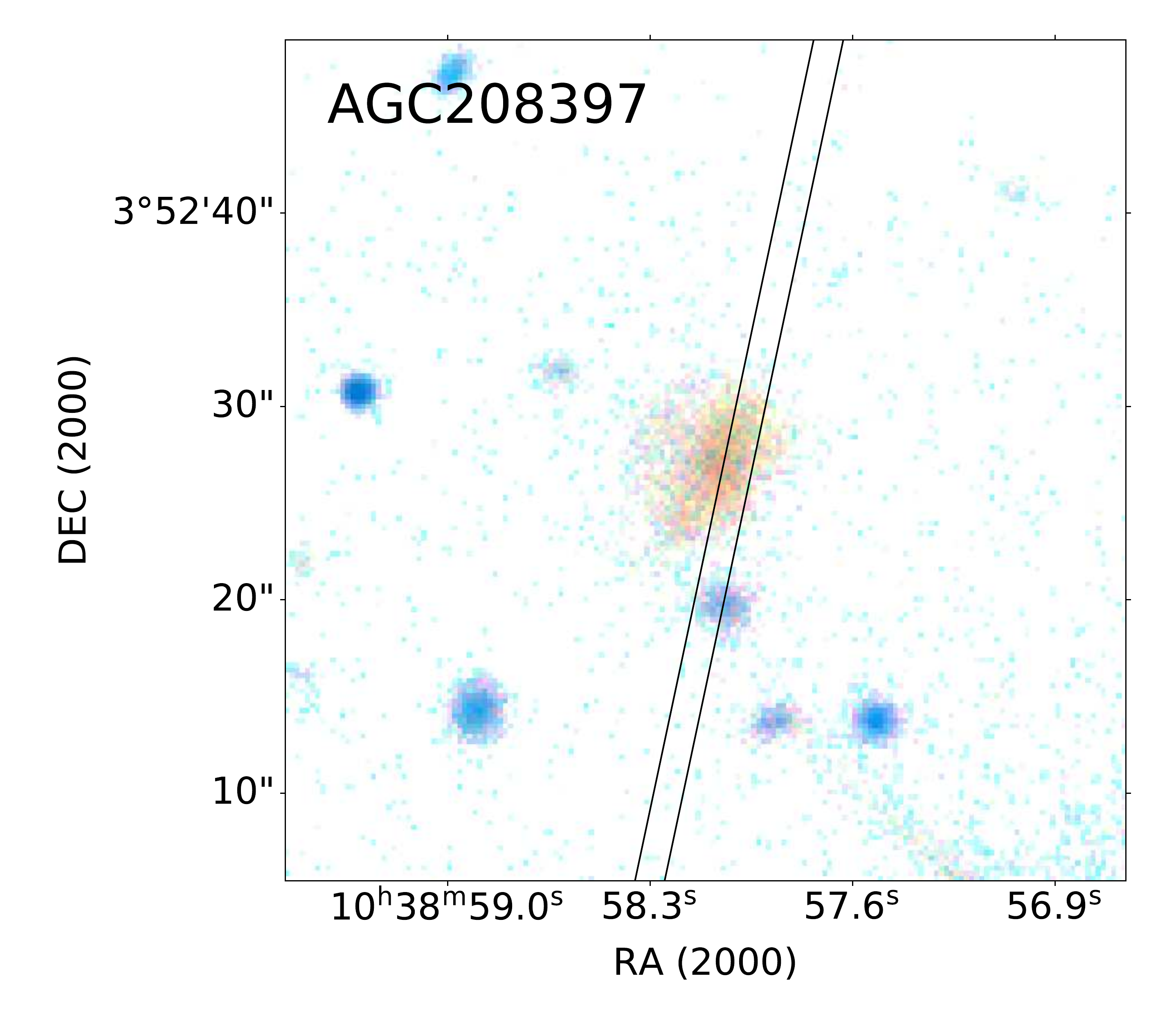}
\includegraphics[width=4.0cm,angle=0,clip=]{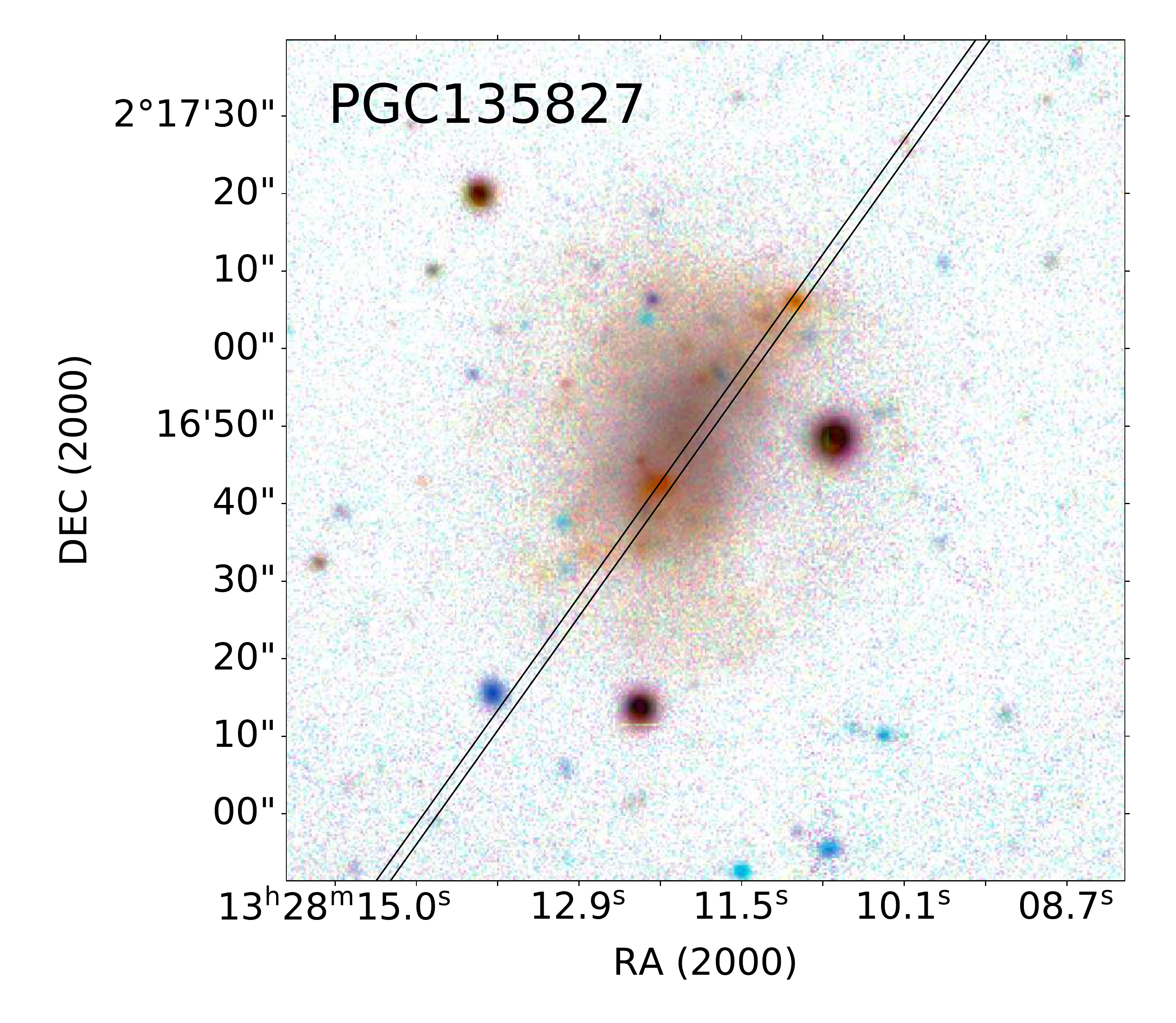}
\includegraphics[width=4.0cm,angle=0,clip=]{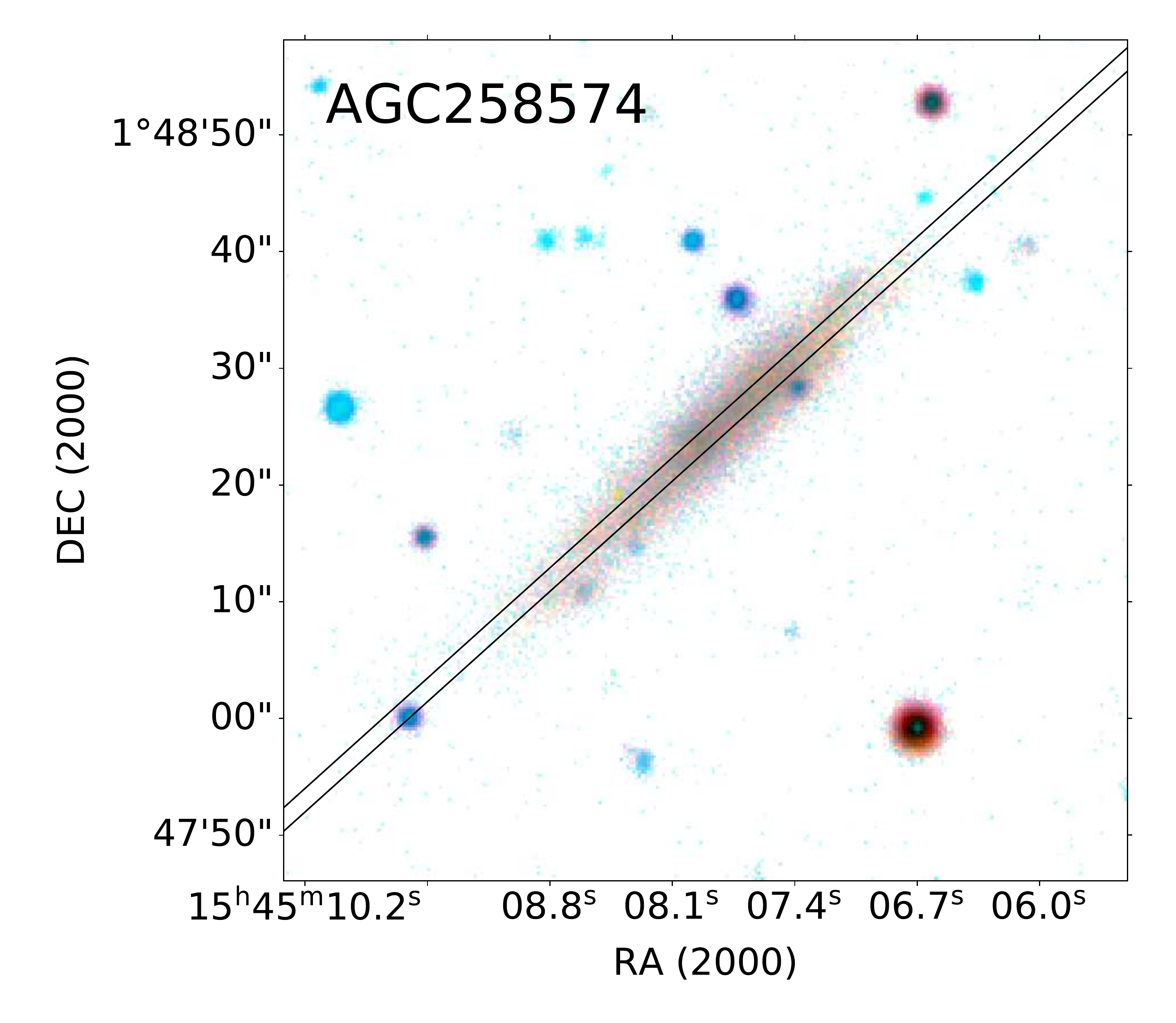}
\includegraphics[width=4.0cm,angle=0,clip=]{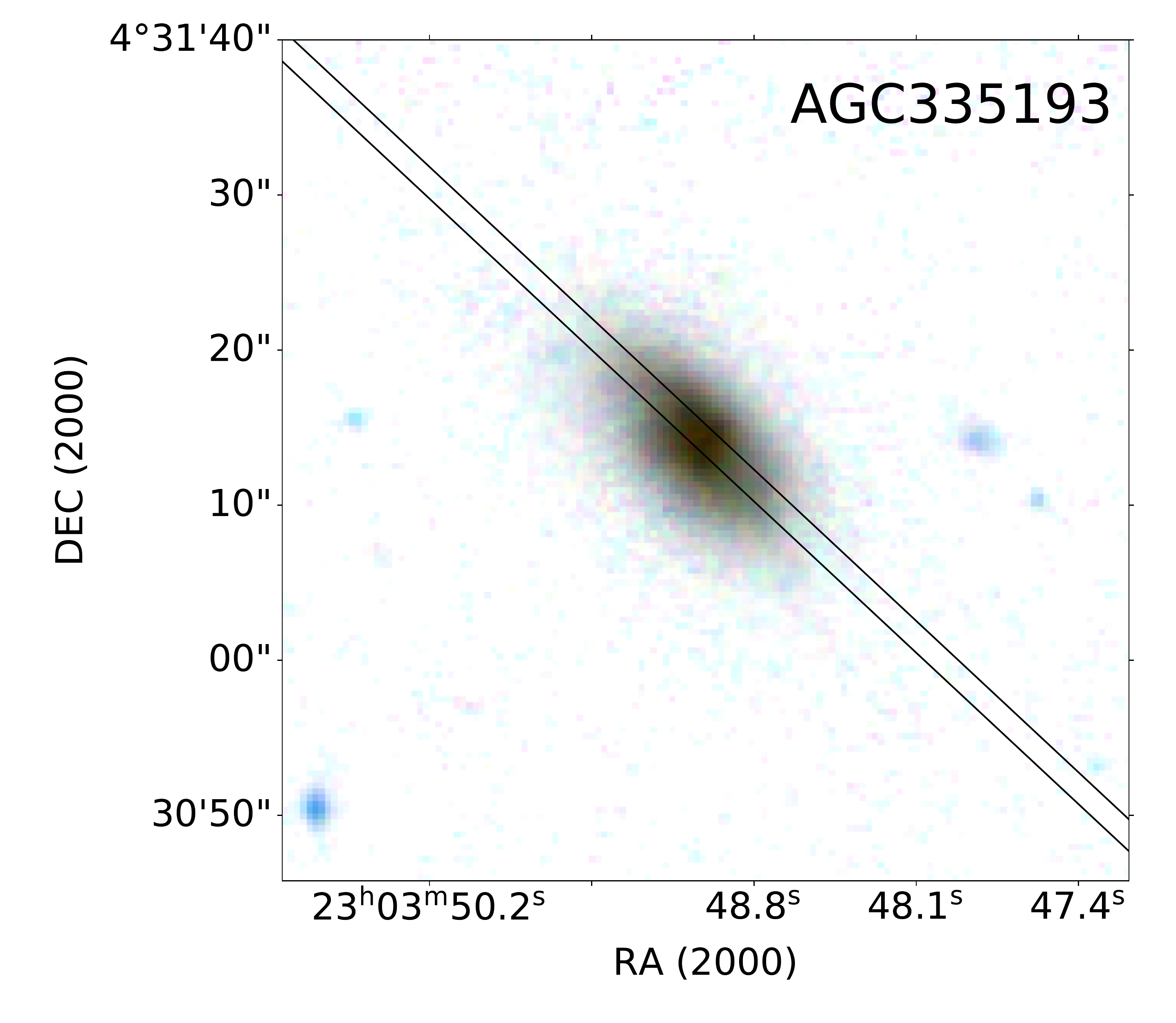}
\caption{Images of observed galaxies with slit positions superimposed.
They are prepared from galaxy images available in the SDSS, PanSTARRS PS1 and
DECaLS public databases. Their colours are inverted to better
emphasize low surface brightness of the majority observed objects.
Galaxy names are printed at the tops of the images. See Table 2 for more
details.
	}
\label{fig:F.charts1}
\end{figure*}

\begin{figure*}
\includegraphics[width=4.0cm,angle=0,clip=]{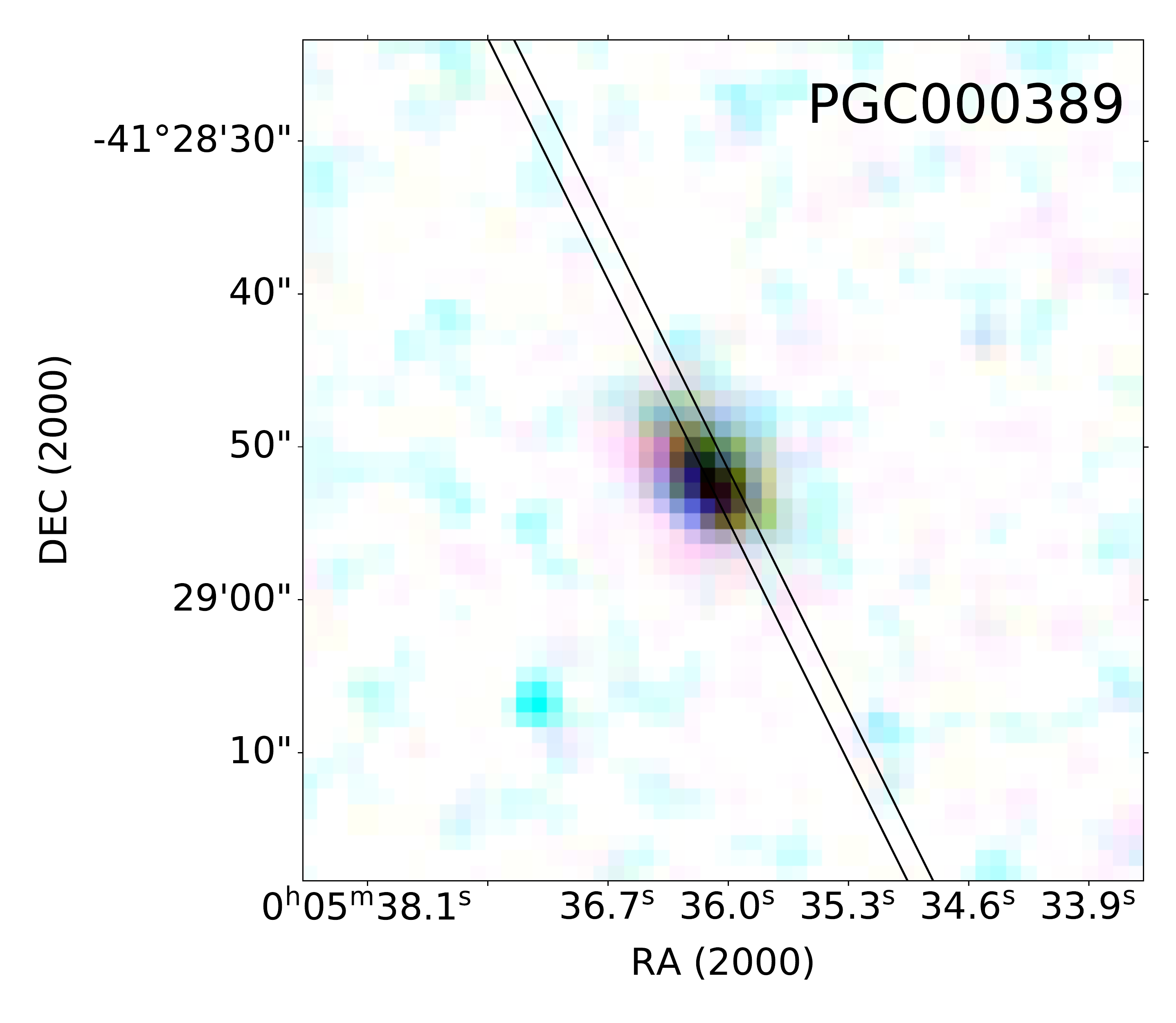}
\includegraphics[width=4.0cm,angle=0,clip=]{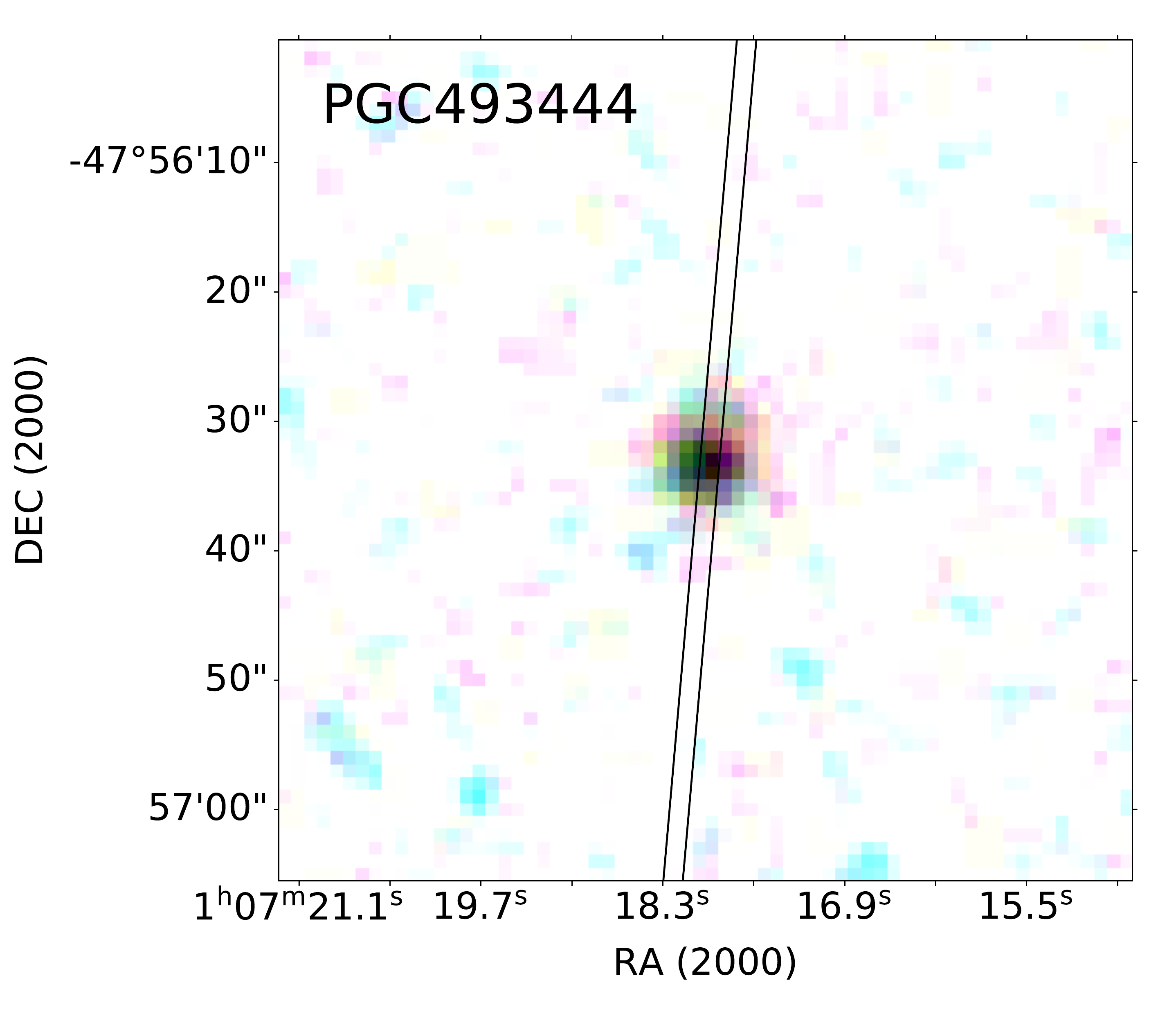}
\includegraphics[width=4.0cm,angle=0,clip=]{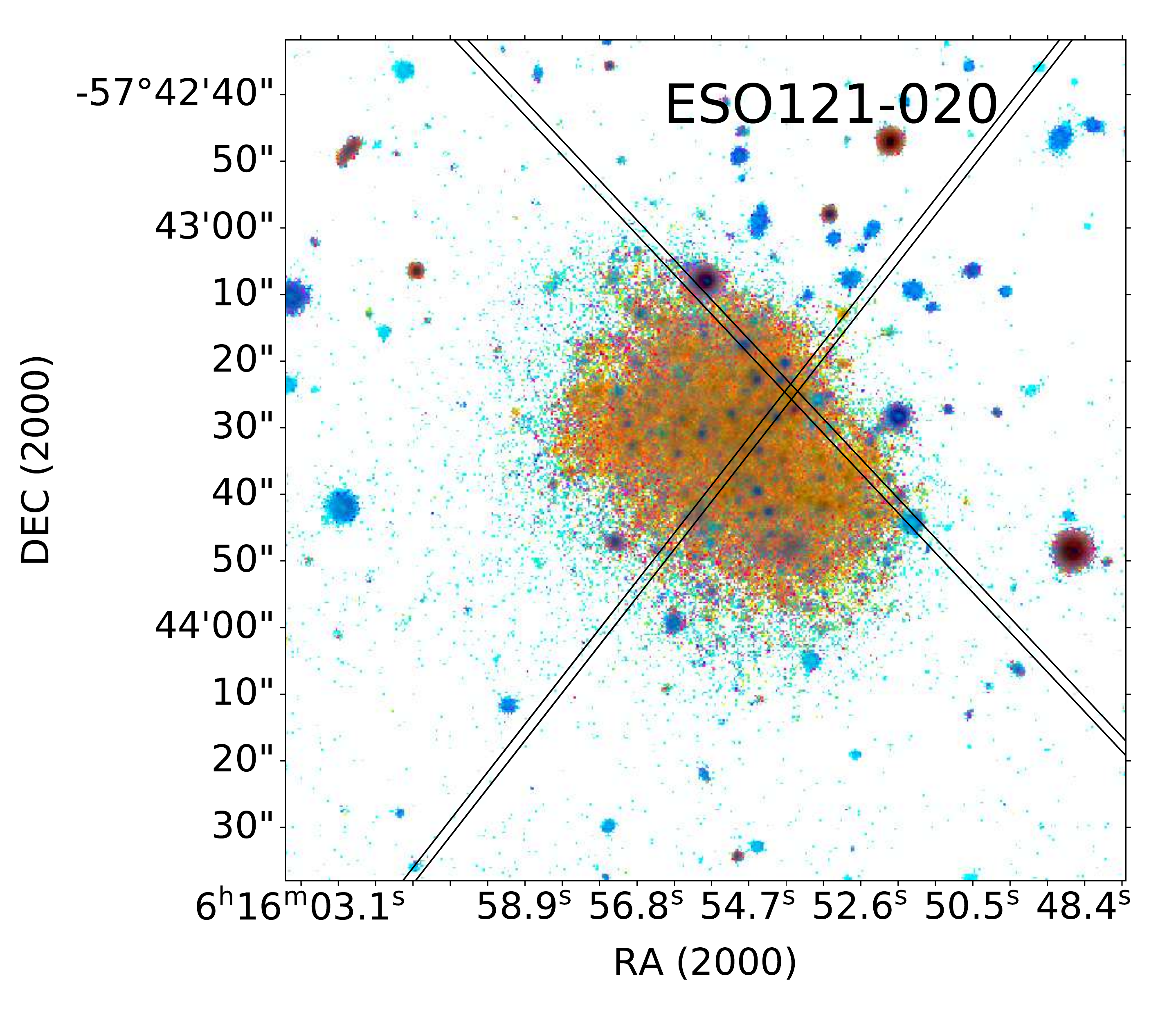}
\includegraphics[width=4.0cm,angle=0,clip=]{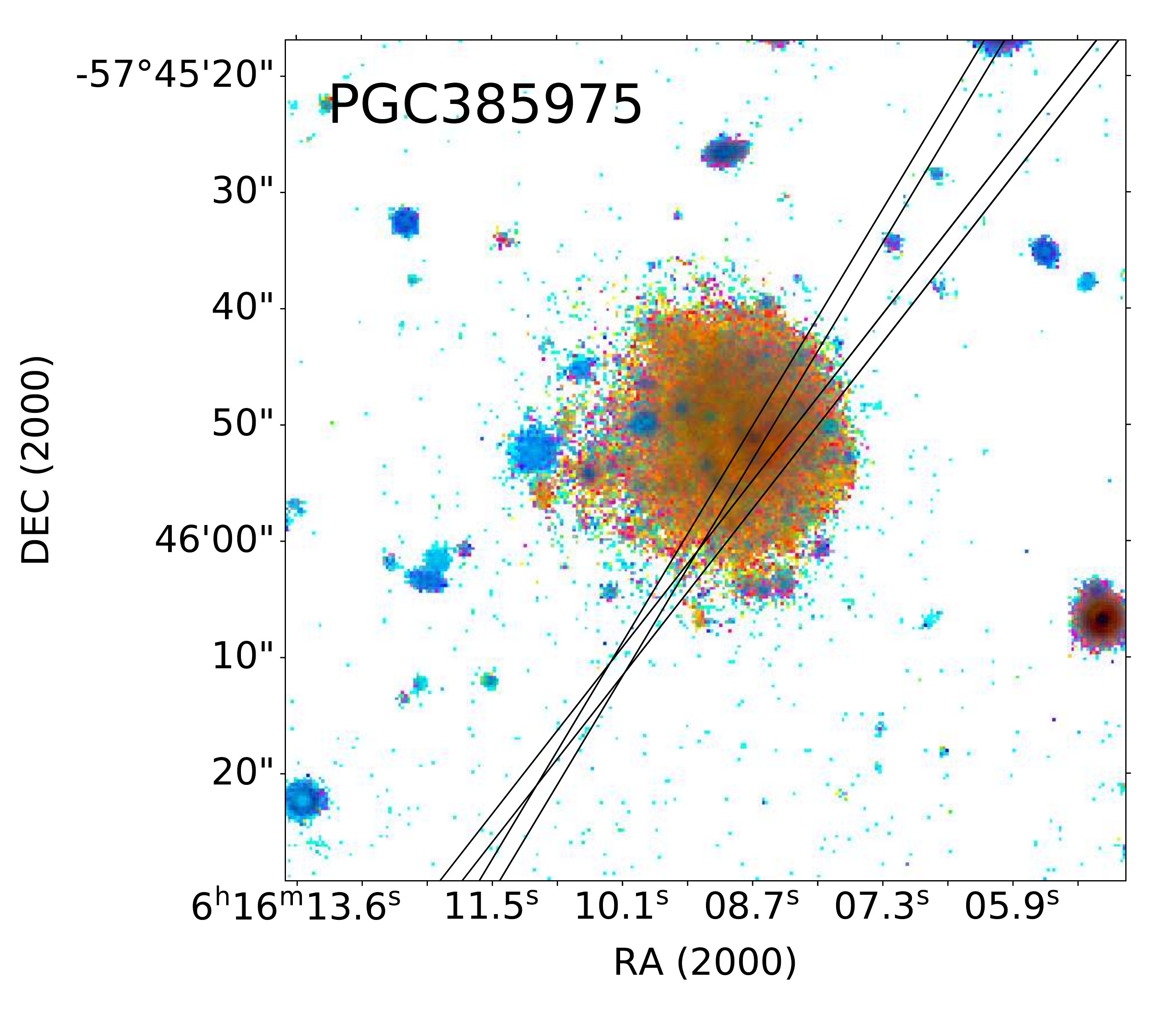}
\includegraphics[width=4.0cm,angle=0,clip=]{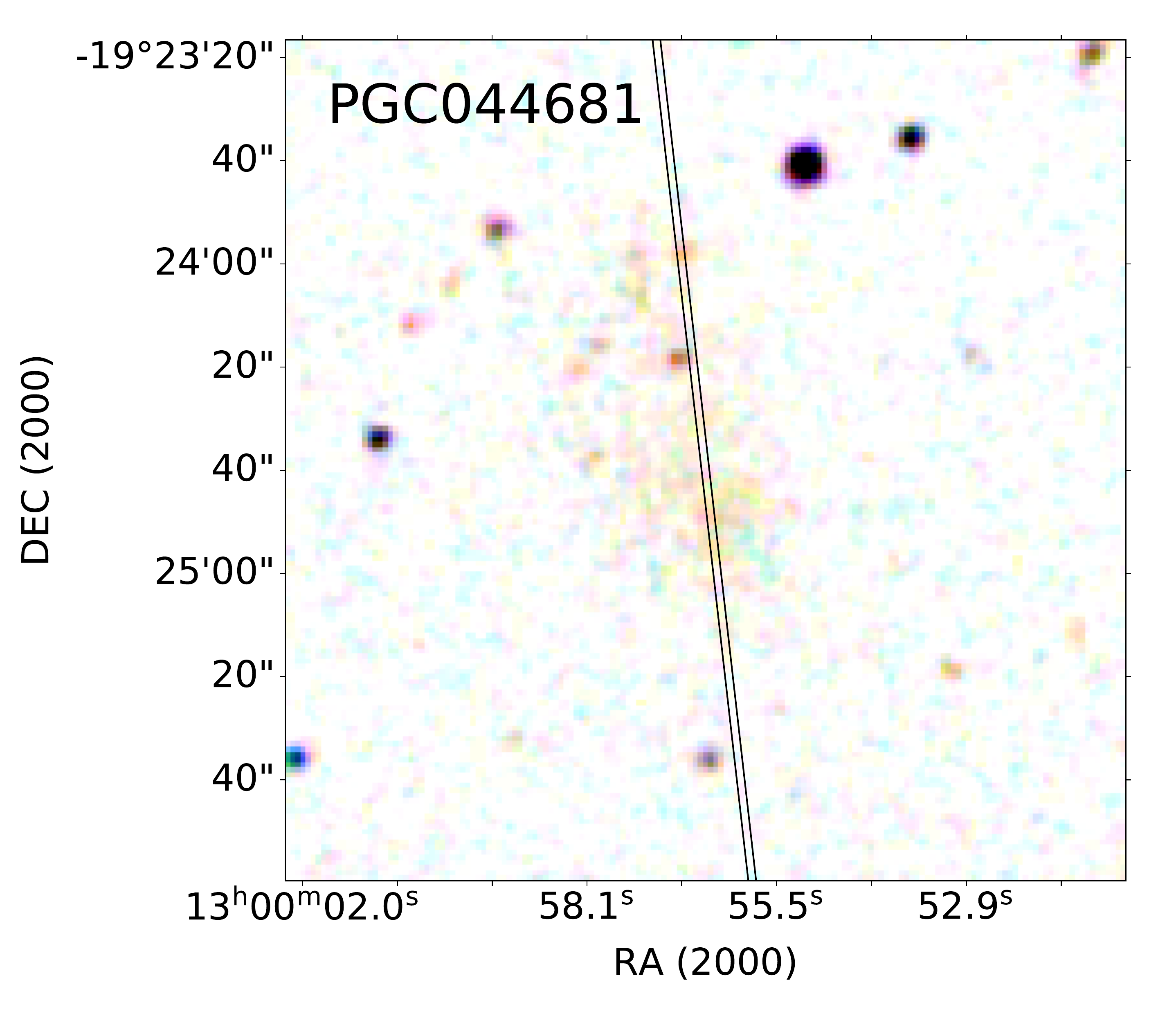}
\includegraphics[width=4.0cm,angle=0,clip=]{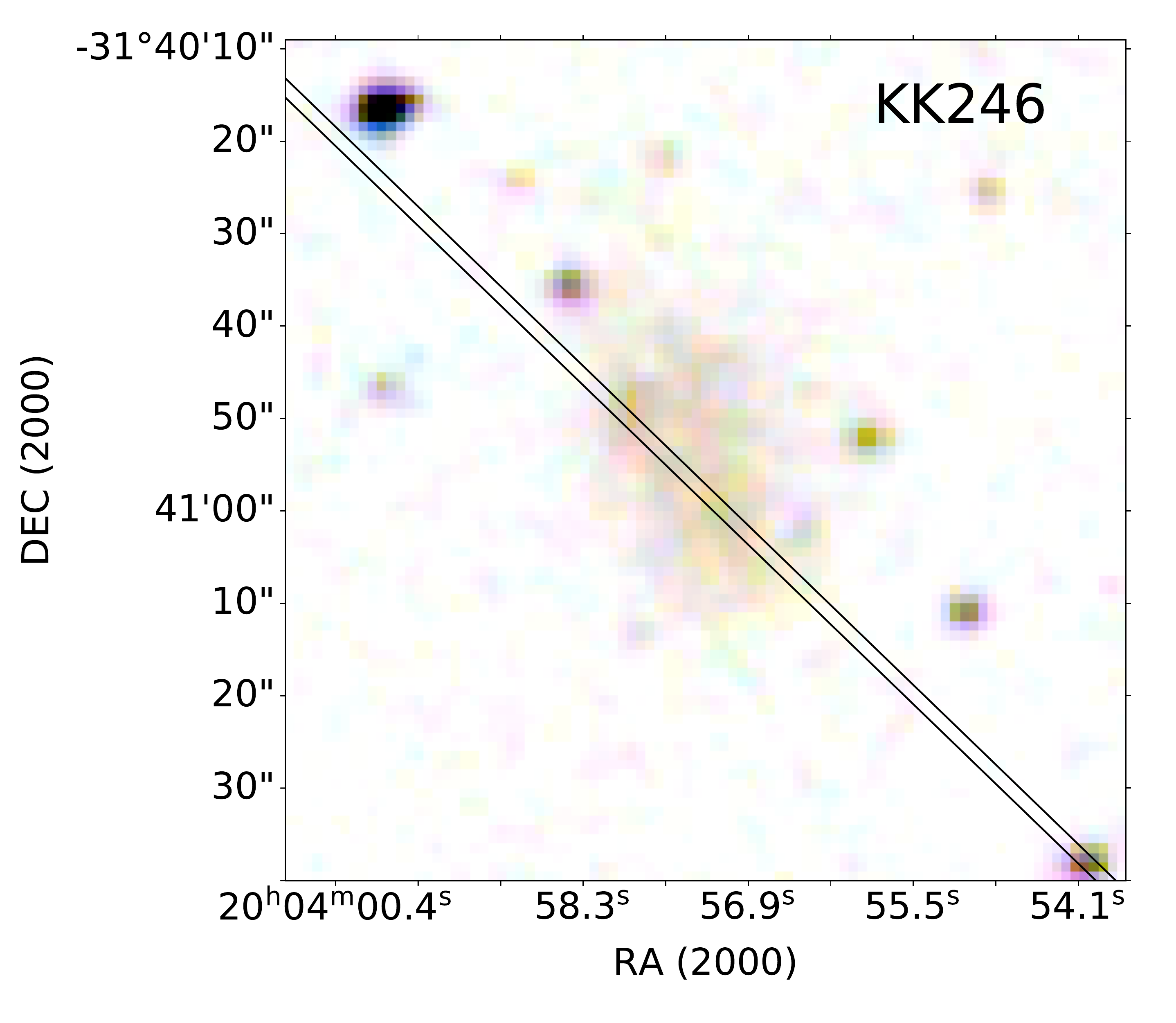}
\caption{Images of the southern-most observed galaxies with slit positions
superimposed. They are prepared from galaxy images available in the ESO DSS
(digitized photographic
sky surveys based on POSS and SERC) or in DECaLS. Their colours are inverted
to better emphasize low surface brightness of the majority observed objects.
Galaxy names are printed at the tops of the images. See Table 2 for more
details.
	}
\label{fig:F.charts2}
\end{figure*}

As described above, the majority of observed candidate XMP dwarfs are faint
and/or of low surface brightness. Therefore an attempt to obtain their
independent spectra and to point to the right region can be a problem.
To provide an opportunity for independent checks of our data, we present
in Table~\ref{tab:journal} the long slit position angles (PA) and in
Figures~\ref{fig:F.charts1},\ref{fig:F.charts2} their images with
superimposed long slit positions.
These images are taken from the SDSS \citep{DR14}, PanSTARRS PS1
\citep{PS1-database}, DECaLS \citep{DECaLS} databases
and from the ESO Online DSS archive (https://archive.eso.org/dss/dss).

The primary data reduction was done with the SALT science pipeline
\citep{Cra2010} which includes bias and overscan subtraction and gain
correction for each CCD amplifier, cross-talk correction and finally
mosaicing. The long-slit reduction, 1D spectra extraction of individual
\HII-regions was done in the way described in the recent paper by
\citet{Eridanus}.

\section[]{Line measurements and O/H determination}
\label{sec:OH}

The emission line fluxes obtained from 1D spectra were measured as
described in detail in \citet{Kniazev04}. Here we summarize briefly the main
procedures. They include the robust determination of the continuum and its noise
level, the subsequent \mbox{MIDAS}\footnote{MIDAS is an acronym for the
European Southern Observatory package -- Munich Image Data Analysis
System.}-based
programs for determination of parameters of emission lines. This
uses the procedure of Gauss fitting of emission lines in the
continuum-subtracted spectrum. Some lines were fitted as a
blend of two or three Gaussians. The quoted errors of line intensities include
three components: first -- the fitting error from the MIDAS program, related
to the Poisson statistics of line photon counts; second --  the error
resulting from the creation of the underlying continuum; and third -- errors related
to the accuracy of the sensitivity curve used to transfer counts to relative
flux units. The errors of the sensitivity curve are typically no more than
(1--2)~\% so that their contribution to the total error budget is small.

After the line fluxes were measured, we performed an iterative procedure
described by \citet{ITL94}, which accounts for dust extinction and the
absorptions in the Balmer lines from the underlying young stellar clusters.
This results in the simultaneous estimate of the equivalent width of
absorption Balmer lines $EW(abs)$ and the extinction coefficient $C(H\beta)$.
The relevant equation (1) from \citet{ITL94} was used:
$$ I(\lambda)/I(H\beta) = [EW_{e}(\lambda)+EW_{a}(\lambda)]/EW_{e}(\lambda) \times  $$
$$ EW_{e}(H\beta)/[EW_{e}(H\beta)+EW_{a}(H\beta)] \times $$
$$ F(\lambda)/F(H\beta)~exp[C(H\beta)f(\lambda)]  $$

Here $I(\lambda)$ is the intrinsic line flux corrected for the overall
extinction (both
in the Milky Way and internal to a particular galaxy) and the underlying
Balmer absorption, while $F(\lambda)$ is the measured line flux.
$EW_{e}(\lambda)$ are equivalent widths of used emission lines.
$EW_{a}(\lambda)$ is the adopted value of the underlying Balmer absorptions.
This term is used to estimate intrinsic fluxes only for Balmer emission lines.
$f(\lambda)$ is the reddening function adopted from \citet{W1958} and
normalized so that $f(H\beta)$ = 0.  For the adopted reddening function,
there is a relation between the excess E($B-V$) and $C(H\beta)$:
E($B-V$) = 0.68 $\times$ $C(H\beta)$.

Due to the low fluxes in the emission lines in the majority of observed
\HII-regions, and due to their low metallicities, the principal weak line,
[O{\sc iii}]$\lambda$4363~\AA, used for the determination of the electron
temperature T$_{\rm e}$ in the 'direct' (T$_{\rm e}$) method,
was not detected in most of our targets. 
Therefore, for the estimate of O/H
in the majority of the observed galaxies, we used the
semi-empirical method suggested by \citet{IT07}. It was carefully
checked and calibrated by the authors and later by us \citep{PaperVII}.

This method uses the fitted empirical dependence between the electron
temperature T$_{\rm e}$ and the value of parameter $R_{\mathrm 23}$. This
dependence was derived from the analysis of the grid of models of
\HII-regions in \citet{SI2003}. The models approximate well the apparent
relations of strong line intensities versus EW(H$\beta$) for the large
representative sample of extragalactic \HII-regions, which cover the whole
range of observed O/H. Here $R_{\mathrm 23}$ is the ratio of the sum of fluxes
of strong oxygen lines [O{\sc ii}]$\lambda$3727~\AA,
[O{\sc iii}]$\lambda$4959~\AA, [O{\sc iii}]$\lambda$5007~\AA\ to the flux
of H$\beta$.
When T$_{\rm e}$ is estimated via $R_{\mathrm 23}$, the rest of the
calculations uses the standard equations of the classic T$_{\rm e}$ method.
The O/H estimates derived by this method are marked as (se)
in Column~9 of Table~\ref{tab:prop_summary}.

For \HII-regions in the lowest metallicity regime,
\citet{Izotov19DR14} suggested recently an improved empirical method,
which uses the relative fluxes of strong Oxygen lines with respect of
H$\beta$. Namely, their Equation (1) reads as:
$$12+\log({\rm O/H}) = 0.950 \times \log(R_{\mathrm 23} - 0.08\times O_{\mathrm 32}) + 6.805 $$
Here $O_{\mathrm 32}$ is the flux ratio of the line
[O{\sc iii}]$\lambda$5007~\AA\ to that of [O{\sc ii}]$\lambda$3727~\AA.
This relation, calibrated on the large number of \HII-regions with
O/H derived via the direct (T$_{\rm e}$) method, empirically accounts for
the large
scatter in the ionization parameter $U$ in various \HII-regions and thus
reduces the relatively large internal scatter of other methods based
on the strong oxygen lines down to only $\sim$0.05~dex.

However, its applicability is limited only to the range of
12+$\log$(O/H)$\lesssim$7.4. This corresponds to the limit of the combination
$R_{\mathrm 23} - 0.08\times O_{\mathrm 32} \lesssim $ 4.0.
Since we have among our observed galaxies a dozen objects satisfying this
condition, we use this method for them and attach (s) to the derived
value of O/H
in Column~9 of Table~\ref{tab:prop_summary}, and denote them hereafter
as O/H(s).

We notice further that as is evident from the \citet{Izotov19DR14} plots in
Fig.3b, there is a small offset in the zero-point of O/H(s) relative to
that of O/H(T$_{\rm e}$), of $\sim$0.03 dex.
Therefore, we need to subtract 0.03~dex from O/H(s) ratios
to compare them directly with the estimates of O/H(T$_{\rm e}$)
for other galaxies.

The errors of O/H(s), $\sigma_{\rm log(O/H)}$, due to observational
uncertainties in the strong line fluxes are rather small. For our highest
signal-to-noise (S/N) spectra, with
$\sigma_{\rm R23}$/$R_{\mathrm 23} <$0.05, the
total error is close to the internal error of the method,
$\sigma_{\rm log(O/H)} \sim$0.05~dex.  For our lowest S/N spectra,
with $\sigma_{\rm R23}$/$R_{\mathrm 23} \sim $0.14,
the final error of $\log$(O/H)(s) increases to 0.08~dex.
For intermediate values of S/N of $R_{\mathrm 23}$, the total
error is $\sigma_{\rm log(O/H)}$ = 0.06 -- 0.07 dex.

\setcounter{qub}{0}
\begin{table*}
\caption{Observed candidate void XMP dwarfs and new O/H data}
\begin{tabular}{r|l|l|r|r|r|r|l|l|p{3.0cm}} \hline\hline
No.&\MC{1}{c}{Name}&\MC{1}{c}{J2000 Coord}&\MC{1}{c}{V$_{\rm h}$}& D &B$_{\rm t}$   &$M_{\rm B}$&M(\HI)/ & 12+$\log$(O/H) &\MC{1}{c}{Notes} \\
   &              &                  &\kms   & Mpc  &mag  &mag    &$L_{\rm B}$    & $\pm$err.       &       \\
 1 &\MC{1}{c}{2}  &\MC{1}{c}{3}&\MC{1}{c}{4}   &   5  & 6   &   7   &     8       &\MC{1}{c}{9} &\MC{1}{c}{10}   \\
\hline
\qq&PGC000389     & J000535.9--412856& 1500& 17.5&18.23&--13.06& ...   & 7.74$\pm$0.10 (se) &  \\ 
\qq&PGC736507     & J000936.2--285138& 7594&104  &19.14&--16.00& ...   & 7.58$\pm$0.08 (se) & 2dF: wrong V$_{\rm h}$=898  \\ %
\qq&HIPASSJ0021+08& J002041.7+083655 &  693& 9.9 &17.22&--13.34& 1.25  & 7.51$\pm$0.07 (se) & \\ 
\qq&AGC104227     & J005823.7+041825 & 1198&16.9 &18.12&--13.12& 2.11  & ...                & Faint H$\alpha$ at V$_{\rm h}$(\HI) \\ 
\qq&PGC493444     & J010718.0--475633& 7050&95.4 &19.22&--15.70& ...   & ...                & 2dF: wrong V$_{\rm h}$=837 \\ 
\qq&PGC1190331    & J010910.1+011727 & 1094&15.4 &17.54&--13.49& 2.24  & 7.48$\pm$0.08 (se) & \\ 
\qq&AGC411446     & J011003.7--000036& 1137&15.9 &19.82&--11.32& 6.55  & 7.05$\pm$0.05 (s)  & \\ 
\qq&AGC114584     & J011250.5+015207 & 1089&15.4 &18.08&--13.02& 1.70  & 7.15$\pm$0.05 (s)  & \\ 
\qq&AGC123223     & J024709.3+100516 & 767 & 12.4&18.16&--13.27& 2.84  & 7.47$\pm$0.09 (s)  & \\ 
\qq&AGC124629     & J025605.6+024831 & 794 & 12.4&19.46&--11.58& 3.61  & 6.95$\pm$0.06 (s)  & \\ 
\qq&AGC132121     & J030644.1+052008 & 678 & 11.0&17.27&--13.68& 1.58  & 7.30$\pm$0.06 (s)  & \\ 
\qq&ESO121-020    & J061554.3--574332& 582 &  6.1&15.73&--13.37& 3.30  & 7.26$\pm$0.05 (s)  & aver. 2 knots \\ 
\qq&PGC385975     & J061608.5--574551& 554 &  6.1&17.01&--11.91& 2.22  & 7.29$\pm$0.07 (s)  & aver. 2 measur. \\ 
\qq&AGC174605     & J075021.7+074740 & 351 &  9.9&18.68&--11.40& 2.90  & ...                & Only H$\alpha$ emission \\ 
\qq&AGC188955     & J082137.0+041901 & 758 & 12.8&17.70&--12.94& 0.85  & 7.73$\pm$0.08      & aver (Te,se). Broad component \\ 
\qq&AGC198454     & J092811.3+073237 &1373 & 21.0&18.51&--13.32& 1.20  & 7.52$\pm$0.09 (se) & \\ 
\qq&PGC1314481$\dagger$& J094805.9+070743 & 526 &  9.2&16.98&--12.96& 1.63  & 7.75$\pm$0.15 (se) & \\ 
\qq&J1001+0846    & J100109.5+084656 &1265 & 19.2&18.10&--13.56& 0.30  & 7.60$\pm$0.08 (se) & \\ 
\qq&PGC1230703    & J100425.1+023331 &1126 & 17.1&18.47&--12.80& 0.71  & 7.66$\pm$0.08 (se) & \\ 
\qq&PGC1178576    & J102138.9+005400 & 701 & 11.0&17.27&--13.15& 1.30  & 7.25$\pm$0.06 (s)  & \\ 
\qq&AGC208397     & J103858.1+035227 & 763 & 11.9&19.95&--10.59& 5.60  & 7.13$\pm$0.05 (s)  & \\ 
\qq&PGC044681     & J125956.6--192441& 827 &  7.3&17.00&--12.73& 3.08  & 7.20$\pm$0.08 (s)  & Faint [O{\sc iii}] lines \\ 
\qq&PGC135827     & J132812.2+021642 &1008 & 13.5&16.51&--14.25& 3.77  & 7.74$\pm$0.11 (Te) & \\ 
\qq&AGC258574     & J154507.9+014822 &1523 & 19.3&17.72&--13.09& 2.60  & 7.23$\pm$0.07 (s)  & \\ 
\qq&KK246         & J200357.4--314054& 431 &  7.1&17.07&--13.49& 2.33  & 7.67$\pm$0.08 (se) & \\ 
\qq&AGC335193     & J230349.0+043113 & 1125& 16.1&17.12&--14.20& 0.48  & 7.57$\pm$0.07 (se) & \\ 
\hline
\multicolumn{10}{p{17.5cm}}{Table~2 content is described in detail in the third paragraph of Sect.~\ref{sec:results}. Here we give brief information. Col.~2: target name from NVG.} \\
\multicolumn{10}{p{17.0cm}}{Col.~3: galaxy coordinates adopted from NVG. Col.~4: radial velocity in \kms. Columns~5, 6 and 7: the adopted distance, } \\
\multicolumn{10}{p{17.0cm}}{total blue magnitude and absolute blue magnitude. Col.~8: mass ratio of \HI\ to blue luminosity, in solar units; Col.~9: derived O/H} \\
\multicolumn{10}{p{17.0cm}}{as 12+$\log$(O/H) and its error, in dex. In parentheses we indicate the method used: (T$_{\rm e}$), direct method; (se), semi-empirical } \\
\multicolumn{10}{p{17.0cm}}{method of \citet{IT07}, with a small correction from \citet{PaperVII}; (s), the new empirical strong line O/H} \\
\multicolumn{10}{p{17.0cm}}{estimator of \citet{Izotov19DR14}, with subtracted 0.03 dex to account for a small offset relative to O/H(T$_{\rm e}$).} \\
\multicolumn{10}{p{17.0cm}}{In Col.~10 we show brief notes with more detailed information, when necessary, presented in Sec.~\ref{sec:individual}. } \\
\multicolumn{10}{p{17.0cm}}{$\dagger$ in Col~2, O/H(se) based on SALT and SDSS \citep{DR7,DR14} spectra. } \\
\end{tabular}
\label{tab:prop_summary}
\end{table*}

\section[]{Results of spectral observations and O/H estimates}
\label{sec:results}

The 1D SALT spectra of the XMP dwarf candidates are presented in Appendix~A,
Figs. \ref{fig:SALT_1Dp1} and \ref{fig:SALT_1Dp2}.
The measurements of emission line fluxes as well as derived parameters:
the extinction  coefficient C(H$\beta$), the adopted equivalent width
of Balmer absorption in
the underlying stellar continuum EW(abs), and the equivalent width of
the H$\beta$ emission line EW(H$\beta$) are presented in Appendix~B,
Tables~B1-B8. For some of the spectra where Balmer absorption was
clearly visible in the UV, we modelled the underlying continuum with
the ULySS package \citep[http://ulyss.univ-lyon1.fr,][]{Koleva2009}.
This model continuum fitted Balmer absorption, and thus
corrected  to a first approximation the flux
of H$\beta$ emission. For these objects, the EW(abs) derived in the next
step via iterations as described above with the procedure from \citet{ITL94},
relates in fact to the residual EW(abs), which is already mainly
accounted for by the ULySS fitting.
We do not give the absolute flux in the emission H$\beta$ since
as explained in Section~\ref{sec:observing}, due to the nature of
SALT observations, absolute flux calibration is impossible.

At the bottom of these tables we also give the derived T$_{\rm e}$ in the
zones of emission of [O{\sc iii}] and [O{\sc ii}], the relative numbers of
ions O$^{+}$, O$^{++}$ and the total abundance of Oxygen
relative to Hydrogen, O/H. Finally,
we present for each galaxy the derived parameter 12+$\log$(O/H) obtained
with the direct method (for a couple objects where applicable), with the
semi-empirical method of \citet{IT07} (for all our objects) and with the
empirical strong-line method of \citet{Izotov19DR14} (for the 10 lowest O/H
objects when applicable).  We note that 12+$\log$(O/H) derived both with the
semi-empirical method and with this new strong-line method are corrected
by several 0.01~dex to  make zero-points consistent with that for the
direct T$_{\rm e}$ method. See Appendix in \citet{PaperVII} and
Sec~\ref{sec:OH} above for details.

In Table~\ref{tab:prop_summary} we summarize all adopted O/H estimates along
with some other important galaxy parameters.
The columns include the following information: Col.~1 -- the target number,
the same as in Table~\ref{tab:journal};
Col.~2 -- the galaxy name as it appears in the NVG catalog, mainly adopted
from the HyperLEDA database\footnote{http://leda.univ-lyon1.fr};
Col.~3  -- J2000 epoch coordinates; Col.~4 -- heliocentric velocity in
\kms; Col.~5 -- Distance in Mpc, either measured
with the Tip of the RGB method,
or with the use of the peculiar velocity correction according to the velocity
field from \citet{Tully08} as adopted in the Nearby Void Galaxies catalog
(PTM19); Col.~6 -- an estimate of the total $B$-band magnitude;
Col.~7 -- the corresponding absolute magnitude $M_{\rm B}$ with the
MW extinction
correction from \citet{Schlafly11}; Col.~8 -- the \HI\ gas-mass to luminosity
ratio M(\HI)/$L_{\rm B}$, in solar units;
Col.~9 -- the parameter 12+$\log$(O/H) with its 1-$\sigma$ uncertainty and the
method used (in parentheses); in Col.~10 we provide notes for some of the
program galaxies.

Of 26 observed targets, the Oxygen emission lines were detected for 23
galaxies, displaying either Emission-Line type spectra (ELG), or Emission
and Absorption line spectra (E+A). However, one of these 23 galaxies
appears to be a distant object at $D \sim$ 100~Mpc. One  other distant
galaxy appeared among the selected XMP
void candidates with only absorption lines visible in its spectrum. Both
these cases are discussed in more detail in Sect.~\ref{sec:individual}.
In two more faint LSB dwarfs only H$\alpha$ emission was detected in the
SALT spectra. While a more careful check for the possible presence
of other \HII-regions can prove useful, the
available data indicate a mostly terminated star-formation episode
in these two void LSB dwarfs.

For the remaining 22 Nearby Void XMP dwarf candidates, we obtained 
estimates of O/H with good to acceptable quality. Their derived values of
12+$\log$(O/H) lie in the range of 6.95 to 7.75~dex. Four of them, with
12+$\log$(O/H) in the range of 6.95 -- 7.15~dex, appear to be new
XMP dwarfs near
the edge of the galaxy gas-metallicity distribution,
(Z\sunn/50 $ \lesssim$ Z $\lesssim $ Z\sunn/30), adding a substantial
fraction to about a dozen known galaxies with such a low O/H
residing within the nearest cell of the Universe of $R \lesssim 25$ Mpc.
 For only four of
these 22 objects does O/H marginally (that is within the cited uncertainties)
exceed the level of (O/H)\sunn/10, corresponding to 12+$\log$(O/H) =
7.69~dex. A more detailed discussion of the presented results  follows
in  Sec.~\ref{sec:discussion}

\section[]{Discussion}
\label{sec:discussion}

\begin{figure*}
\includegraphics[width=12.0cm,angle=-90,clip=]{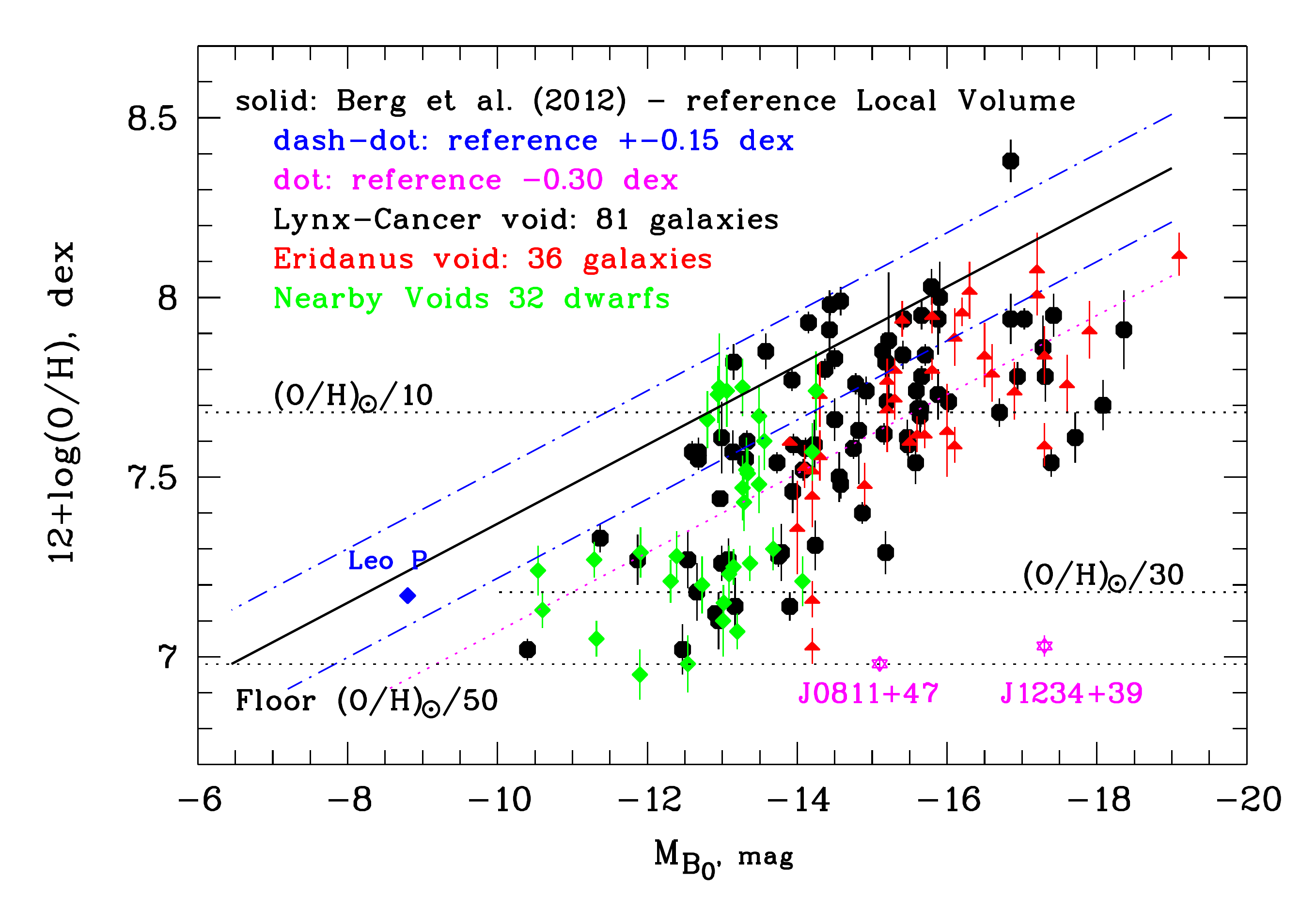}
\caption{Positions of 22 NVG XMP candidates studied with SALT,
plus one known from the literature (J2104--0035) and nine
candidates from the BTA part of the program (a paper in preparation),
in the diagram 12+$\log$(O/H) versus absolute blue magnitude
$M_{\rm B}$ (green rhombs) along with similar data for the Lynx-Cancer
(black octagons, Pustilnik et al. 2016) and Eridanus (red triangles,
Kniazev et al. 2018) voids. The solid line shows the relation between
12+$\log$(O/H) and $M_{\rm B}$ for the reference sample of the Local Volume
late type galaxies from \citet{Berg12}, extrapolated for
$M_{\rm B} >$ --10.5. The dotted line running 0.3~dex lower in
$\log$(O/H), separates the region in the diagram below
--2 $\sigma_{\rm scatter}$ from the reference line. Points in this region
are treated as significant outliers.
Two outstanding distant record-low metallicity XMP BCGs J0811+4730 and
J1234+3901 from the recent papers by
Izotov et al. (2018, 2019) are shown by purple stars for comparison.
See a more detailed discussion in the text.
}
\label{fig:ZvsMB}
\end{figure*}

As discussed in the previous paper of this series (PEPK19), the prototype
void gas-rich XMP dwarfs are atypical compared to other void dwarfs as
well as to the general dwarf galaxy population. The four new
gas-rich XMP dwarfs of this study residing in Nearby Voids share all the
unusual
properties of the prototype group. Their O/H is reduced for their
$M_{\rm B}$ by a factor of $\sim$2 to $\sim$4 with respect to the reference
relation of \citet{Berg12}, or, in other words, their blue luminosity
is elevated for their O/H by $\sim$10--100 times with respect to this
reference relation. The current SF activity in these XMP dwarfs
(with the possible exclusion of AGC114584) is low and does not seem
to shift them significantly
to the brighter $M_{\rm B}$, which is partly the case for the prototype
XMP blue compact galaxy IZw18. As we show in
the accompanying paper on these XMP dwarfs' photometric properties, their
mass in stars comprises $\sim$0.01--0.02 of the total baryonic mass.
Also, the colours of unresolved stars in their outer parts appear rather blue
(as opposed  to the majority
of the main void galaxy population) and are consistent with ages of one to
a few Gyr.

The very atypical properties of XMP dwarfs in the Nearby Voids
emphasize the need for advanced model simulations which can
reproduce the observed population and, thus, help to understand their
formation and specific evolutionary scenarios.
One of the ways of confronting the observations and models is related to
the Very Young Galaxies (VYGs) defined recently by \citet{Tweed18} as
galaxies that formed most of their stellar mass during the last
$\sim$1~Gyr.

In simulations predicting the existence of VYGs \citep{Tweed18},
neither their gas mass-fraction, nor their type of environment are noticed.
The small group of void XMP gas-rich dwarfs, both the prototype
galaxies and the four new XMP dwarfs found in this work, due to their
observational properties, represent the most natural proxy for
the simulated VYGs.
At the moment, we do not know whether simulations predict
VYGs with less extreme properties. If they do, it seems it will be hard
to find them in a systematic manner.

At the same time, very gas-rich XMP dwarfs can be systematically found
within the Nearby Voids. Therefore, the use of them as good proxies of
simulated VYGs gives the possibility to estimate their lower number fraction.
Such an estimate in turn can be helpful for comparison with predictions
of VYG number fractions in cosmological models with Cold (CDM) and Warm (WDM)
Dark Matter. As \citet{Tweed18} emphasize, the simulated number of VYGs in WDM
cosmology appears many times larger than in the CDM one.

On the other hand, the statistical analysis of already discovered XMP dwarfs
and understanding their diversity and the boundaries of their parameter
space should help in designing more effective searches for these
very rare galaxies.

\subsection{General statistics of O/H in void galaxies}

 To compare our new data on O/H for XMP candidate galaxies from the
NVG sample, we plot them in the diagram
12+$\log$(O/H) versus $M_{\rm B}$ (Fig.~\ref{fig:ZvsMB})\footnote{We
note that in order to have the most reliable O/H estimate for the case of
unavailable O/H via the direct T$_{\rm e}$ method, we employ for all earlier
measurements with 12+$\log$(O/H)(se) $\lesssim$7.4, the new strong-line method
of \citet{Izotov19DR14} discussed above. Therefore, some of the old
published values of 12+$\log$(O/H) for this O/H range are slightly changed.
In particular, their placement in Fig.~\ref{fig:ZvsMB}  is
slightly different compared to similar published diagrams in
\citet{PaperVII,Eridanus}. }
along with our earlier results for galaxies residing in the Lynx-Cancer
(81 filled octagons) \citep{PaperVII} and Eridanus (36 filled triangles)
\citep{Eridanus} voids. To O/H data for 22 NVG dwarfs from this study
we added O/H of 9 dwarfs from BTA results (in preparation) and J2104-0035
with O/H from \citet{Izotov19DR14}. They are shown by green rhombs.

Due to the implied absolute magnitude cut of $M_{\rm B} > -14.3$ for the
selected XMP void candidates, our new data occupy the region of the diagram
containing about
a half of the Lynx-Cancer void sample and $\sim$20\% of the Eridanus sample.
As one can see, our new O/H data for void galaxies match the results
of the previous studies well. Their higher incidence in the range,
12+$\log$(O/H) $\lesssim$7.2, presumably reflects both factors: the
$M_{\rm B}$ cut and the additional selection criteria for XMP candidates
described in Section~\ref{sec:intro} and PEPK19.

It is of interest to compare the updated results on void galaxy metallicities
with the reference relation 12+$\log$(O/H) versus $M_{\rm B}$ from
\citet{Berg12} derived for the Local Volume galaxies residing in denser
environments, mostly in typical groups. Their relation is shown in
Fig.~\ref{fig:ZvsMB} by a solid line. Two dash-dot lines show the 1~$\sigma$
scatter of their data points around their linear regression. The dotted line
runs 2~$\sigma$ below the reference relation.

As we concluded in \citet{PaperVII}, the Lynx-Cancer void sample has as a
whole a reduced gas metallicity, with the average O/H deficiency with
respect to the reference relation of $\sim$0.18~dex. If we limit the
comparison of the reference and void samples with currently available O/H
by the cut, $M_{\rm B} > -12.5$, we find that the void sample of a dozen
of the lowest luminosity dwarfs has a significantly larger O/H deficiency
of $\sim$0.5~dex.
This is an interesting finding. However, to date it is difficult to
decide whether this is the effect of the special XMP candidate
selection or the reflection of stronger environmental effects.
To clear up this issue, we
need an unbiased measurement of gas metallicity for all available void
dwarfs with the lowest luminosities.

The current spectral results were obtained for about a half of the 60
preselected XMP candidates from PEPK19. We expect to find in the remaining
part of the original candidate list several other similar objects. The final
list of XMP gas-rich void galaxies should be valuable material for a
first statistical analysis of such outstanding galaxies. They are interesting
by themselves since the combination of their very low stellar mass
fraction, non-cosmological ages of the oldest visible stellar population
outside regions of current/recent star formation\footnote{See the
results for several such objects from the Lynx-Cancer void in \citet{PaperIV}.
For four new void XMP dwarfs from the current work  similar results will be
presented in the accompanying paper devoted to their photometric properties.}
and extremely low gas
metallicity indicate their unusual evolutionary status.

In addition, if they can be identified as VYGs, their statistics can provide
limits on modern cosmological models as emphasized by \citet{Tweed18}.
To this end, apparently more advanced model simulations are necessary that
probe a lower mass range of dwarfs, better matching the observed mass range
of VYG candidates. It is also important to understand how these nearby
XMP dwarfs with low and very low star formation rate (SFR) are connected
with the outstanding
starbursting XMP dwarfs found by \citet{Izotov18,Izotov19}.

The increase in the number of known unusual void dwarfs (with
properties summarized in the beginning of Sec.~\ref{sec:discussion})
and the advance in the study of their group properties hopefully will
provide deeper insights into predicted properties of galaxies with both
low baryon mass and tiny fraction of stellar mass.

\subsection{New void dwarfs with the lowest metallicities}
\label{sec:extremeXMP}

It is worth describing in more detail the four most extreme
XMP dwarfs with $Z \lesssim Z$\sunn/30 (12+$\log$(O/H) $\lesssim$7.2).
As seen in Column~9 of Table~\ref{tab:prop_summary}, these are the following
dwarfs with the respective values of 12+$\log$(O/H) in parentheses:
J0110--0000 (7.05), J0112+0152 (7.15); J0256+0248 (6.95) and J1038+0352
(7.13).
All these galaxies are identified as faint \HI-sources in the
blind ALFALFA survey \citep{ALFALFA}.

The galaxy, J1259--1924, with the current estimate of O/H of
12+$\log$(O/H)=7.20$\pm$0.08, may also belong to this small group, but its
O/H uncertainty is too large and the [O{\sc iii}]4959,5007\AA\ lines are
extremely faint.

The blue absolute magnitudes M$_{\rm B}$ of
these dwarfs vary between --10.6 and --13.0 (factor of $\sim$9). The
respective values of hydrogen mass M(\HI), in the same R.A. order,
in units of 10$^7$~M\sunn, are as follows: 3.16, 3.92, 1.56, 1.40. That is,
the gas mass varies by a factor of $\sim$2.2.

Despite the rather small statistics, it is useful to compare parameters of
the new void XMP dwarfs with those for the prototype XMP void group, compiled
in Table~1 of PEPK19 paper. The ten prototype XMP dwarfs have a four times
broader range of blue luminosities,
with M$_{\rm B}$ between --9.6 and --14.1 (factor of 40). Respectively,
their range of M(\HI) (in the same units) is much wider, from 1.6
to 32 (factor of 20). If we combine these 4 new void XMP dwarfs with
the prototype XMP dwarfs, they appear to be among the six lowest M(\HI)
galaxies of the total of 14 such objects.

While such a situation can occur by chance due to the rather small number
statistics of new void XMP dwarfs, this also can hint that our selection
procedure introduces some additional bias on the parameter
space of XMP dwarfs found in this work. We further address this issue
in the abovementioned accompanying paper (in preparation) summarizing
the photometric parameters of new XMP dwarfs. It
includes data on the stellar masses and their fractions of all examined XMP
dwarfs. This also will include additional XMP dwarfs found in
the similar program at the SAO 6-m BTA telescope.

\subsection{Comments on several individual galaxies}
\label{sec:individual}

Among the galaxies presented in Table~\ref{tab:prop_summary} there are several
cases which are worth additional comments:

{\bf PGC736507 = J0009--2851.} This rather distant emission-line galaxy
(V$_{\rm h}$=7594~\kms) appeared in the list of preselected candidates
(PEPK19) because an incorrect radial velocity, V$_{\rm h}$= 898~\kms\,
appears in 2dFGRS \citep{2dF_2001} and subsequently in HyperLEDA and
the NVG sample.
Our SALT spectrum of this object shows high-contrast emission lines.
Therefore,  it is difficult to
understand the origin of this error. Moreover, this is not the only case
of an incorrect radial velocity among our selected void XMP candidates
based on the data of this survey.
Based on our experience,  we caution potential users on the need
for careful checks of 2dFGRS redshifts
for objects with radial velocities of $\lesssim$1000~\kms.

The gas metallicity of this background dwarf with $M_{\rm B} =-16.0$,
corresponding to 12+$\log$(O/H) = 7.58~dex, appears rather low for its
luminosity
in comparison to the reference relation in Fig.~\ref{fig:ZvsMB}.
As one can see, its O/H ratio is near the lower limit for dwarfs
with the same luminosity residing in the Lynx-Cancer and Eridanus voids.
Therefore, it is of interest to examine the  nature of its environment.

A check of nearby galaxies in
NED\footnote{NED is an acronym for the NASA/IPAC Extragalactic Database}
within a projected distance of 100\arcmin\
(3~Mpc at the assumed distance of 104.5~Mpc) reveals the galaxy cluster
ABELL 2734 and its probable members at the projected distance of
$\sim$0.7~Mpc. The difference in radial velocity between PGC736507
and ABELL 2734 of 201~\kms\ is not so large as to completely exclude the
membership of
PGC736507 in ABELL 2734. If this is the case, it is situated at the
cluster periphery, at least at the distance of $\sim$0.7~Mpc. In a more
probable scenario, where PGC736507 is far from the cluster and the
velocity difference is due
to the Hubble flow (with adopted H$_{\mathrm 0} = 73$~\kms~Mpc$^{-1}$),
it is at $\sim$2.8~Mpc from the nearest luminous/massive galaxy, that is,
in a rather low-density environment. The latter case would be more
consistent with its low observed metallicity.

We noticed a faint bluish nebulosity (J0009--2852) at $\sim$50\arcsec\
SSW of the target galaxy, which could be its companion.
Therefore, the slit was positioned across this potential companion. As its
spectrum revealed (see Fig.~\ref{fig:SALT_1Dp1}), it appears to be a distant
ELG with the redshift $z =$ 0.20159. The strong line ratios in the spectrum
are typical of star-forming galaxies. We estimate its 12+$\log$(O/H)
$\sim8.2 \pm$0.1~dex using the lower branch of the strong
line empirical estimator of \citet{PT05}.

{\bf PGC493444 = J0107--4756.} This rather distant absorption-line galaxy has
radial velocity V$_{\rm h}$=7050~\kms, as revealed by several
Balmer lines in its spectrum. It appeared in the list of preselected void XMP
candidates (PEPK19) also due to a mistaken radial velocity V$_{\rm h}$ =
837~\kms\ in 2dFGRS \citep{2dF_2001} and then, in HyperLEDA and the NVG
sample.

{\bf ESO121-020 = J0615--5743.} The slit for this spectrum passed through
two nebulous emission knots and a star-like emission object in between
(as revealed
by our H$\alpha$ image obtained with SALT before the spectral observation).
The very low values of O/H in both nebulosities are the same within
the cited errors. Therefore, we adopt for this galaxy the average of the two
\HII-regions.

The star-like object displays Balmer absorptions in the UV and H$\alpha$ and
H$\beta$ in emission, but no hint of nebular emission lines.
Moreover, H$\alpha$ has broad underlying wings with  a
FWHM of $\sim$ 800~\kms. The estimate of its $B$ magnitude,
$\sim 22.5$, implies an absolute magnitude $M_{\rm B} \sim -6.5$,
characteristic of supergiants.
Having only this rather low S/N spectrum, it is difficult to
make more or less reasonable suggestions on the nature of this object.
However, one of the options is a composite spectrum of a young stellar cluster
with an age of 12--13 Myr estimated from the EW of the narrow component,
EW(H$\alpha$) = 31~\AA\  (according to \citet{Starburst99}),
and a luminous emission-line star of a comparable luminosity. It could be
an LBV in a relatively faint phase.
A more advanced characterization of this emission star-like object requires
more data on possible variability and other emission lines.

{\bf PGC385975 = J0616--5745}. For this companion of ESO121-020,
we obtained two independent spectra of two different \HII-regions.
Their O/H values are consistent with each other within their errors,
so we adopt their mean value.

{\bf AGC188955 = J0821+0419.} There are two emission-line knots in the 2D
spectrum. In the brighter knot all suitable lines are well measured and
its O/H is derived with the direct method (see Table~B4). In the fainter knot
the [O{\sc iii}]$\lambda$4363~\AA\ line is undetected and its O/H is estimated
via the method of \citet{IT07}. The 12+$\log$(O/H) values differ for the
two knots: 7.81$\pm$0.08 for the brighter and 7.64$\pm$0.08 for the fainter.
However, since the difference is only $\sim$1.5~$\sigma$ of the combined
error of O/H, we adopt for this galaxy the average O/H of the two knots.
The brighter knot displays a low-contrast broad (FWHM $\sim$ 1050~\kms)
underlying component in the strong lines, best visible in H$\alpha$ and
[O{\sc iii}]$\lambda$5007~\AA. For the O/H estimate the flux in the broad
component was not included.
In the [O{\sc iii}]$\lambda$5007~\AA\ line, the flux
of the broad component comprises $\sim$0.07 of the narrow component. The
nature of this broad component is unclear. However, in low-metallicity
galaxies with active star formation the appearance of broad components
of strong emission lines is not rare \citep[e.g.][]{ITG07}. If this
phenomenon is not related to the short phase of WR stars, then the collective
effect of many SNR and the related fast shells could make the main
contribution to the observed broad components.

{\bf PGC1314481 = J0948+0707}. Our SALT spectrum is rather noisy and
certainly of worse quality in comparison to the spectrum of this object
presented in the SDSS spectral database.
For an optimal estimate of O/H in this galaxy, we use its spectral data
from SDSS and add the flux of
[O{\sc ii}]$\lambda$3727~\AA\ from our SALT spectrum.
To combine this line flux correctly with fluxes of other lines in
the SDSS spectrum, we carefully checked
the similarity of the relative fluxes of the strongest Balmer and
[O{\sc iii}] lines in the SDSS and SALT spectra.

{\bf KK246 = J2003--3104 = ESO461-036 = SIGRID68}. This Local Volume gas-rich
galaxy, with a Milky Way (MW) $B$-band extinction
of $A_{\rm B} = 1.10$ \citep{Schlafly11},
has been studied many times, including by \citet{Kreckel11}
and by \citet{KK246_OH}. Its O/H was first derived by
\citet{KK246_OH} with a spectrum without detected
[O{\sc iii}]$\lambda$4363~\AA.
Using their Mapping IV model analysis \citep{Dopita2013}, these authors derive
12+$\log$(O/H)$\sim$8.2~dex. That is, our (7.67$\pm$0.08~dex) and their O/H
determinations differ significantly and need further examination.
The first factor of the difference, which \citet{KK246_OH} emphasize, is
due to the known systematics of $\sim$0.2~dex between the two methods.
However, there still remains an additional difference of $\sim$0.3~dex.

It appears that there are problems with the primary uncorrected spectrum
of this object obtained with the WiFeS IFU spectrograph at the ANU 2.3-m
telescope. Their observed Balmer decrement implies an overall extinction
A$_{\rm V}$ of only 0.032~mag for KK246, equivalent to A$_{\rm B}=0.042$.
At the same time, as mentioned above,
the known MW extinction in this direction is much higher, $A_{\rm B} = 1.10$.

The Balmer decrement of our uncorrected spectrum implies C(H$\beta$) =
0.60$\pm$0.05. This translates, for the standard extinction curve,
to $A_{\rm B} = 1.75$$\pm$0.15, which does not contradict the minimal expected
value of $A_{\rm B} = 1.10$.
In terms of C(H$\beta$), the MW $A_{\rm B} = 1.10$ implies a minimum
expected C(H$\beta$)=0.38. Our measured C(H$\beta$) = 0.60$\pm$0.05,
after subtraction of the MW contribution, implies an internal
C(H$\beta$,inter) = 0.22$\pm$0.05 dex, which is quite typical of
low-metallicity dwarfs \citep[e.g.][and references therein]{Guseva17}.

Another doubtful issue with the data  of \citet{KK246_OH} is the
extremely low level of the underlying continuum in their original spectrum.
In our long-slit spectrum the underlying
continuum is basically visible. The related EW(H$\beta$) = 29~\AA.
This brief examination of the two results indicates that our O/H estimate for
KK246 should be treated as closer to reality.
With our O/H = (O/H)\sunn/10, KK246 in the diagram 12+$\log$(O/H) vs
M$_{\rm B}$ in Fig.~\ref{fig:ZvsMB}  sits well within the main cloud of void
dwarfs in contrast to the result of \citet{KK246_OH} (their Fig.~12).

\section[]{Conclusions}
\label{sec:conclusions}

In the previous paper (PEPK19) we selected 60 candidate XMP void objects
from the fainter
part ($M_{\rm B} \gtrsim -14.3$) of the Nearby Void Galaxies (NVG) sample.
 They were selected based on the similarity
of their properties (available in public databases and in the literature)
to those of ten known XMP very gas-rich void dwarfs. Here we present the
first results of spectral observations of these candidates with SALT.

Summarizing the results presented here and the related discussion, we draw the
following conclusions:

\begin{enumerate}
\item
To date, 26 of the void XMP candidates have been observed with SALT.
For 23 of them, Oxygen lines were detected along with Balmer lines
and estimates of their gas O/H were derived. They appear in the range of
12+$\log$(O/H) between $\sim$6.95 and $\sim$7.8 dex, with the exception
of one emission line galaxy which appeared to be a background object.
The majority of our void galaxies with measured O/H fall in the
12+$\log$(O/H) vs M$_{\rm B}$  diagram within the O/H range
typical of void galaxies for their luminosity. Of the 22 Nearby
Void galaxies in this study, ten have the parameter
12+$\log$(O/H)$<$7.39 dex, or Z$_{\rm gas}$ below Z\sunn/20.
Such galaxies are still quite rare and their addition will improve
their statistics and seemingly increase the diversity of their
properties.
\item
Of these ten objects, four XMP dwarfs have  12+$\log$(O/H) $<$7.19~dex,
or $Z < Z$\sunn/30. One of them, AGC124629=J0256+0248, shows one of the
lowest gas-metallicities in the Local Universe (12+$\log$(O/H)=6.95$\pm$0.06)
and is the first LSB dwarf with that record-low Z$_{\rm gas}$ found.
Looking ahead, based on an accompanying paper on their photometric
parameters (in preparation) and the discussion in Sec.~\ref{sec:discussion},
these new XMP dwarfs have unusual properties: blue colours of the outer
parts, corresponding to non-cosmological ages of the oldest visible stellar
population, and extremely large gas-mass fraction, $\sim$0.98--0.99.
Thus, they appear to be very similar to those of the original small
XMP group in the Lynx-Cancer and other nearby voids. These four new XMP
void dwarfs add to the group of eight nearby prototype  XMP void
dwarfs (those with known O/H from Table~1 of PEPK19).
Thus, the number of {\bf void} candidate Very Young Galaxies has grown
to a dozen. This also
allows us to better study their similarity and diversity as
well as their finer properties related to their origin and evolution.
\item
The results of SALT spectroscopy of the selected XMP candidate list in the NVG
sample gives a reasonably high detection rate for the objects that  we are
primarily looking for and qualifies the search method as an efficient one.
This could also be a feasible method to use for an XMP dwarf search
in the farther parts of the Local Supercluster.
\end{enumerate}

\section*{Acknowledgements}
This work is based on observations obtained with the
Southern African Large Telescope (SALT), program \mbox{2017-2-MLT-001}
(PI: Kniazev) and \mbox{2017-2-DDT-002} (PI: Pustilnik).
The reported study was funded by RFBR according to the research project
 No.~18-52-45008-IND$\_{\rm a}$.
AYK acknowledges support from the National Research Foundation (NRF) of
South  Africa.
The authors thank the referee J.~Sanchez~Almeida for helpful report which
allowed to improve presentation and clear up some points.
We thank J.~Menzies for general check and improvement of the paper
language.
The use of the HyperLEDA database 
is greatly acknowledged.
This research has made use of the NASA/IPAC Extragalactic Database (NED)
which is operated by the Jet Propulsion Laboratory, California Institute
of Technology, under contract with the National Aeronautics and Space
Administration. We also acknowledge the great effort of the ALFALFA team
which opened access to the nearby Universe gas-rich dwarfs with low or
moderate SFR and thus helped us to identify the majority of very low
metallicity galaxies of this study.

We acknowledge the use of the SDSS database.
Funding for the Sloan Digital Sky Survey (SDSS) has been provided by the
Alfred P. Sloan Foundation, the Participating Institutions, the National
Aeronautics and Space Administration, the National Science Foundation,
the U.S. Department of Energy, the Japanese Monbukagakusho, and the Max
Planck Society. The SDSS Web site is http://www.sdss.org/.
The SDSS is managed by the Astrophysical Research Consortium (ARC) for the
Participating Institutions.


\appendix

\section{1D spectra}

\begin{figure*}
\includegraphics[width=4.0cm,angle=-90,clip=]{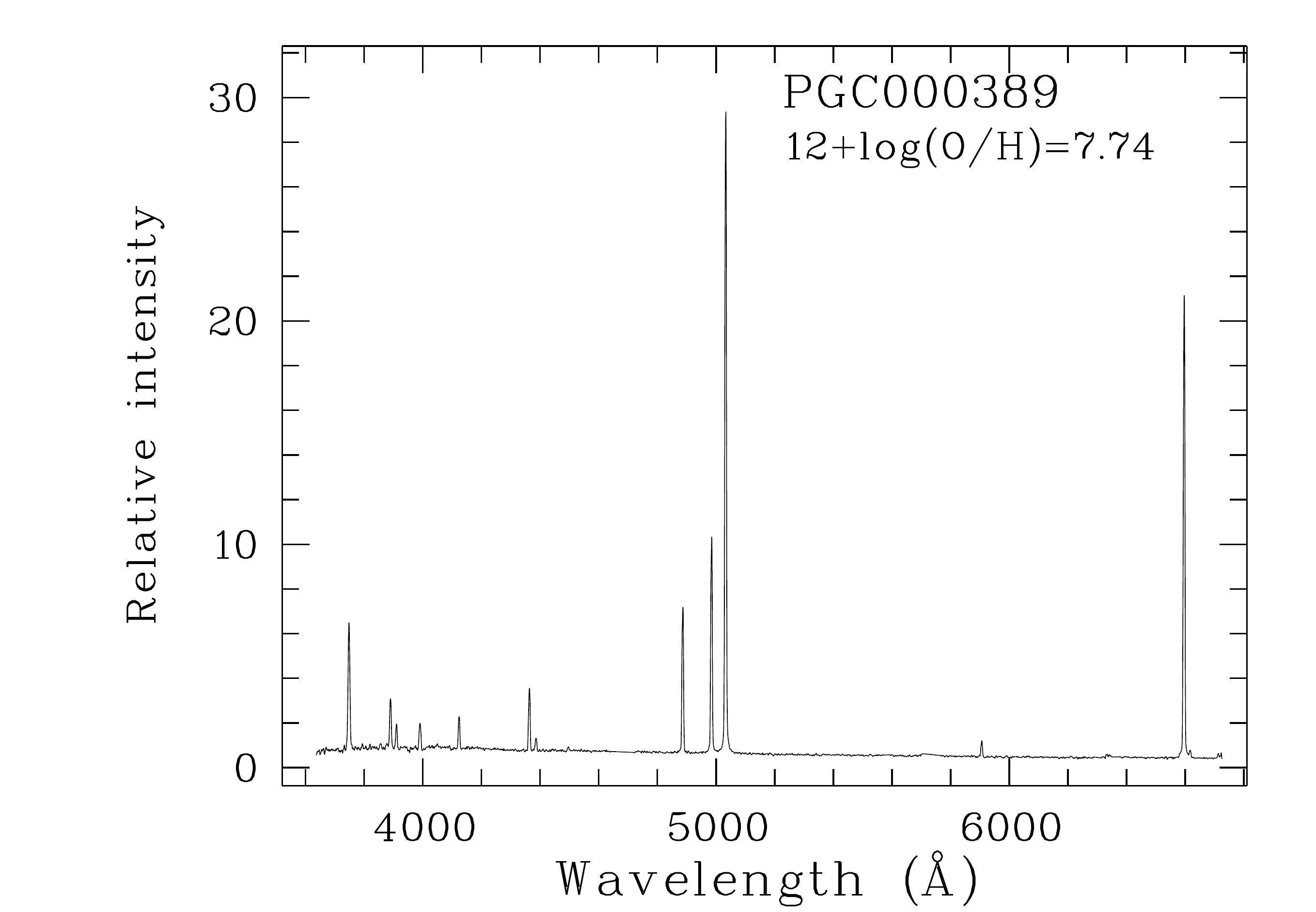}
\includegraphics[width=4.0cm,angle=-90,clip=]{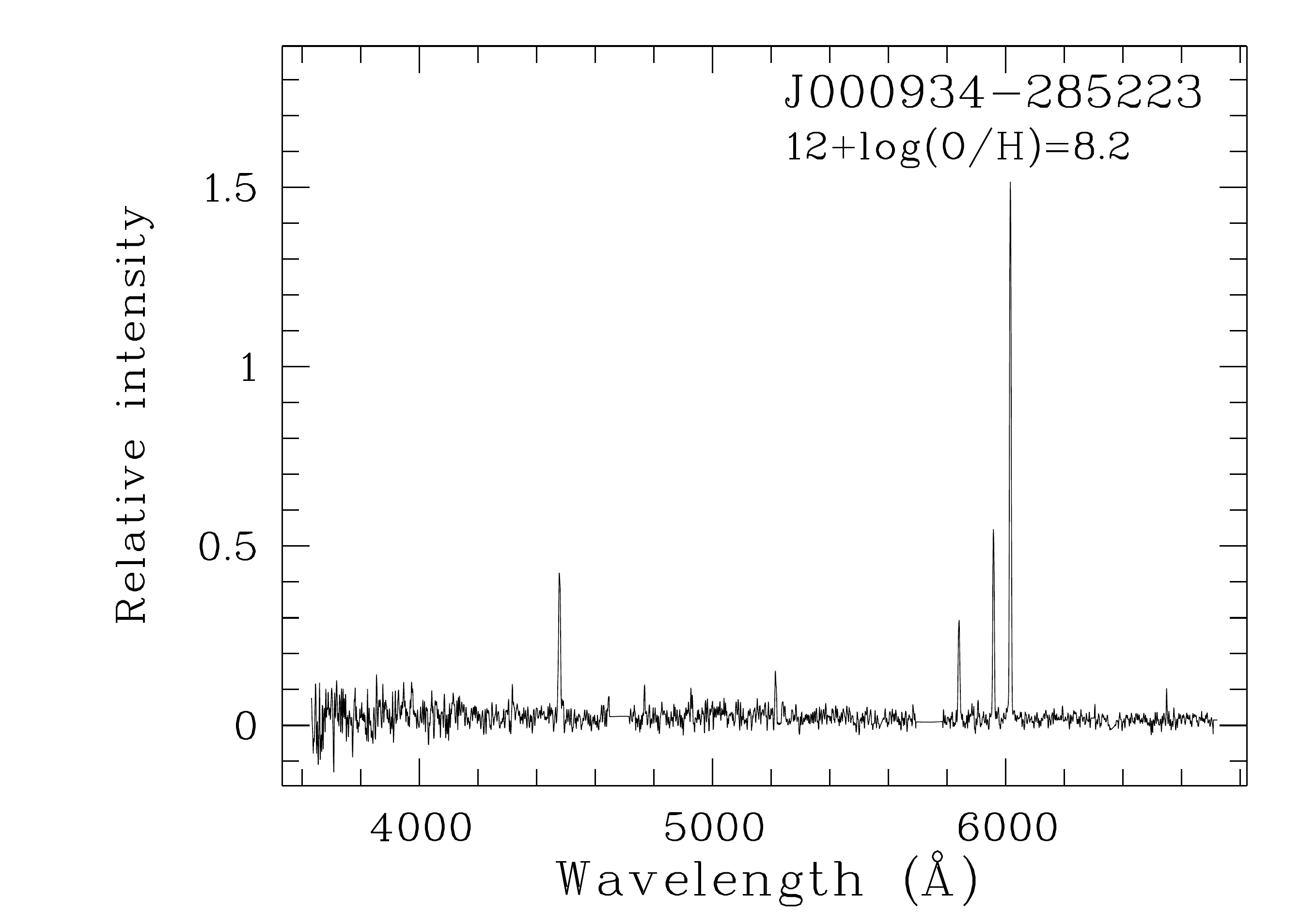}
\includegraphics[width=4.0cm,angle=-90,clip=]{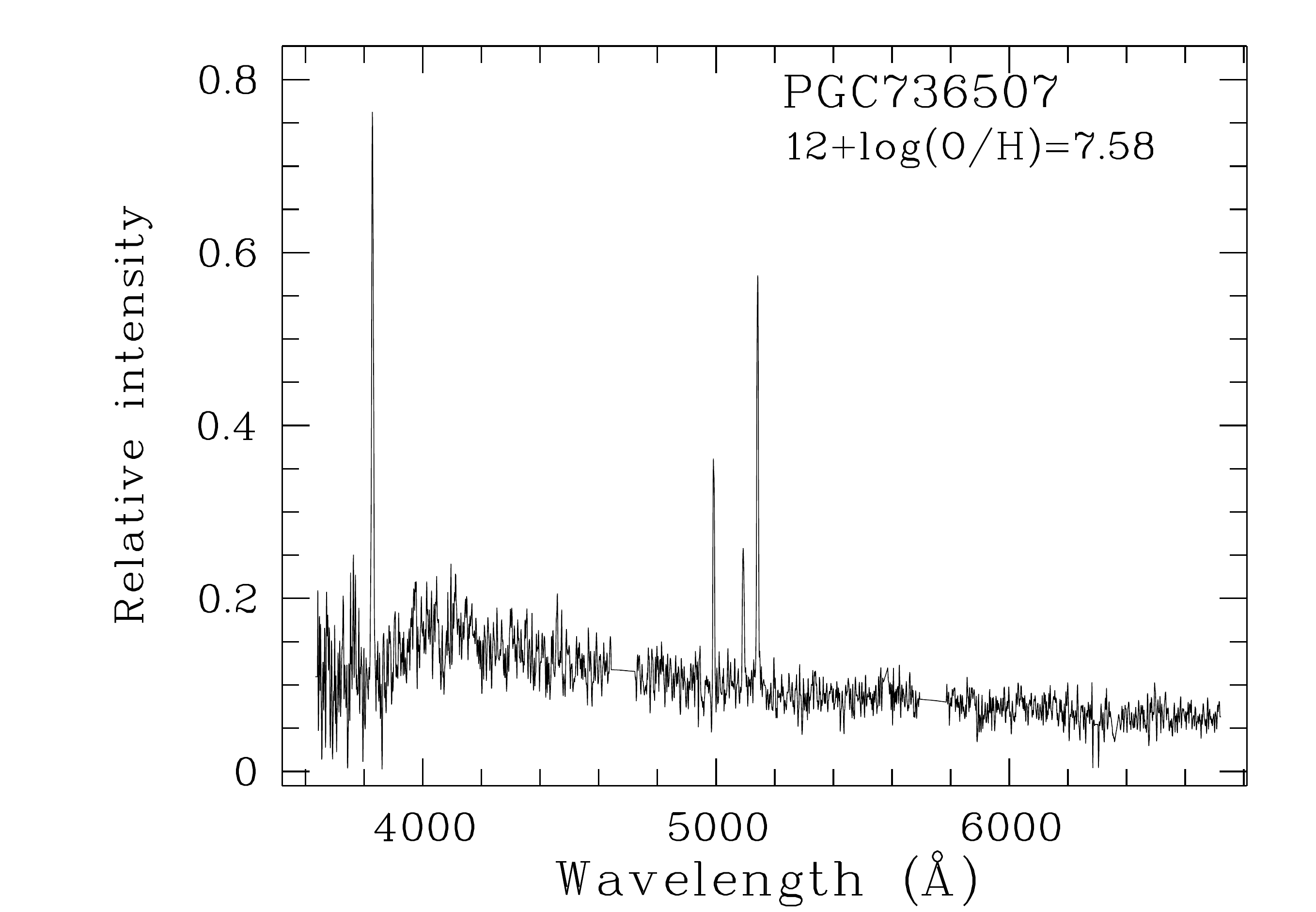}
\includegraphics[width=4.0cm,angle=-90,clip=]{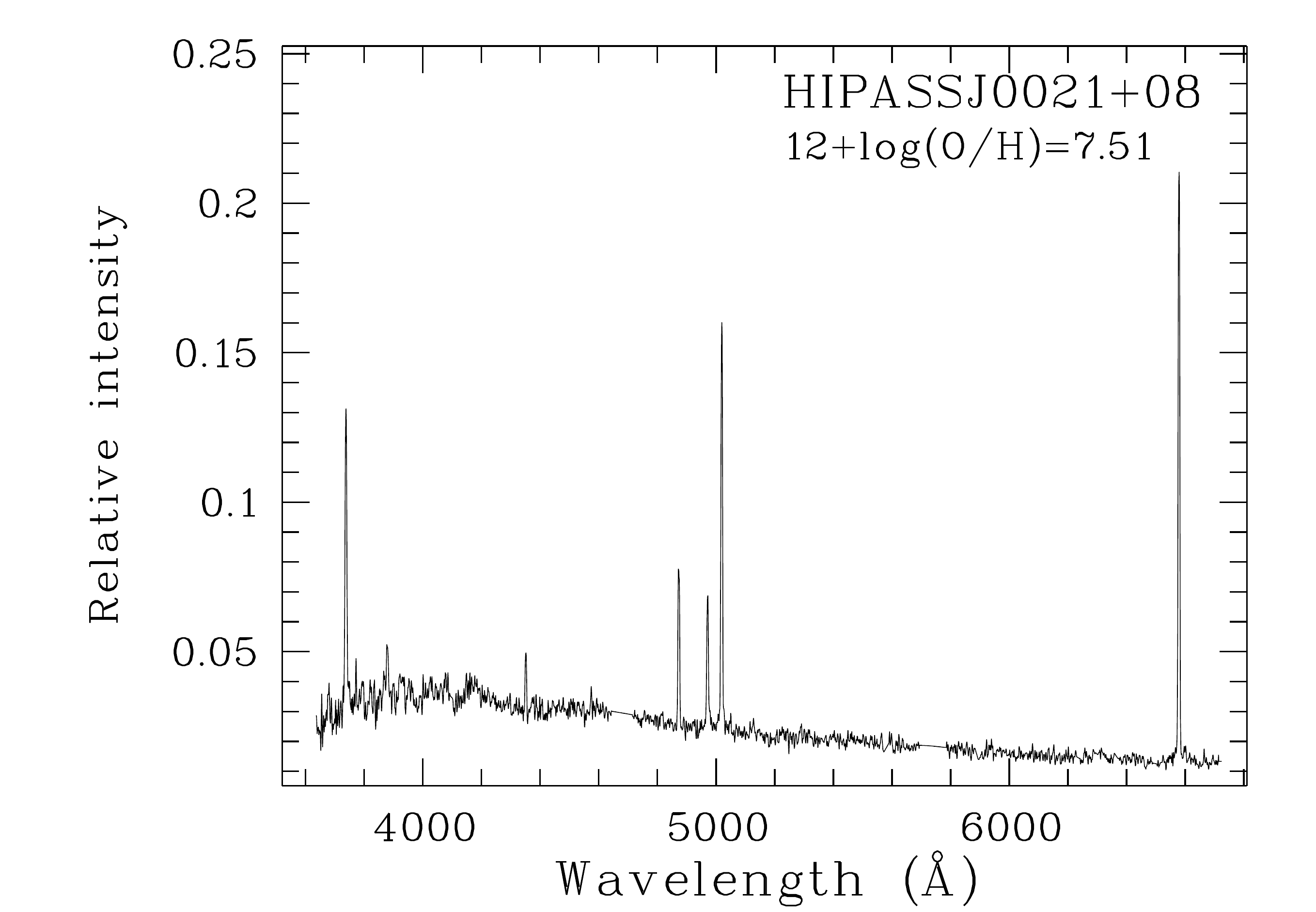}
\includegraphics[width=4.0cm,angle=-90,clip=]{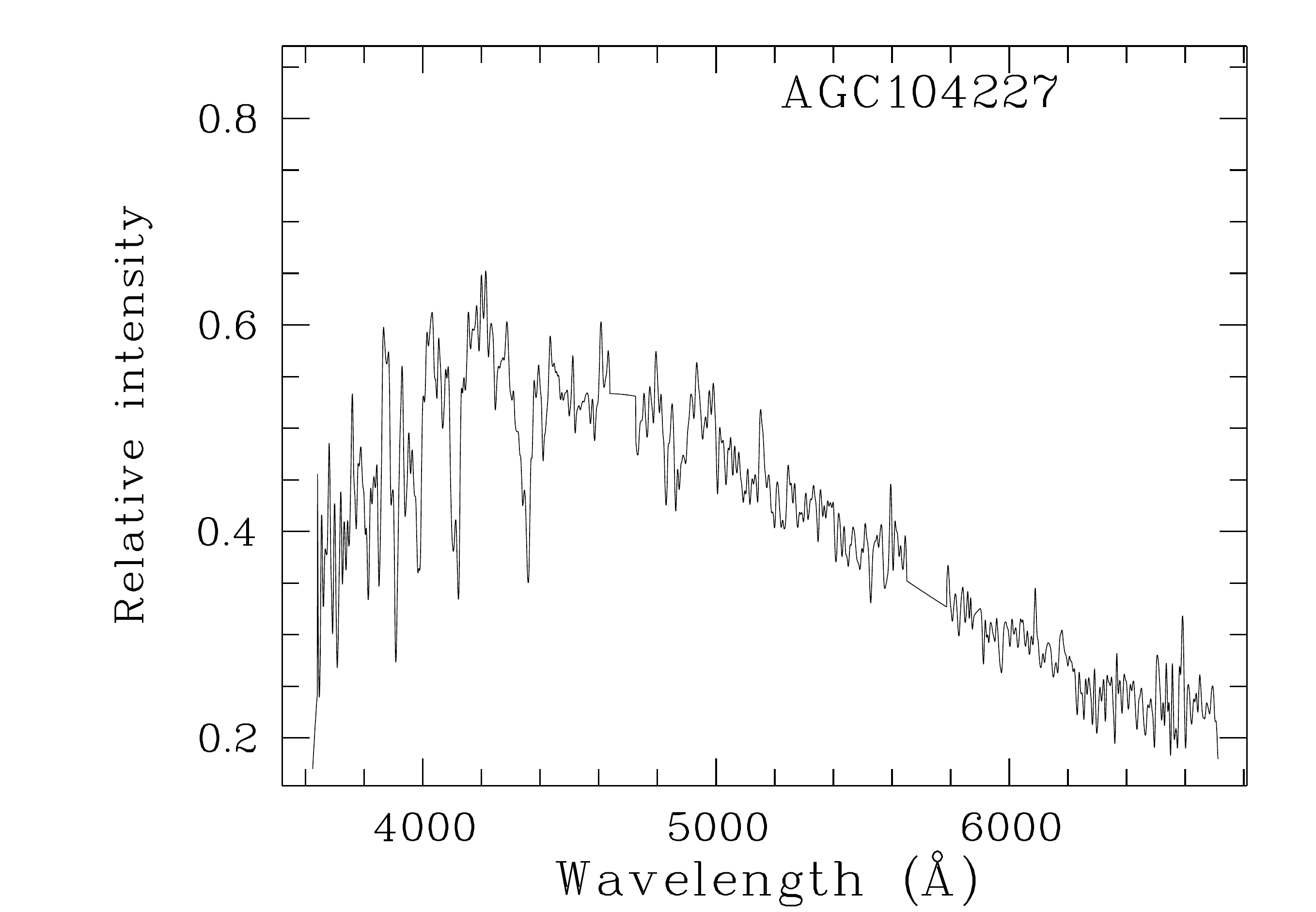}
\includegraphics[width=4.0cm,angle=-90,clip=]{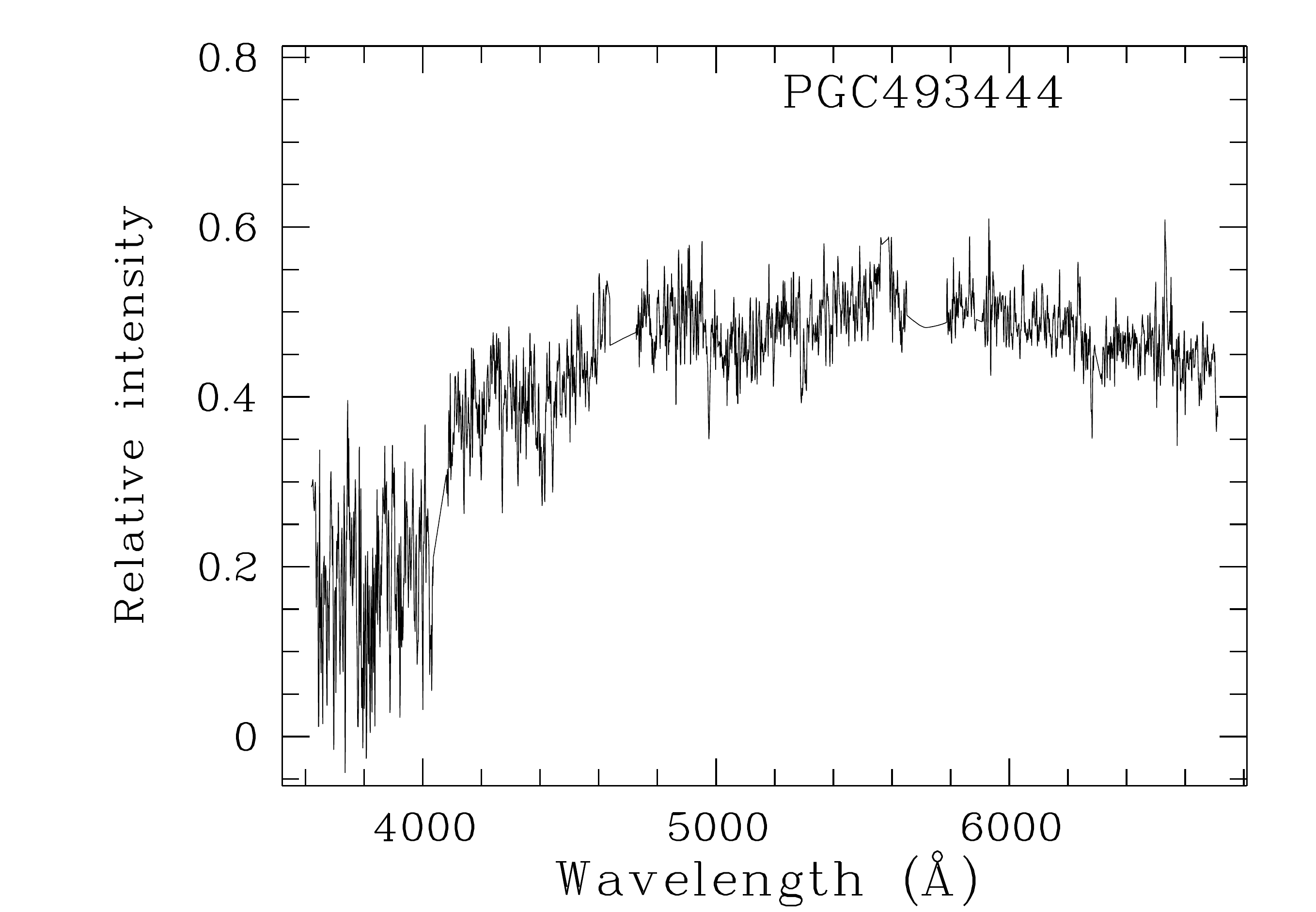}
\includegraphics[width=4.0cm,angle=-90,clip=]{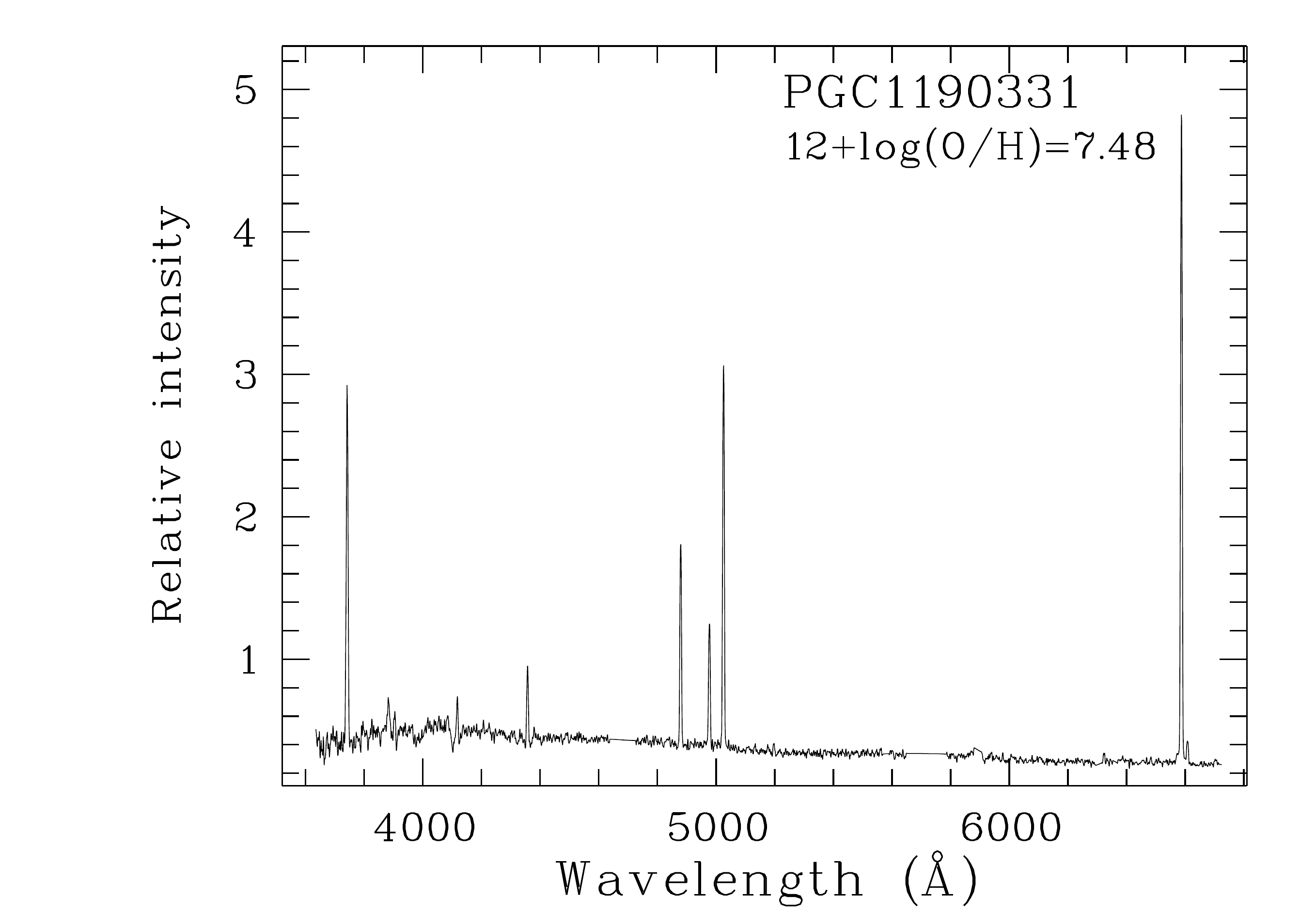}
\includegraphics[width=4.0cm,angle=-90,clip=]{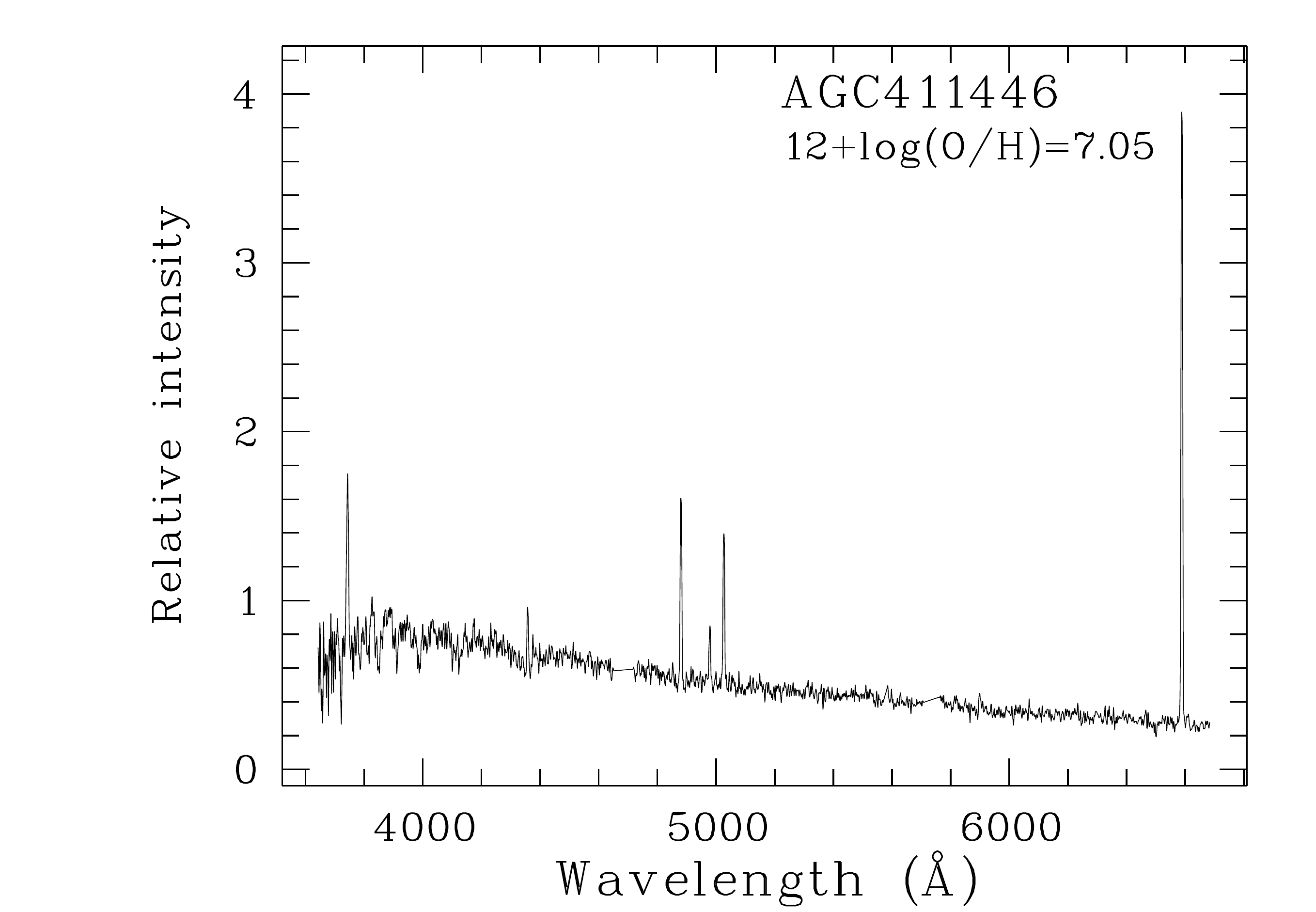}
\includegraphics[width=4.0cm,angle=-90,clip=]{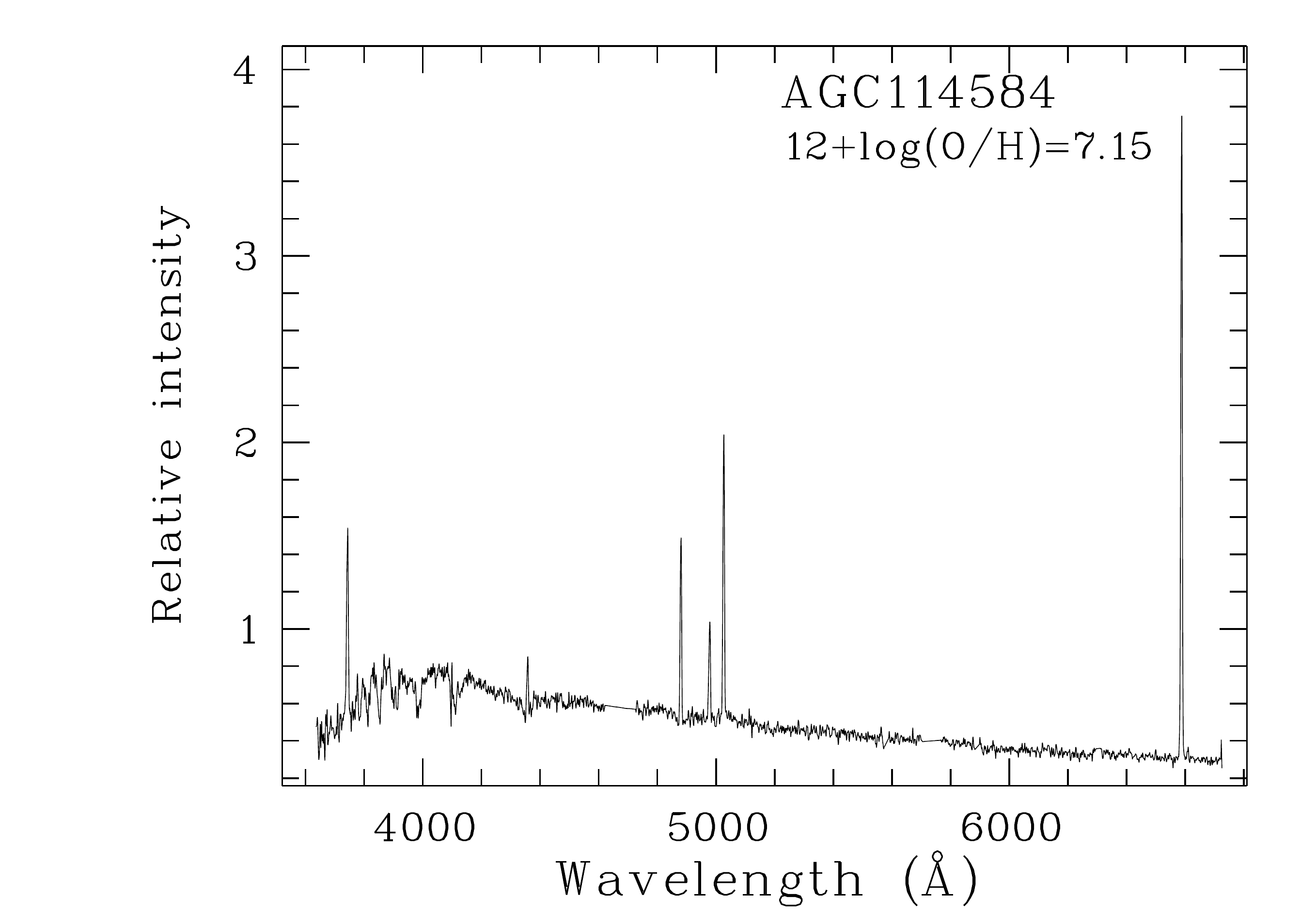}
\includegraphics[width=4.0cm,angle=-90,clip=]{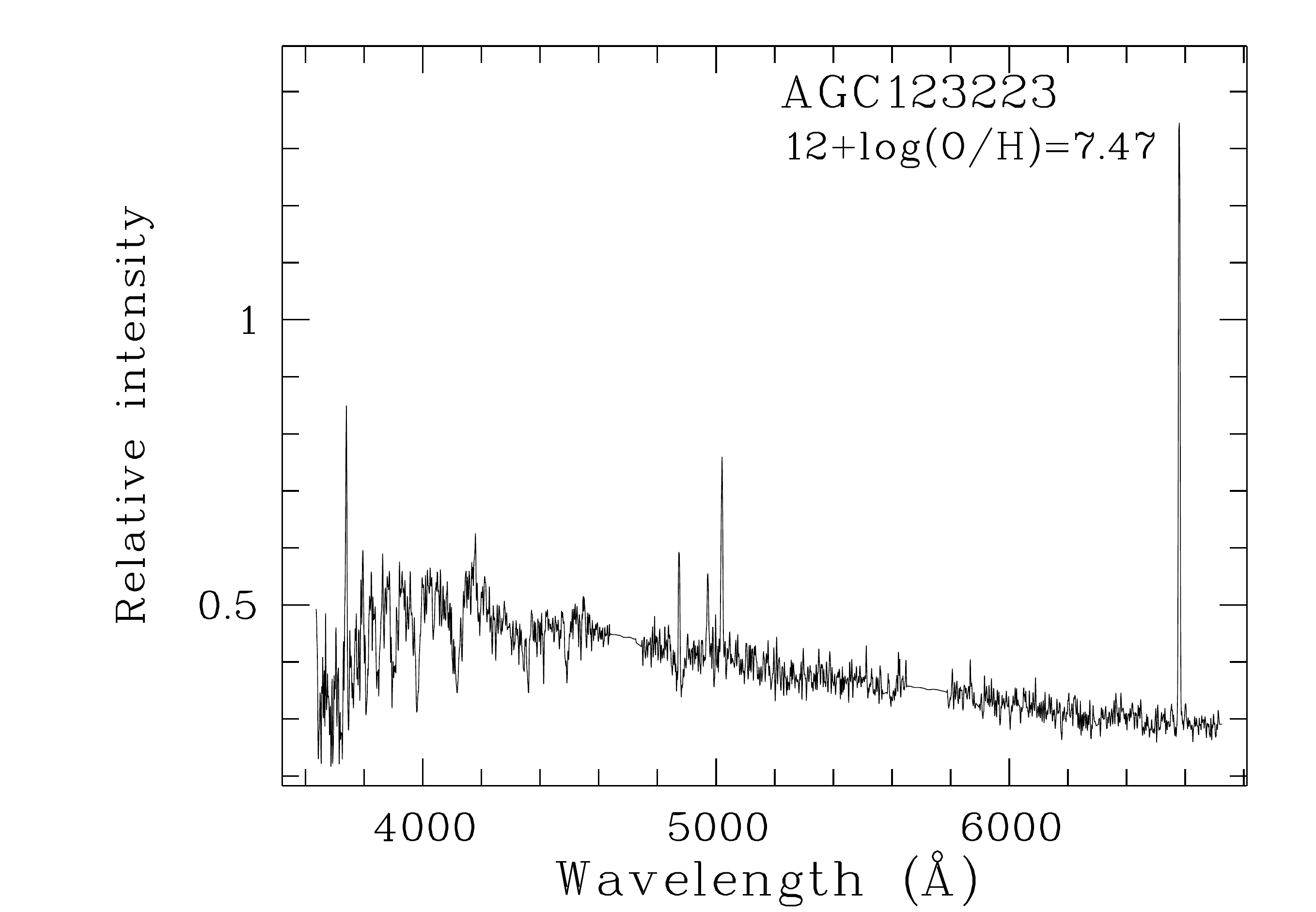}
\includegraphics[width=4.0cm,angle=-90,clip=]{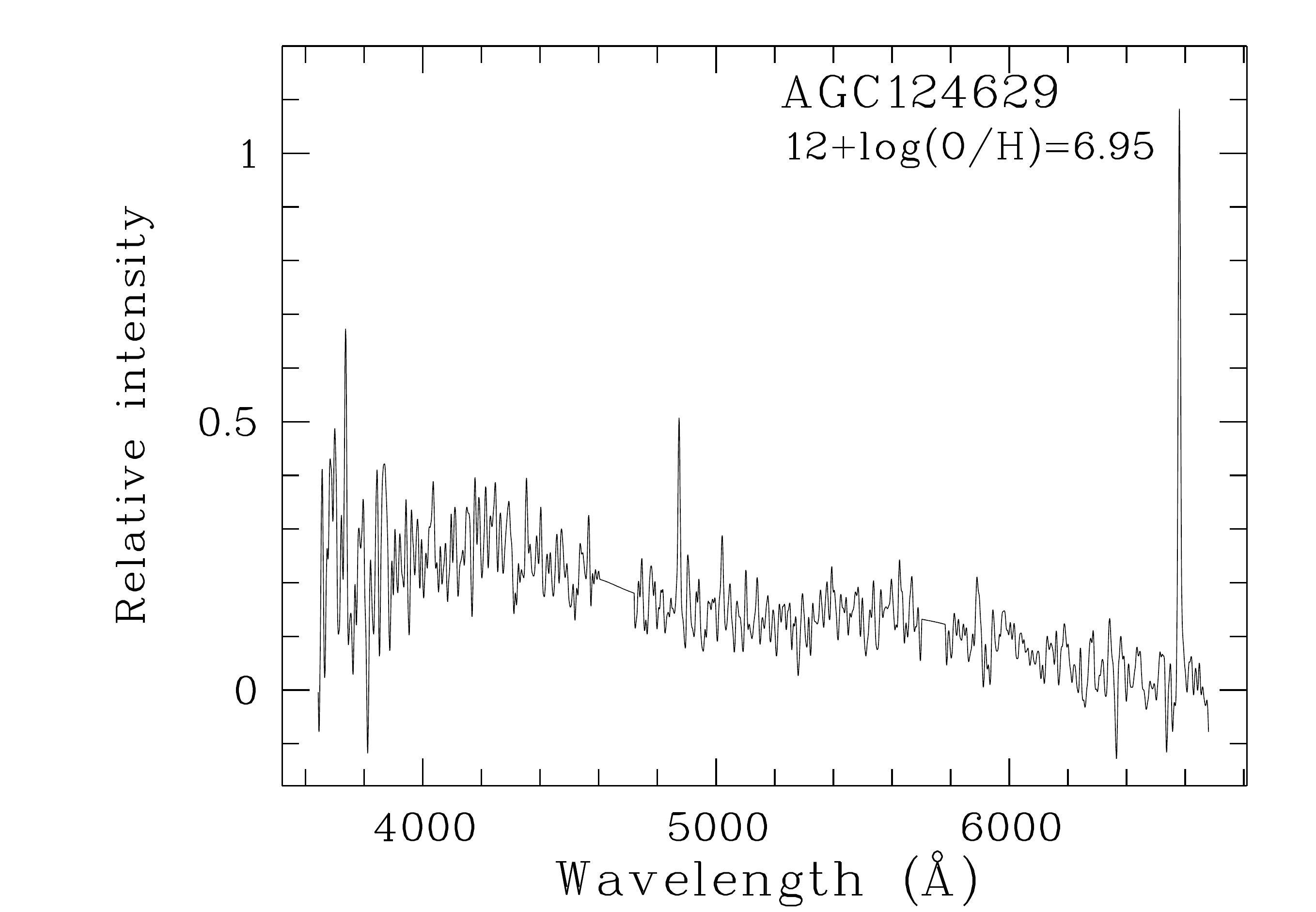}
\includegraphics[width=4.0cm,angle=-90,clip=]{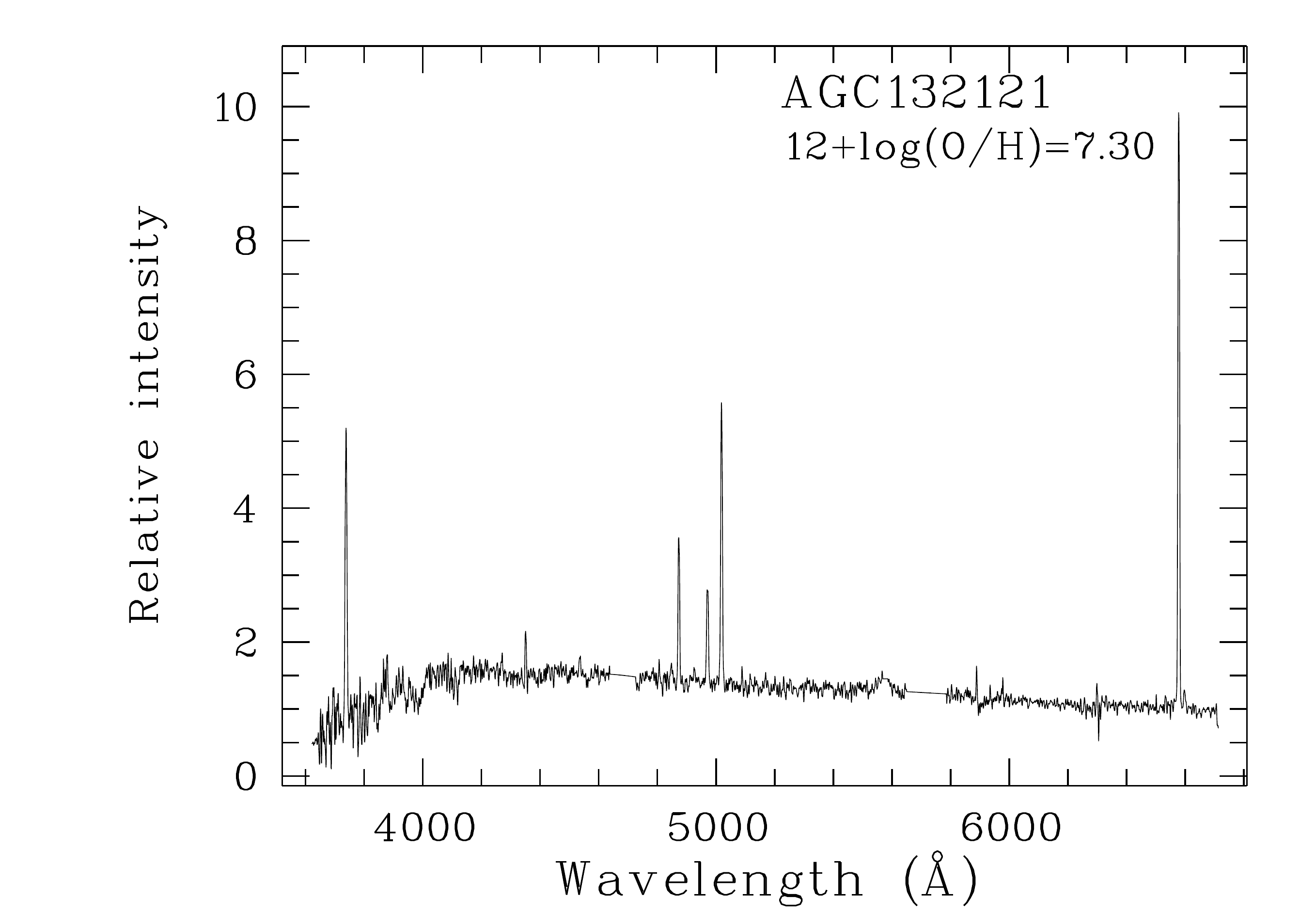}
\includegraphics[width=4.0cm,angle=-90,clip=]{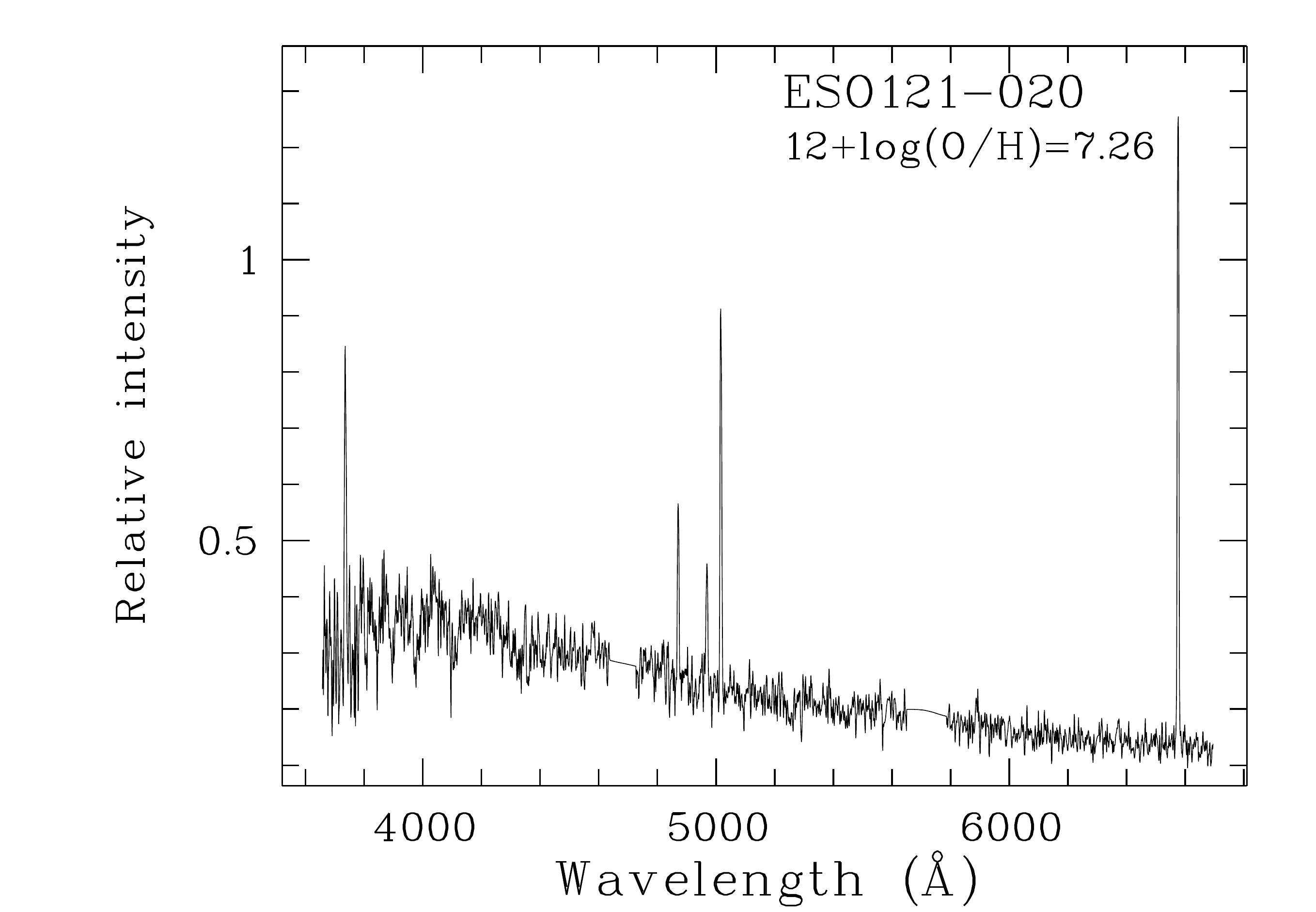}
\includegraphics[width=4.0cm,angle=-90,clip=]{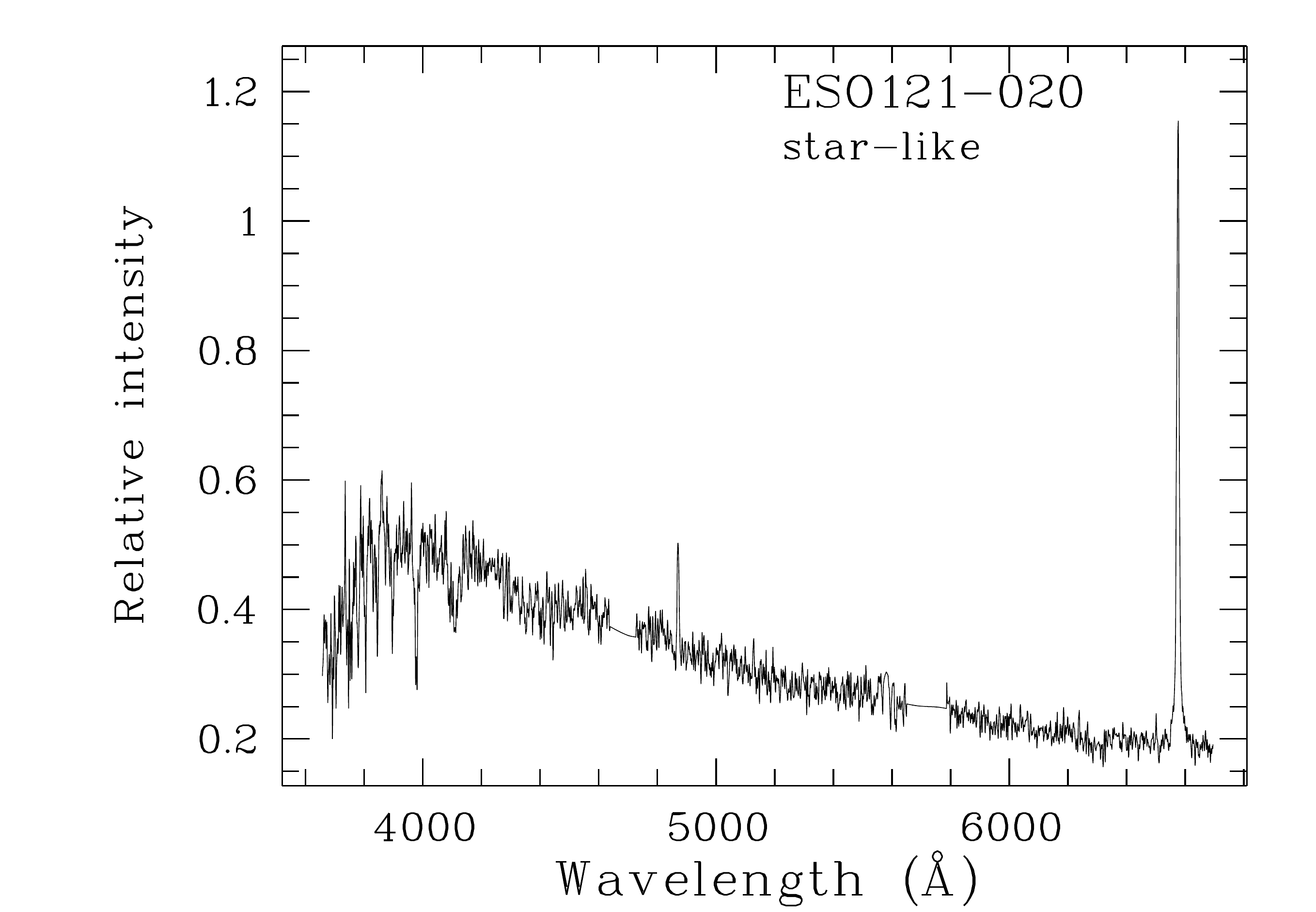}
\includegraphics[width=4.0cm,angle=-90,clip=]{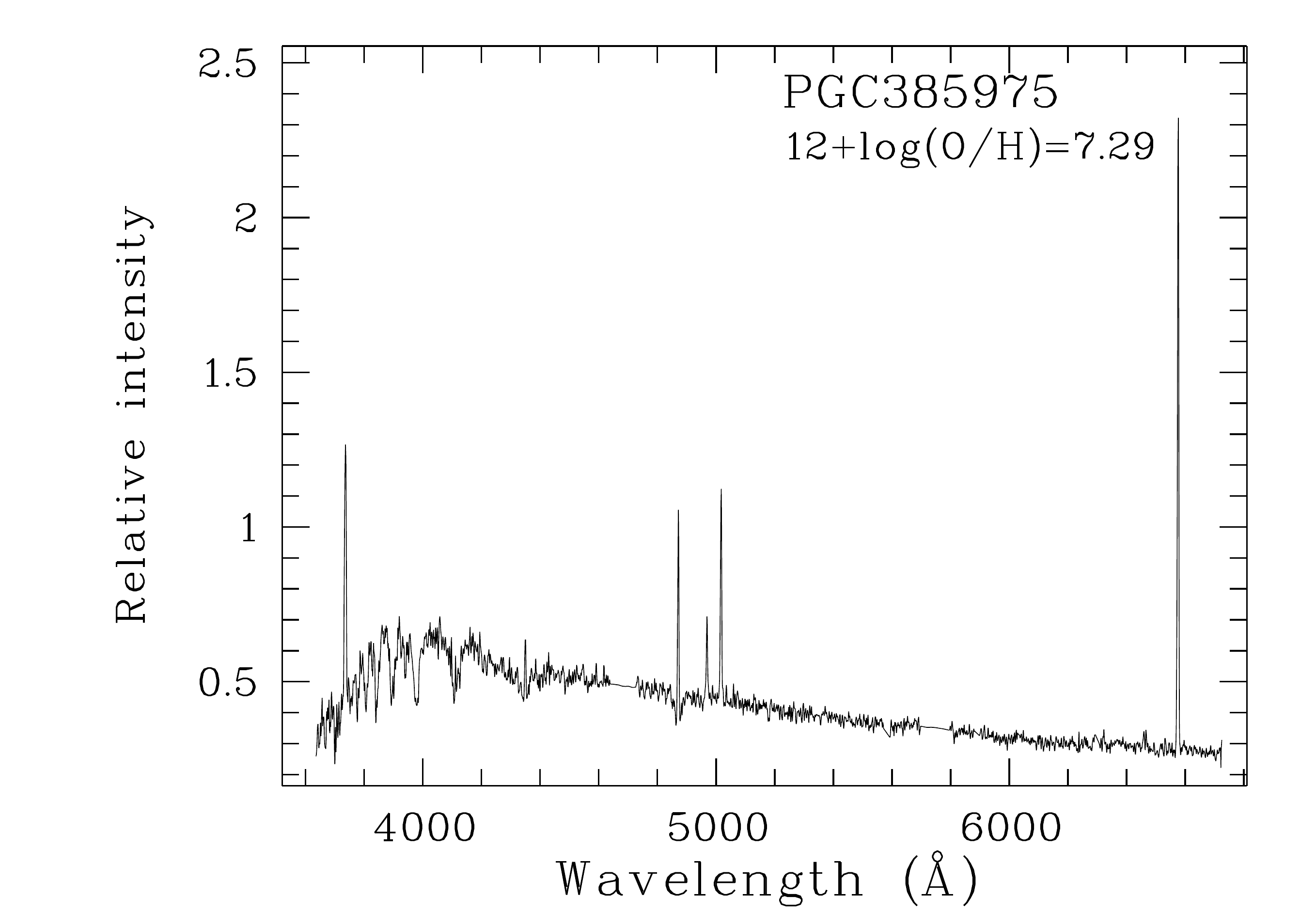}
\caption{1D spectra of void XMP candidates obtained with SALT. The
wavelengths are not in the rest-frame. The galaxy name and derived
value of
12+$\log$(O/H) are shown at the top of each box. See the text for
discussion of individual objects.
}
\label{fig:SALT_1Dp1}
\end{figure*}

In this Appendix we present plots of the 1D SALT spectra of the
XMP candidates discussed in this paper. The wavelengths on the $X$ axis
are observed, not in the rest-frame.
For the candidate galaxy PGC736507, we also obtained on the slit
the spectrum of a nearby bluish galaxy J0009--2852 (see comments in
Sec.~\ref{sec:individual}). We present it in Fig.~\ref{fig:SALT_1Dp1}
along with a spectrum of PGC736507.

For the program galaxy, ESO121-020, we give two
spectra: the first one of an \HII-region, the second of a
star-like object with only H$\alpha$ and H$\beta$ in emission
(see Sec.~\ref{sec:individual} for details).

\begin{figure*}
\includegraphics[width=4.0cm,angle=-90,clip=]{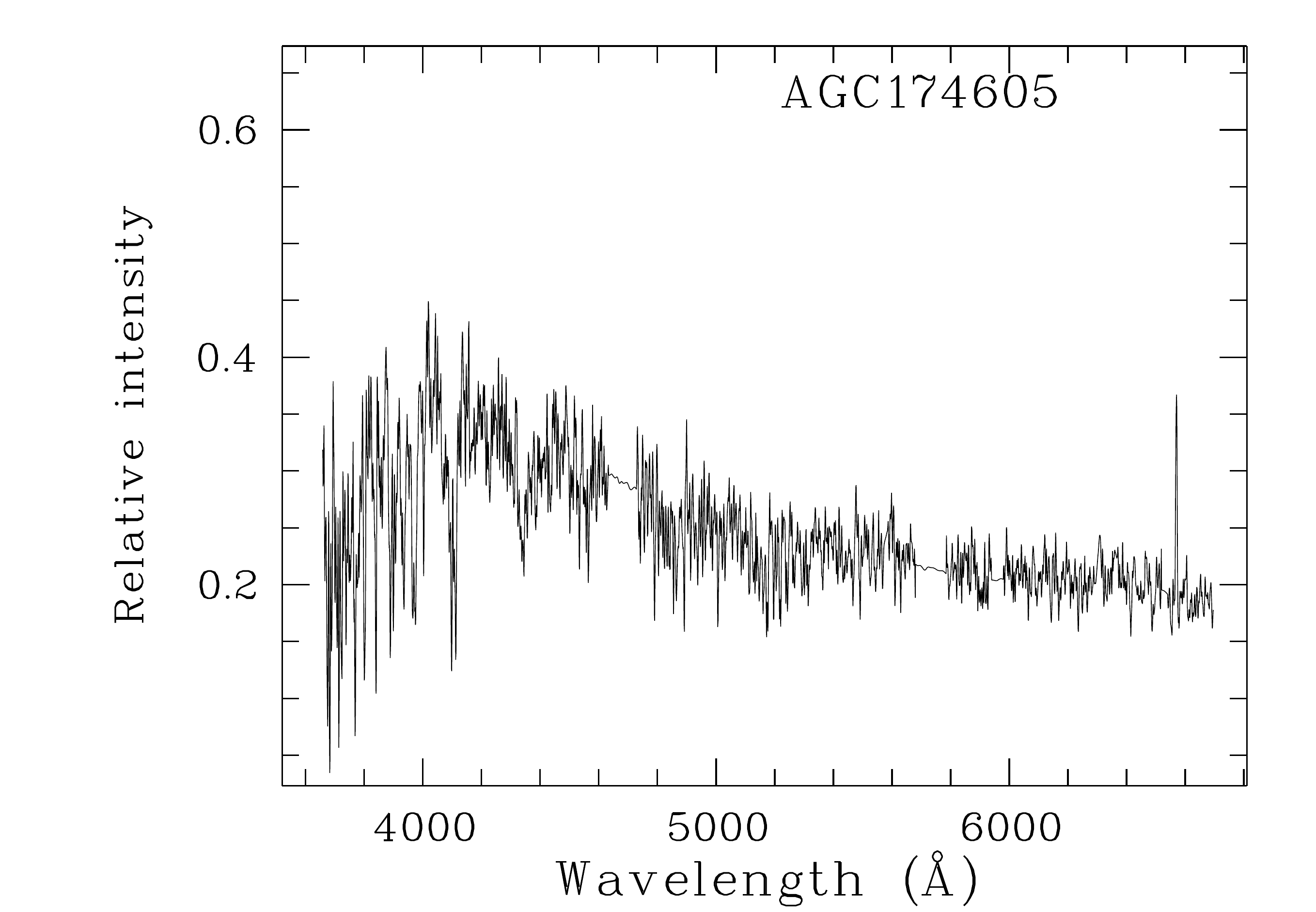}
\includegraphics[width=4.0cm,angle=-90,clip=]{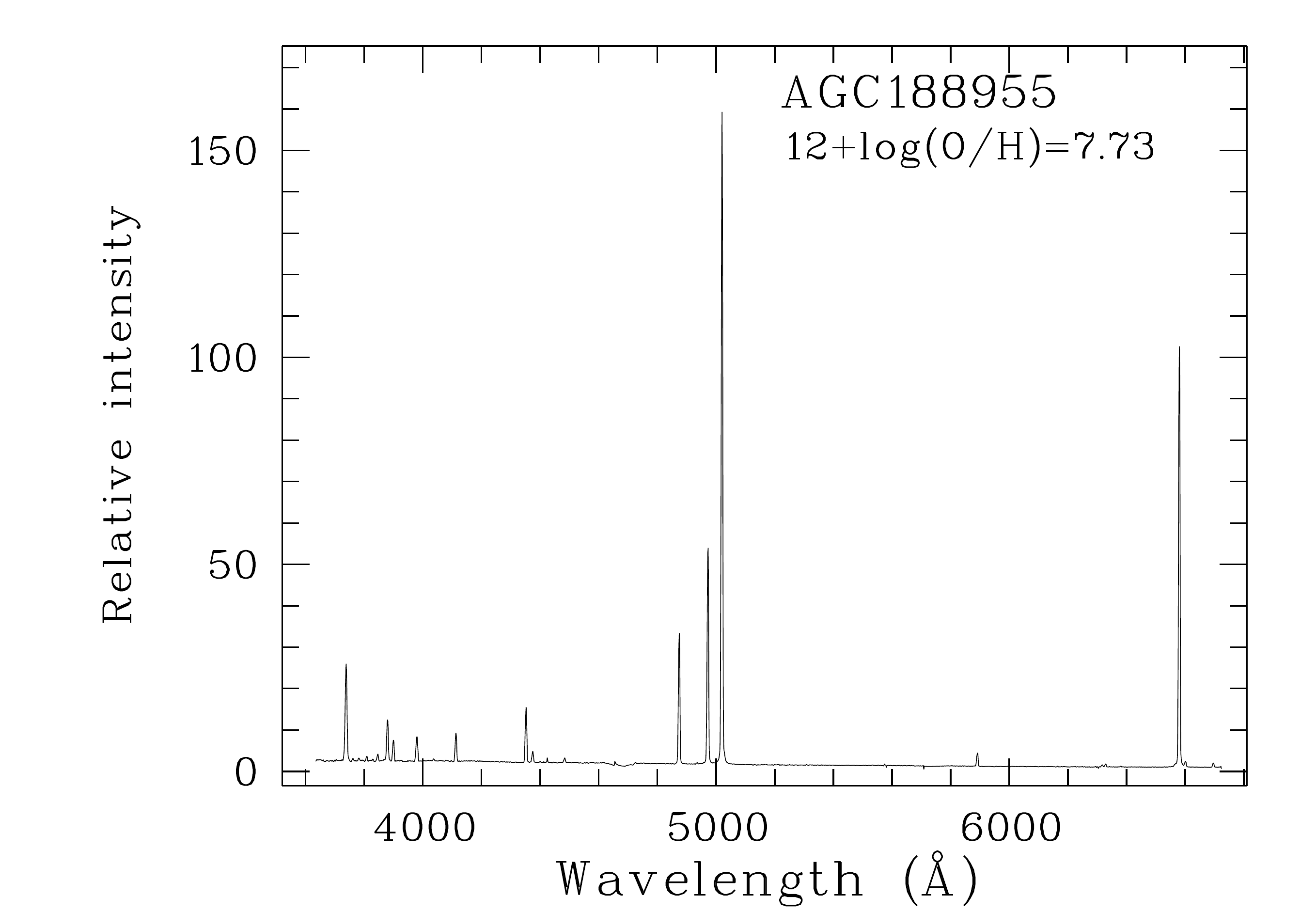}
\includegraphics[width=4.0cm,angle=-90,clip=]{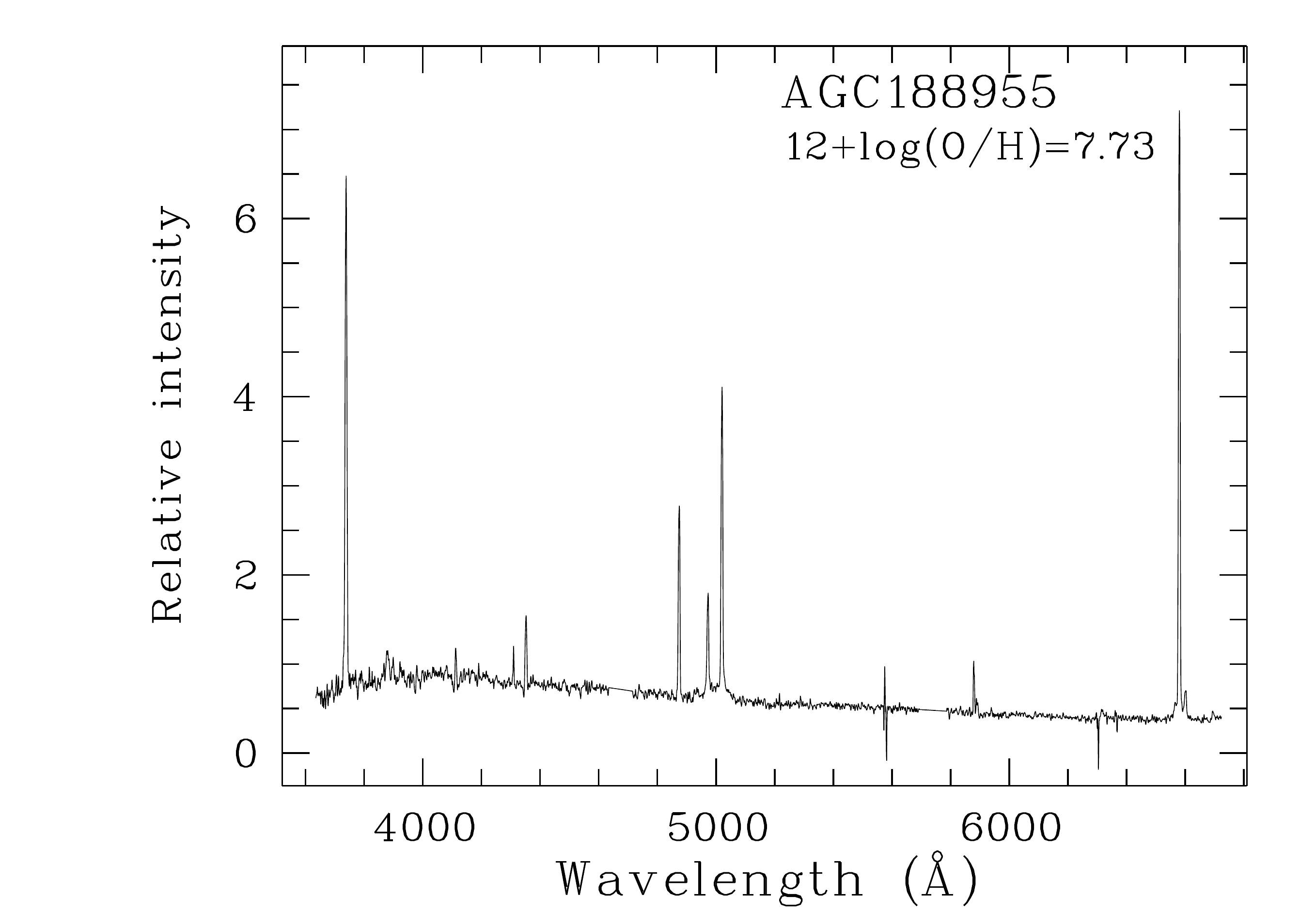}
\includegraphics[width=4.0cm,angle=-90,clip=]{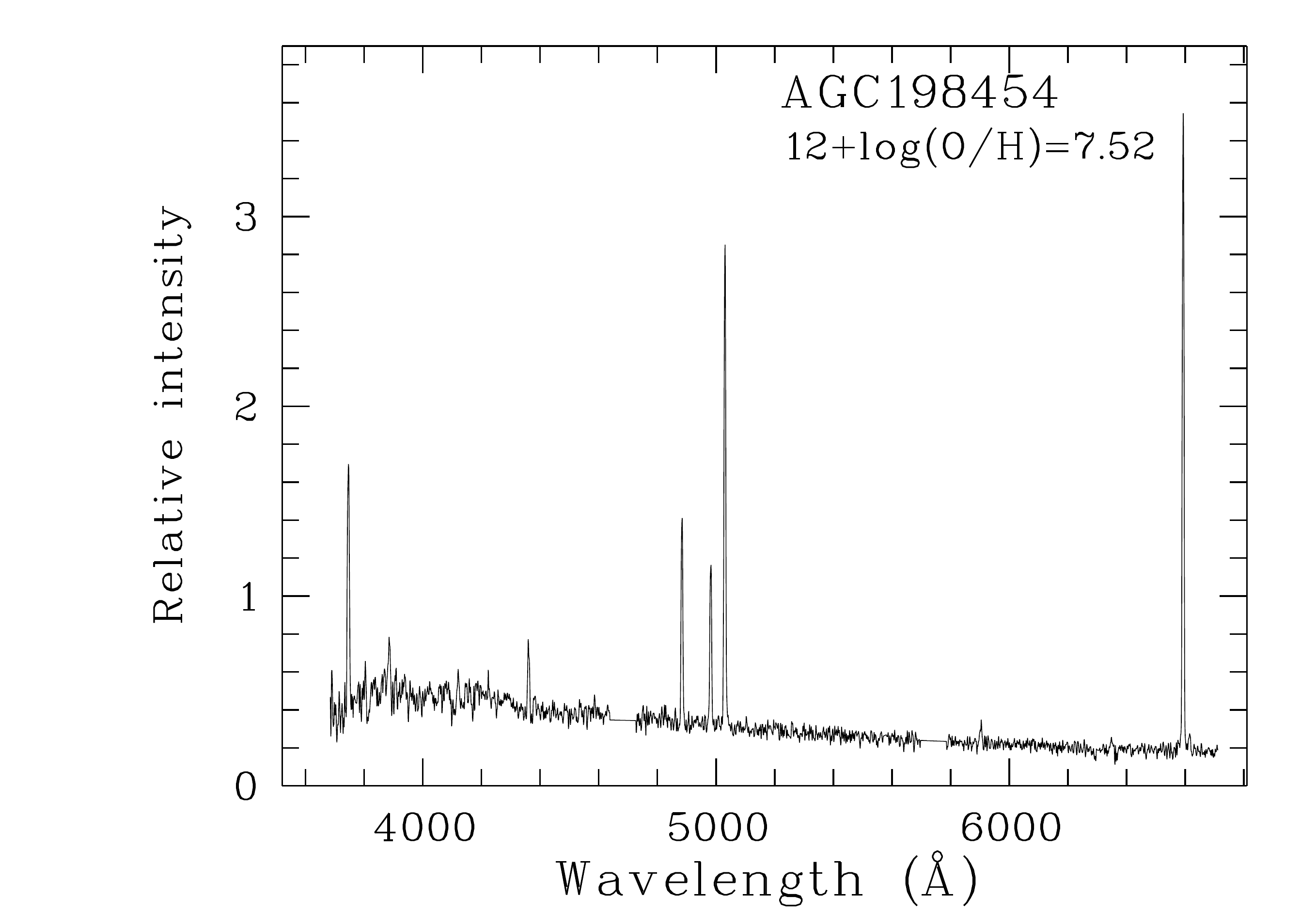}
\includegraphics[width=4.0cm,angle=-90,clip=]{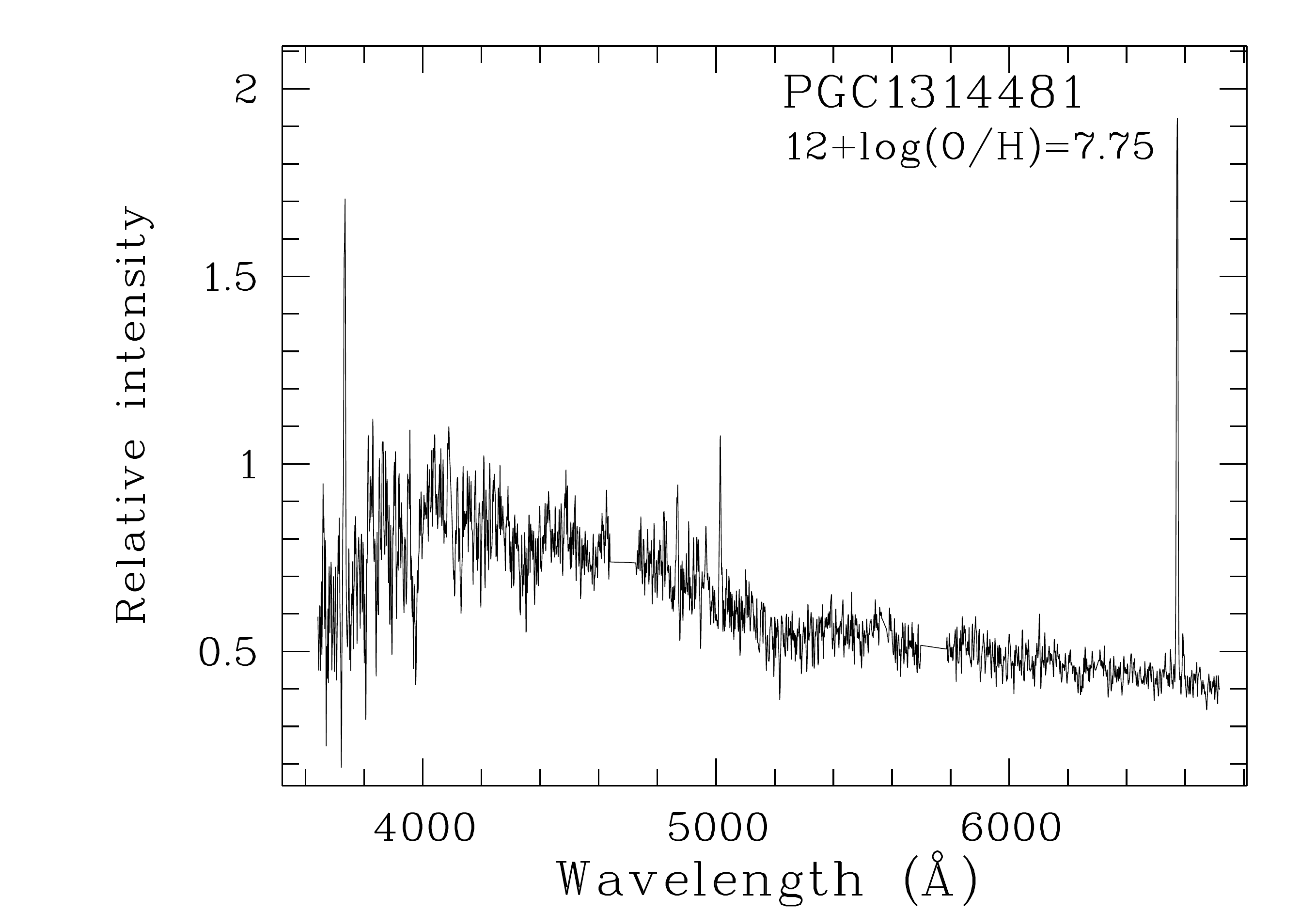}
\includegraphics[width=4.0cm,angle=-90,clip=]{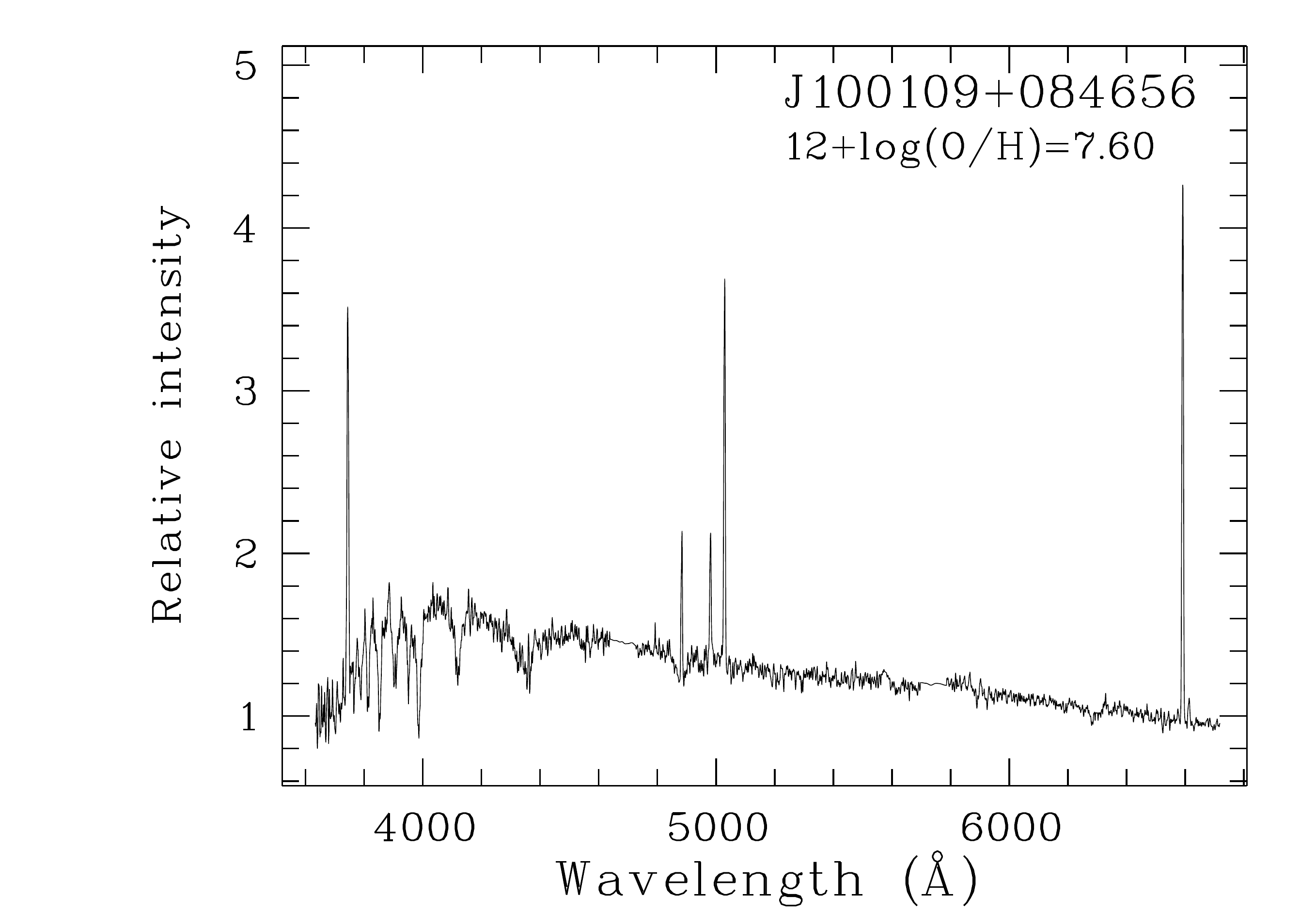}
\includegraphics[width=4.0cm,angle=-90,clip=]{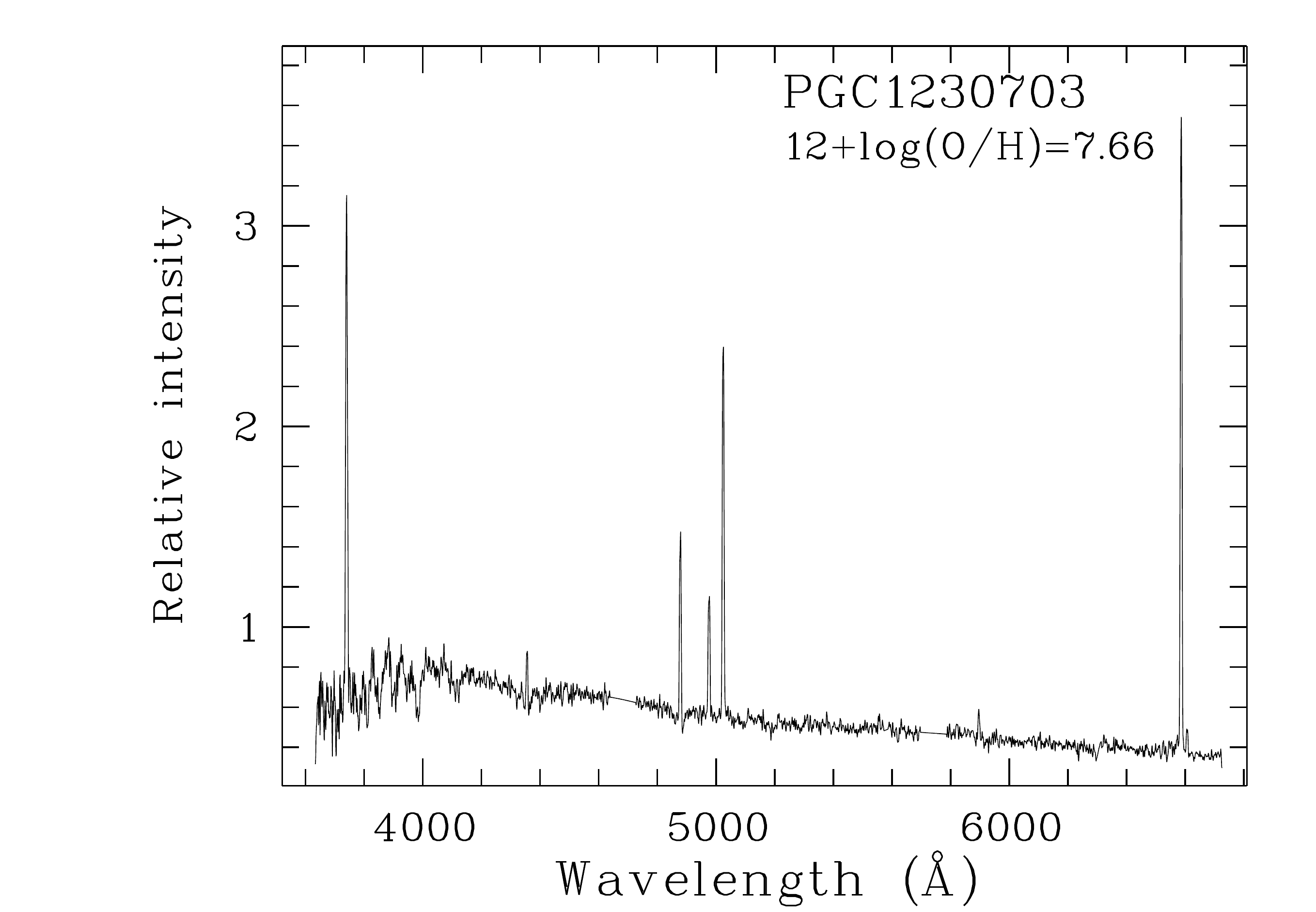}
\includegraphics[width=4.0cm,angle=-90,clip=]{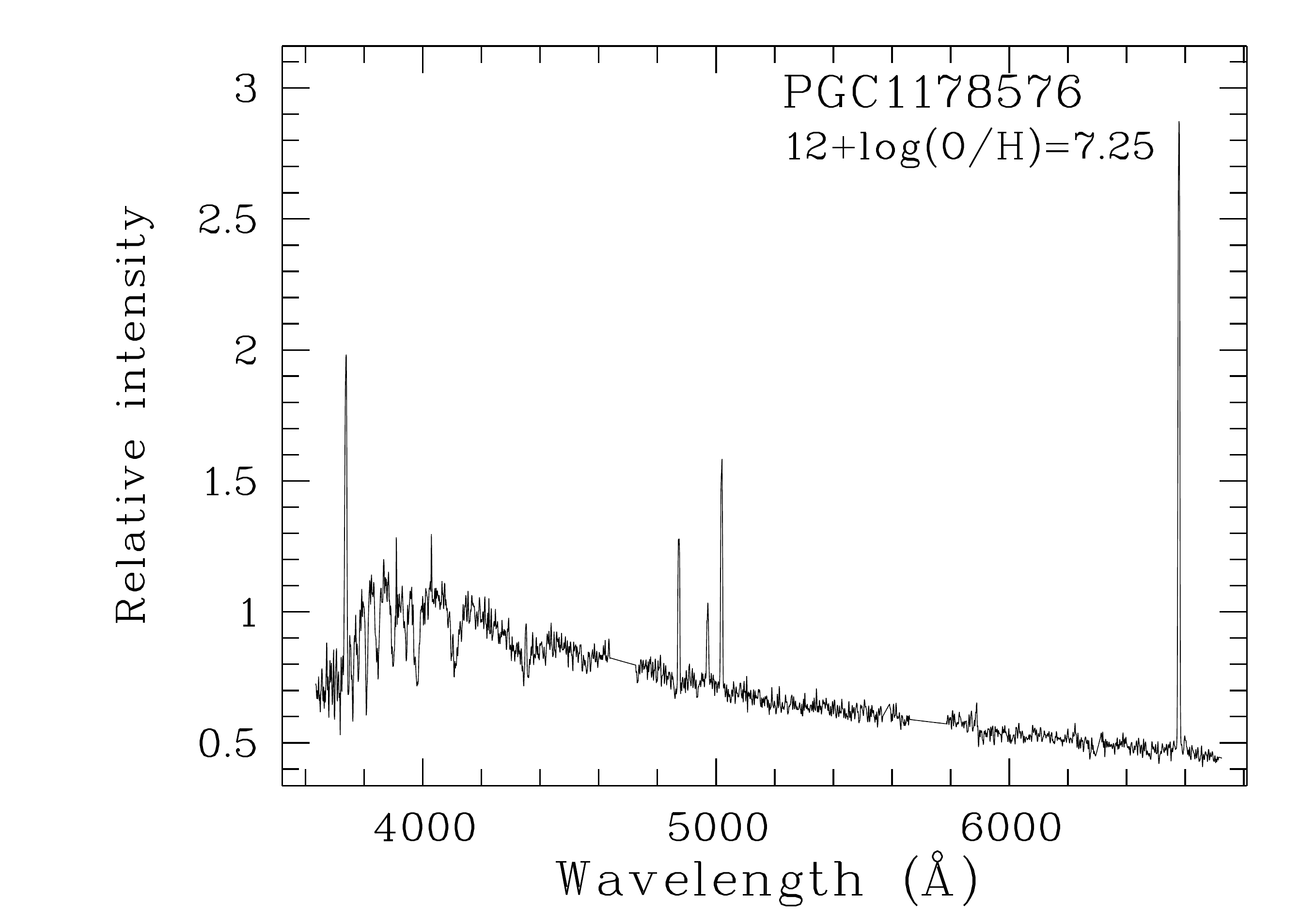}
\includegraphics[width=4.0cm,angle=-90,clip=]{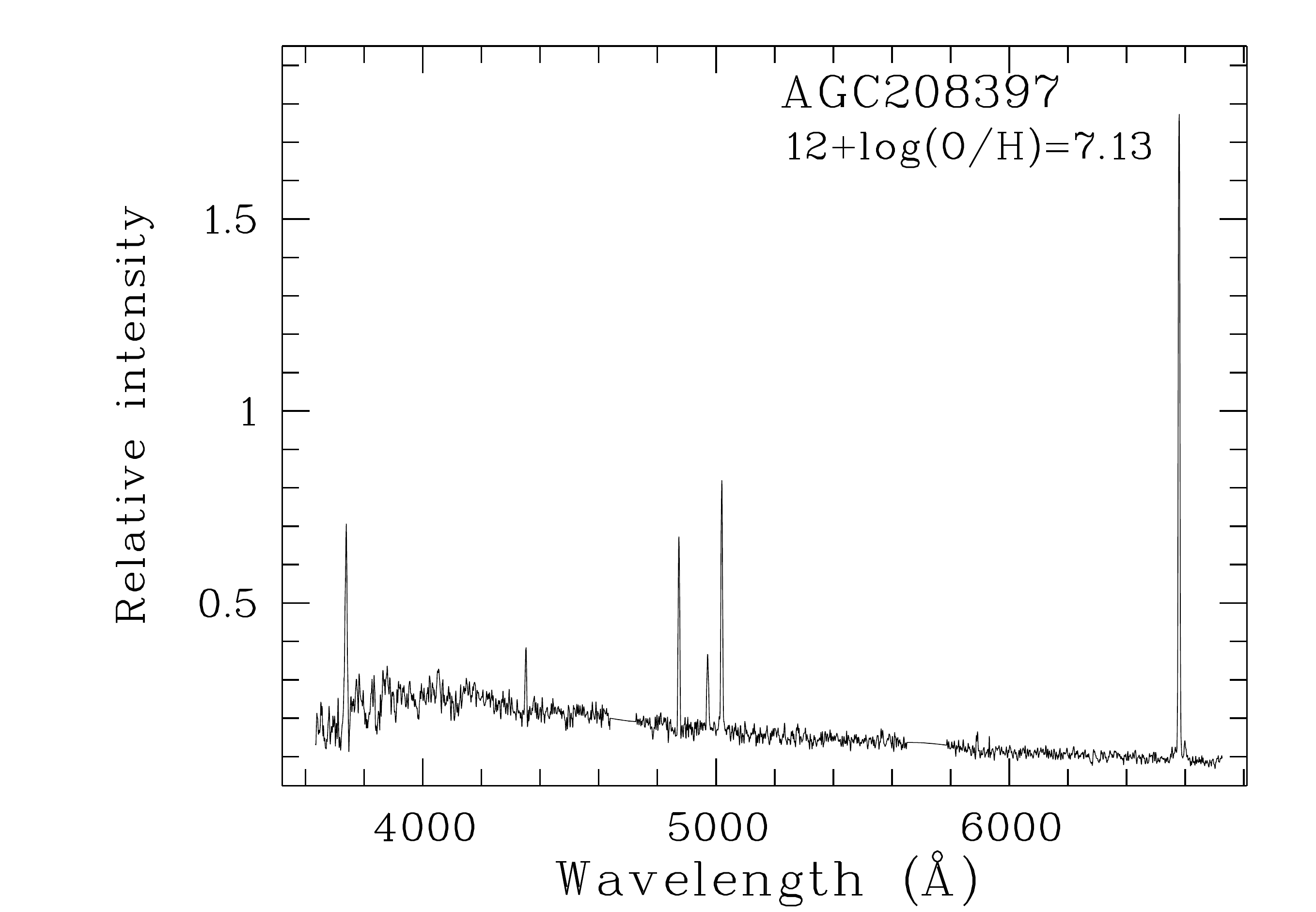}
\includegraphics[width=4.0cm,angle=-90,clip=]{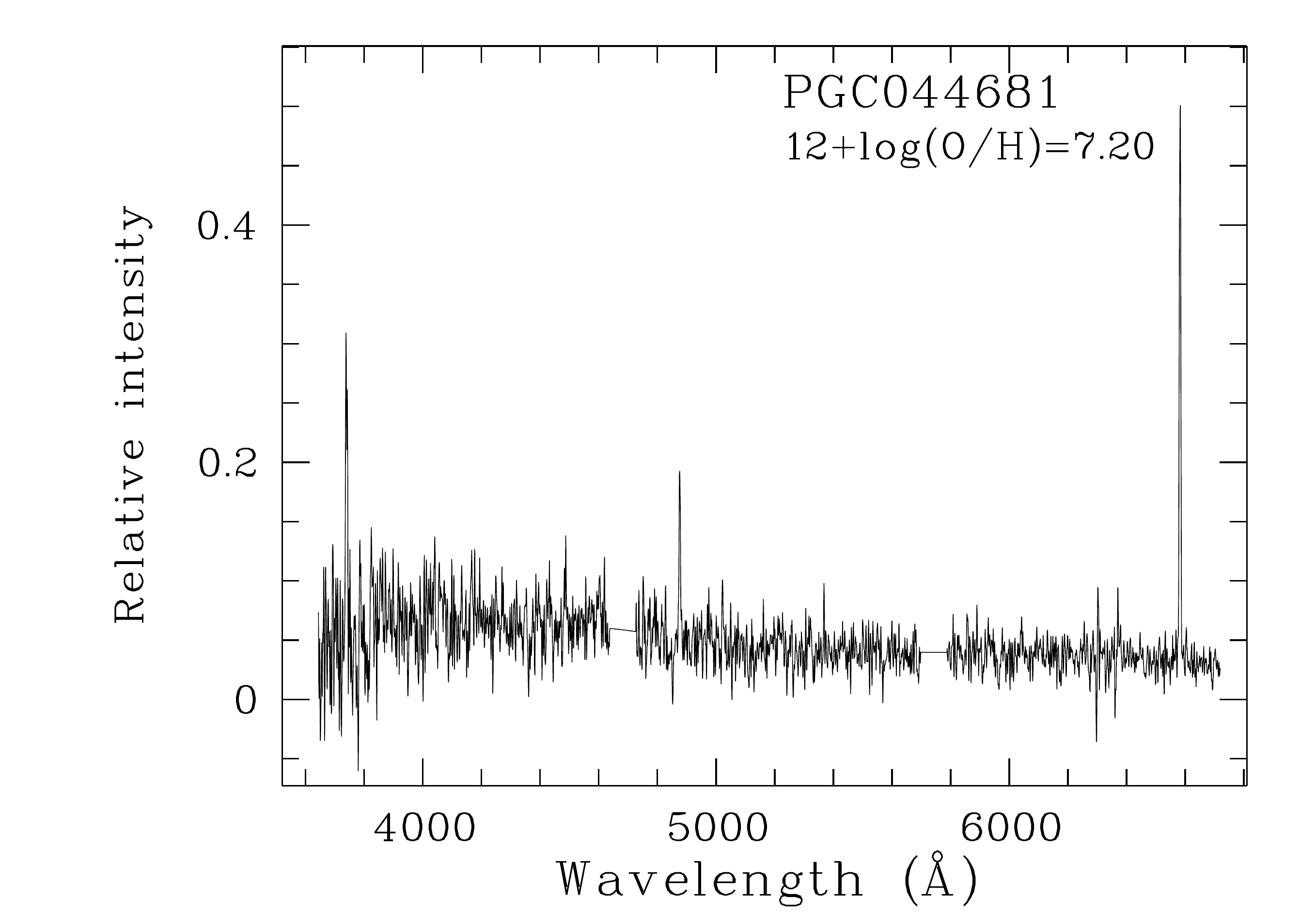}
\includegraphics[width=4.0cm,angle=-90,clip=]{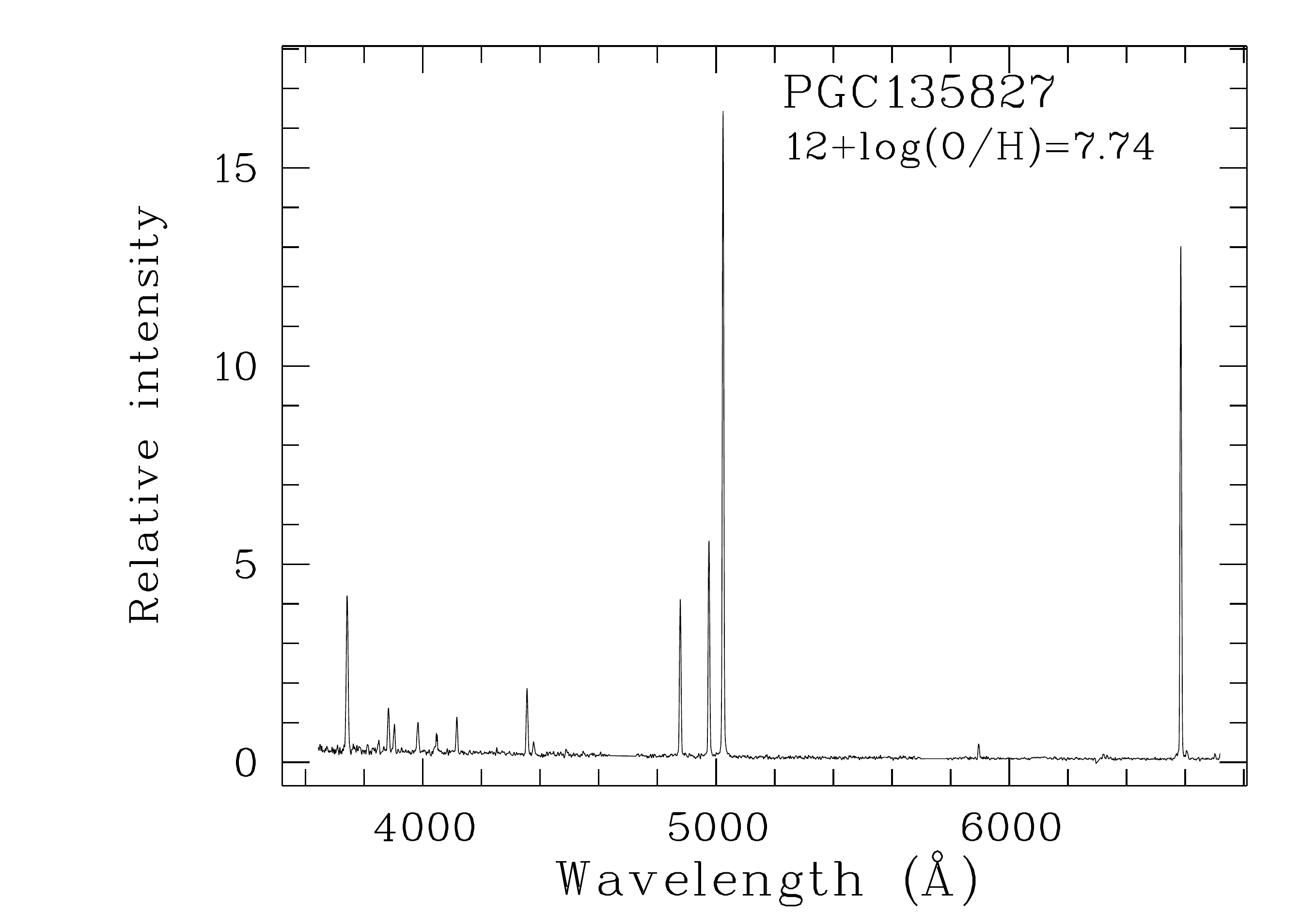}
\includegraphics[width=4.0cm,angle=-90,clip=]{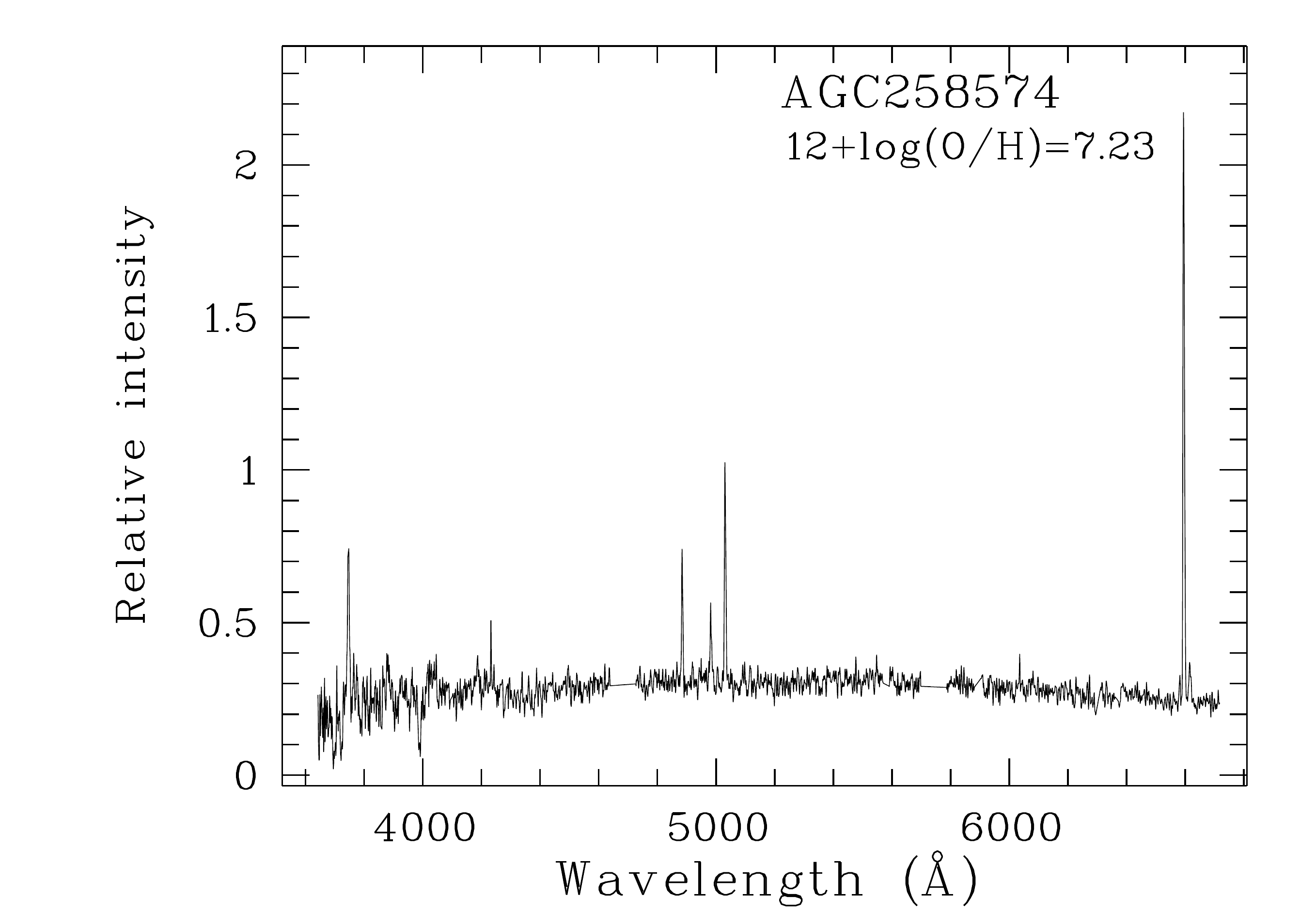}
\includegraphics[width=4.0cm,angle=-90,clip=]{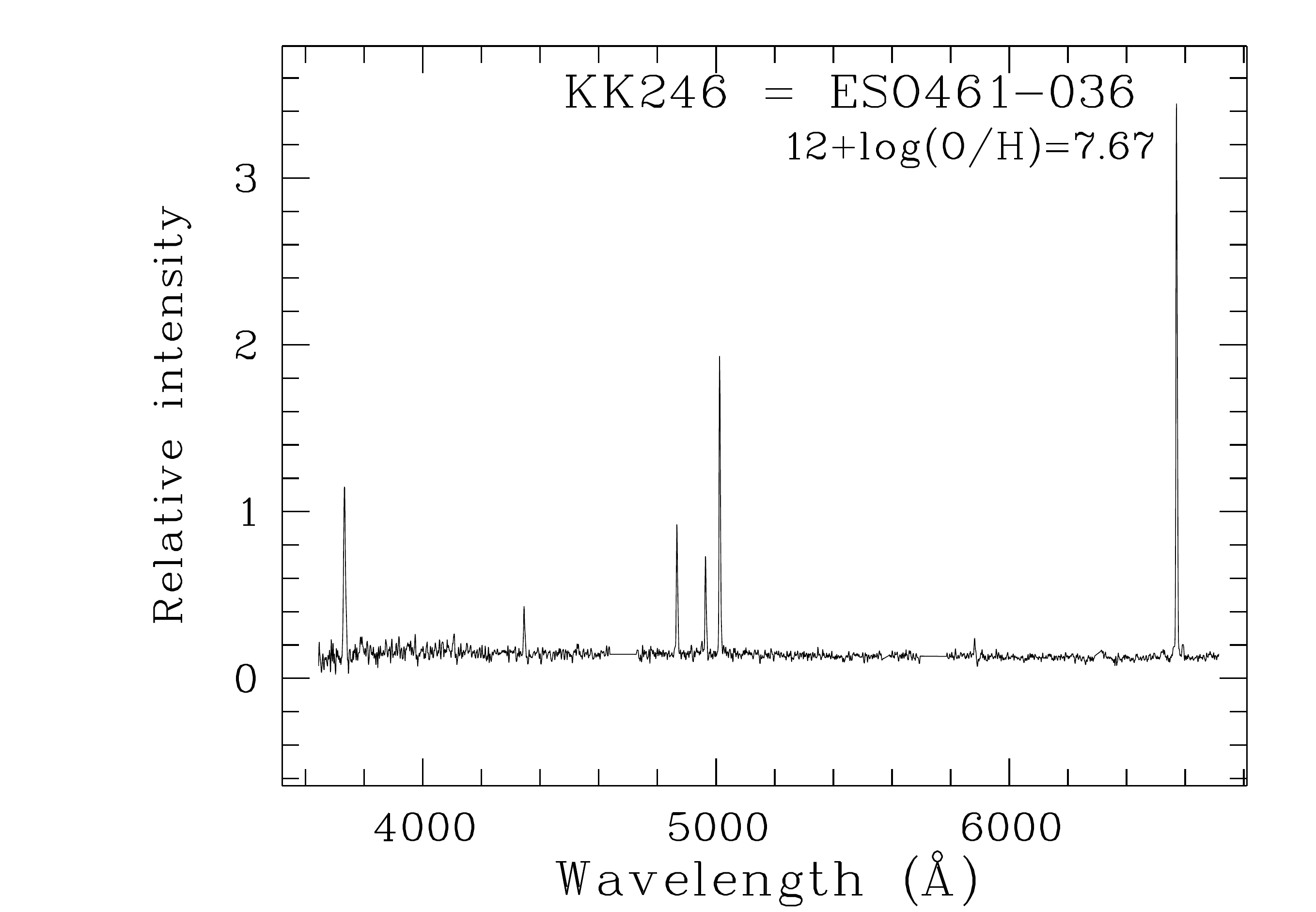}
\includegraphics[width=4.0cm,angle=-90,clip=]{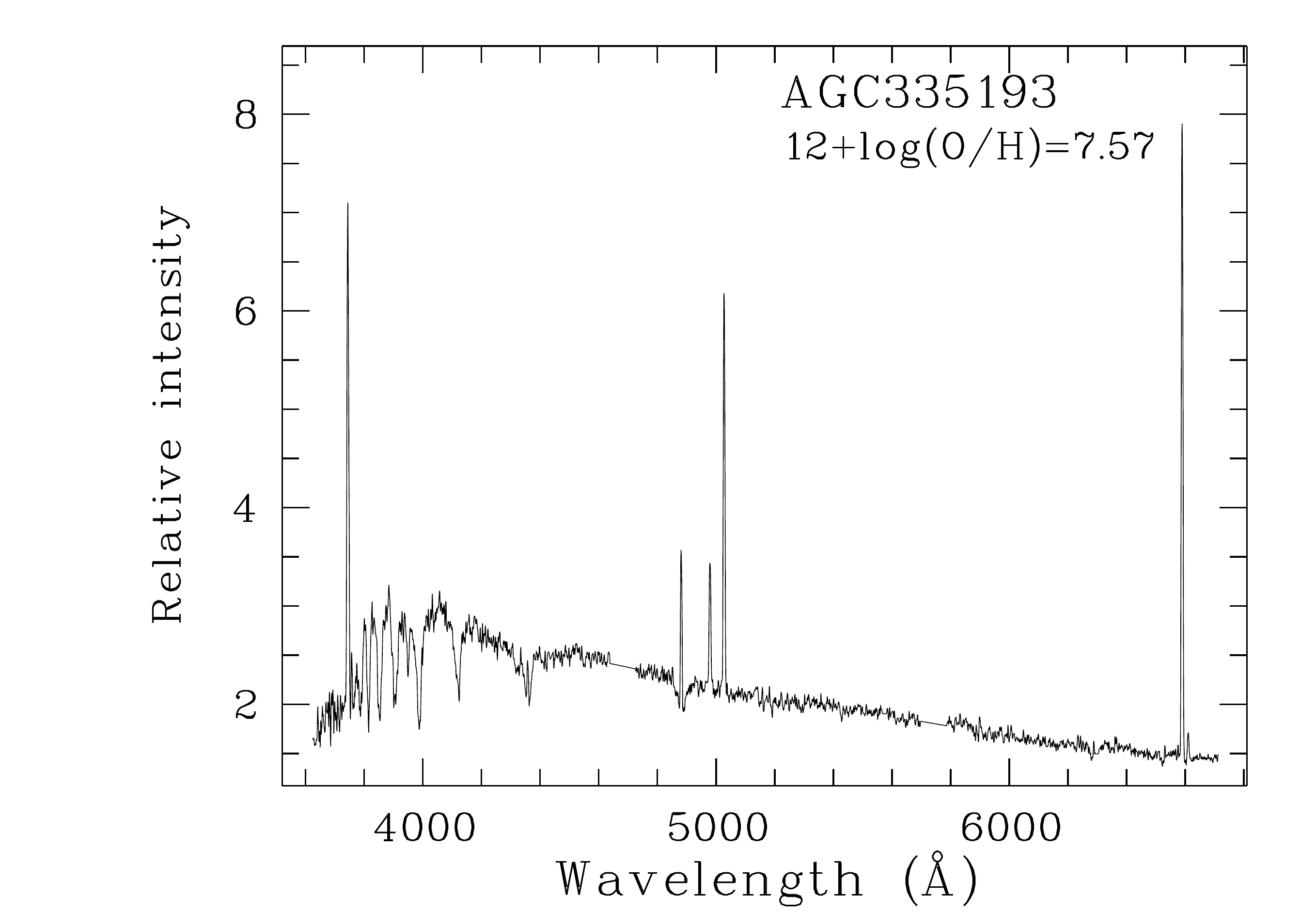}
\caption{1D spectra of void XMP candidates obtained with SALT.
The wavelengths are not in the rest-frame.
The galaxy name and derived value of
12+$\log$(O/H) are shown at the top of each box. See the text for
discussion of individual objects.
}
\label{fig:SALT_1Dp2}
\end{figure*}

For the galaxy, AGC188955, we obtained quite different spectra for two knots
which we also discuss in Section~\ref{sec:individual}. We show both
of them in Fig.~\ref{fig:SALT_1Dp2}.

\section{Tables with line fluxes and derived parameters}

The tables in this Appendix include the measured line fluxes $F(\lambda)$
(divided by the flux of H$\beta$)  and the line fluxes $I(\lambda)$,
corrected for extinction and underlying stellar Balmer absorption,
with their errors. The error for $F(H\beta)$ reflects its
measurement uncertainty.
The tables also include the measured $EW$ of H$\beta$ emission
and the derived parameters: the extinction coefficient $C(H\beta)$,
the equivalent width of the underlying stellar Balmer absorptions $EW(abs)$,
the electron temperatures T$_{\rm e}$(OIII) and T$_{\rm e}$(OII) in two zones
of Oxygen ionization. We also present the derived Oxygen abundances
in two stages of ionization and the total value of O/H, including its
value in units of 12+$\log$(O/H).
Electron densities $n_{\rm e}$ in \HII-regions could not be estimated
because the doublet [S{\sc ii}]$\lambda\lambda$6716,6730~\AA\
was outside the observed wavelength range. In this case we adopted
$n_{\rm e}$ to be  10 per cm$^3$, typical of \HII-regions in dIrr galaxies.

When the faint auroral line [O{\sc iii}]$\lambda$4363~\AA\ was found
in the spectra, O/H was estimated
via the direct method and marked as 12+$\log$(O/H)(T$_{\rm e}$). Otherwise
we used
two methods described in Sec.~\ref{sec:OH}. O/H values derived via the
semi-empirical method of \citet{IT07} are marked as (se) with subscript (c),
which means the small correction to the original value of O/H derived with
the method. The correction of 0.0-0.05 dex was derived in \citet{PaperVII} to
put estimates of O/H(se) onto the zero-point of the direct method.
For the lowest O/H objects, we employed the new empirical method by
\citet{Izotov19DR14} based on the relative fluxes of strong Oxygen lines.
The values of O/H estimated by this method are marked (s). The subscript (c)
for these estimates also means a small correction (--0.03 dex)
was applied to the original method value, which puts them
on the zero-point of the direct method.

We do not include the following objects from Table~\ref{tab:prop_summary}
in Tables B1-B8 with emission line data: a) PGC736507 and PGC493444 because
they are not included in the NVG sample and are interlopers due to an
incorrect radial velocity in the original 2dFGRS catalog; b) AGC104227
and AGC174605 -- since they show only faint H$\alpha$ emission in the spectra.

For the galaxy, AGC188955 (J0821+0419), for two knots we have very
different line ratios but rather similar  O/H ratios.
Therefore we include data for both knots.

\begin{table*}
\centering{
\caption{Line intensities and derived parameters of PGC000389, HIPASSJ0021+08 and PGC1190331}
\label{t:Intens}
\begin{tabular}{lcccccc} \hline
\rule{0pt}{10pt}
& \MC{2}{c}{PGC000389=J0005--4128}     
& \MC{2}{c}{HIPASSJ0021+08} 
& \MC{2}{c}{PGC1190331=J0109+0117}    \\ \hline
\rule{0pt}{10pt}
$\lambda_{0}$(\AA) Ion                  & $F(\lambda)/F(H\beta)$ & $I(\lambda)/I(H\beta)$ & $F(\lambda)/F(H\beta)$ & $I(\lambda)/I(H\beta)$ & $F(\lambda)/F(H\beta)$ & $I(\lambda)/I(H\beta)$  \\ \hline
3727\ [O\ {\sc ii}]\                           &  0.973$\pm$ 0.021 &  0.944$\pm$ 0.027      &  2.026$\pm$ 0.075 &  1.836$\pm$ 0.085    &  1.889$\pm$ 0.048 &  1.849$\pm$ 0.054  \\ 
3967\ [Ne\ {\sc iii}]\ +\ H7\                  &  0.166$\pm$ 0.010 &  0.199$\pm$ 0.015      &       ...         &        ...           &       ...         &        ...         \\ 
4101\ H$\delta$\                               &  0.197$\pm$ 0.007 &  0.228$\pm$ 0.011      &       ...         &        ...           &  0.145$\pm$ 0.016 &  0.168$\pm$ 0.023  \\ 
4340\ H$\gamma$\                               &  0.356$\pm$ 0.019 &  0.378$\pm$ 0.022      &  0.342$\pm$ 0.028 &  0.475$\pm$ 0.094    &  0.319$\pm$ 0.030 &  0.335$\pm$ 0.035  \\ 
4861\ H$\beta$\                                &  1.000$\pm$ 0.025 &  1.000$\pm$ 0.026      &  1.000$\pm$ 0.037 &  1.000$\pm$ 0.077    &  1.000$\pm$ 0.028 &  1.000$\pm$ 0.030  \\ 
4959\ [O\ {\sc iii}]\                          &  1.477$\pm$ 0.036 &  1.434$\pm$ 0.036      &  0.816$\pm$ 0.034 &  0.701$\pm$ 0.034    &  0.627$\pm$ 0.020 &  0.614$\pm$ 0.020  \\ 
5007\ [O\ {\sc iii}]\                          &  4.436$\pm$ 0.096 &  4.307$\pm$ 0.097      &  2.421$\pm$ 0.069 &  2.076$\pm$ 0.069    &  1.917$\pm$ 0.047 &  1.876$\pm$ 0.047  \\ 
6548\ [N\ {\sc ii}]\                           &  0.014$\pm$ 0.011 &  0.014$\pm$ 0.011      &  0.008$\pm$ 0.007 &  0.007$\pm$ 0.007    &  0.035$\pm$ 0.021 &  0.034$\pm$ 0.021  \\ 
6563\ H$\alpha$\                               &  2.857$\pm$ 0.120 &  2.793$\pm$ 0.131      &  3.289$\pm$ 0.117 &  2.767$\pm$ 0.129    &  2.806$\pm$ 0.091 &  2.761$\pm$ 0.100  \\ 
6584\ [N\ {\sc ii}]\                           &  0.043$\pm$ 0.029 &  0.042$\pm$ 0.029      &  0.026$\pm$ 0.026 &  0.021$\pm$ 0.025    &  0.110$\pm$ 0.029 &  0.107$\pm$ 0.029  \\ 
   &  &  &  &  & & \\
$C(H\beta)$\ dex                                 & \MC {2}{c}{0.00$\pm$0.05}                  & \MC {2}{c}{0.07$\pm$0.05}               & \MC {2}{c}{0.00$\pm$0.04}  \\
$EW(abs)$\ \AA\                                  & \MC {2}{c}{1.85$\pm$0.28}                  & \MC {2}{c}{2.30$\pm$0.92}               & \MC {2}{c}{0.55$\pm$0.26}   \\
$EW(H\beta)$\ \AA\                               & \MC {2}{c}{  62$\pm$ 1}                    & \MC {2}{c}{  14$\pm$ 1}                 & \MC {2}{c}{  25$\pm$ 1}    \\
\hline                                                                                                                                 
$T_{\rm e}$(OIII)(K)\                          & \MC {2}{c}{14966$\pm$1010}                 & \MC {2}{c}{17171$\pm$1029}              & \MC {2}{c}{17536$\pm$1014}   \\
$T_{\rm e}$(OII)(K)\                           & \MC {2}{c}{13888$\pm$517 }                 & \MC {2}{c}{14720$\pm$250 }              & \MC {2}{c}{14801$\pm$201 }   \\
O$^{+}$/H$^{+}$($\times$10$^5$)\               & \MC {2}{c}{1.095$\pm$0.138}                & \MC {2}{c}{1.765$\pm$0.124}             & \MC {2}{c}{1.747$\pm$0.090}   \\ 
O$^{++}$/H$^{+}$($\times$10$^5$)\              & \MC {2}{c}{4.760$\pm$0.815}                & \MC {2}{c}{1.656$\pm$0.231}             & \MC {2}{c}{1.415$\pm$0.186}   \\ 
O/H($\times$10$^5$)\                           & \MC {2}{c}{5.855$\pm$0.827}                & \MC {2}{c}{3.421$\pm$0.262}             & \MC {2}{c}{3.162$\pm$0.207}   \\ 
12+log(O/H)$_{c}$(se)  \                       & \MC {2}{c}{7.74$\pm$0.10}                  & \MC {2}{c}{7.51$\pm$0.07}               & \MC {2}{c}{7.48$\pm$0.08}    \\
\MC{3}{l}{~~} \
\end{tabular}
 }
\end{table*}

\begin{table*}
\centering{
\caption{Line intensities and derived parameters of AGC411446, AGC114584 and AGC123223}
\label{t:Intens}
\begin{tabular}{lcccccc} \hline
\rule{0pt}{10pt}
& \MC{2}{c}{AGC411446=J0110--0000}   
& \MC{2}{c}{AGC114584=J0112+0152}   
& \MC{2}{c}{AGC123223=J0247+1005}     \\ \hline
\rule{0pt}{10pt}
$\lambda_{0}$(\AA) Ion                  & $F(\lambda)/F(H\beta)$ & $I(\lambda)/I(H\beta)$ & $F(\lambda)/F(H\beta)$ & $I(\lambda)/I(H\beta)$ & $F(\lambda)/F(H\beta)$ & $I(\lambda)/I(H\beta)$  \\ \hline
%
3727\ [O\ {\sc ii}]\                           &  1.105$\pm$ 0.057 &  1.067$\pm$ 0.064      &  1.232$\pm$ 0.052 &  0.951$\pm$ 0.056    &  2.154$\pm$ 0.232 &  2.295$\pm$ 0.340  \\ 
3967\ [Ne\ {\sc iii}]\ +\ H7\                  &  0.027$\pm$ 0.005 &  0.197$\pm$ 0.062      &       ...         &        ...           &       ...         &        ...         \\ 
4101\ H$\delta$\                               &  0.098$\pm$ 0.010 &  0.271$\pm$ 0.037      &       ...         &        ...           &       ...         &        ...         \\ 
4340\ H$\gamma$\                               &  0.346$\pm$ 0.035 &  0.480$\pm$ 0.058      &  0.283$\pm$ 0.022 &  0.478$\pm$ 0.057    &  0.231$\pm$ 0.058 &  0.487$\pm$ 0.185  \\ 
4861\ H$\beta$\                                &  1.000$\pm$ 0.024 &  1.000$\pm$ 0.028      &  1.000$\pm$ 0.030 &  1.000$\pm$ 0.041    &  1.000$\pm$ 0.089 &  1.000$\pm$ 0.121  \\ 
4959\ [O\ {\sc iii}]\                          &  0.276$\pm$ 0.025 &  0.239$\pm$ 0.025      &  0.544$\pm$ 0.022 &  0.418$\pm$ 0.022    &  0.643$\pm$ 0.093 &  0.479$\pm$ 0.092  \\ 
5007\ [O\ {\sc iii}]\                          &  0.822$\pm$ 0.026 &  0.711$\pm$ 0.026      &  1.591$\pm$ 0.043 &  1.223$\pm$ 0.043    &  1.769$\pm$ 0.144 &  1.303$\pm$ 0.140  \\ 
6548\ [N\ {\sc ii}]\                           &       ...         &        ...             &  0.015$\pm$ 0.043 &  0.011$\pm$ 0.043    &  0.042$\pm$ 0.119 &  0.022$\pm$ 0.084  \\
6563\ H$\alpha$\                               &  3.368$\pm$ 0.082 &  2.715$\pm$ 0.082      &  3.372$\pm$ 0.144 &  2.729$\pm$ 0.165    &  4.965$\pm$ 0.380 &  2.770$\pm$ 0.304  \\ 
6584\ [N\ {\sc ii}]\                           &       ...         &        ...             &  0.047$\pm$ 0.056 &  0.036$\pm$ 0.056    &  0.147$\pm$ 0.140 &  0.078$\pm$ 0.098  \\
  & & & & &  & \\
$C(H\beta)$\ dex                              & \MC {2}{c}{0.14$\pm$0.03}                  & \MC {2}{c}{0.01$\pm$0.06}                & \MC {2}{c}{0.47$\pm$0.10}         \\
$EW(abs)$\ \AA\                                 & \MC {2}{c}{2.65$\pm$0.05}                  & \MC {2}{c}{3.80$\pm$0.18}                & \MC {2}{c}{1.20$\pm$0.08}          \\
$EW(H\beta)$\ \AA\                            & \MC {2}{c}{  18$\pm$ 1}                    & \MC {2}{c}{  13$\pm$ 1}                  & \MC {2}{c}{   4$\pm$ 1}           \\
\hline
$T_{\rm e}$(OIII)(K)\                          & \MC {2}{c}{22089$\pm$1056}                 & \MC {2}{c}{20596$\pm$1038}               & \MC {2}{c}{17904$\pm$1302}   \\
$T_{\rm e}$(OII)(K)\                           & \MC {2}{c}{16222$\pm$229 }                 & \MC {2}{c}{15636$\pm$14  }               & \MC {2}{c}{14866$\pm$200 }   \\
O$^{+}$/H$^{+}$($\times$10$^5$)\               & \MC {2}{c}{0.765$\pm$0.056}                & \MC {2}{c}{0.760$\pm$0.045}              & \MC {2}{c}{2.140$\pm$0.330}   \\ 
O$^{++}$/H$^{+}$($\times$10$^5$)\              & \MC {2}{c}{0.334$\pm$0.033}                & \MC {2}{c}{0.663$\pm$0.070}              & \MC {2}{c}{0.967$\pm$0.180}   \\ 
O/H($\times$10$^5$)\                           & \MC {2}{c}{1.099$\pm$0.065}                & \MC {2}{c}{1.422$\pm$0.083}              & \MC {2}{c}{3.107$\pm$0.376}   \\ 
12+log(O/H)$_{c}$(se)  \                       & \MC {2}{c}{7.04$\pm$0.08}                  & \MC {2}{c}{7.15$\pm$0.08}                & \MC {2}{c}{7.47$\pm$0.09}    \\
12+log(O/H)$_{c}$(s)         \                 & \MC {2}{c}{7.05$\pm$0.05}                  & \MC {2}{c}{7.15$\pm$0.05}                & \MC {2}{c}{...}    \\
\MC{3}{l}{~~} \
\end{tabular}
 }
\end{table*}

\begin{table*}
\centering{
\caption{Line intensities and derived parameters of AGC124629, AGC132121 and ESO121-020 }
\label{t:Intens}
\begin{tabular}{lcccccc} \hline
\rule{0pt}{10pt}
& \MC{2}{c}{AGC124629=J0256+0248}    
& \MC{2}{c}{AGC132121=J0306+0520}    
& \MC{2}{c}{ESO121-020=J0615--5743}     \\ \hline
\rule{0pt}{10pt}
$\lambda_{0}$(\AA) Ion                  & $F(\lambda)/F(H\beta)$ & $I(\lambda)/I(H\beta)$ & $F(\lambda)/F(H\beta)$ & $I(\lambda)/I(H\beta)$ & $F(\lambda)/F(H\beta)$ & $I(\lambda)/I(H\beta)$  \\ \hline
%
3727\ [O\ {\sc ii}]\                           &  0.972$\pm$ 0.119 &  0.976$\pm$ 0.123      &  2.207$\pm$ 0.117 &  1.673$\pm$ 0.128     &  2.479$\pm$ 0.302 &  1.663$\pm$ 0.318  \\ 
4340\ H$\gamma$\                               &       ...         &        ...             &  0.282$\pm$ 0.025 &  0.477$\pm$ 0.067     &       ...         &        ...         \\ 
4861\ H$\beta$\                                &  1.000$\pm$ 0.039 &  1.000$\pm$ 0.047      &  1.000$\pm$ 0.046 &  1.000$\pm$ 0.067     &  1.000$\pm$ 0.105 &  1.000$\pm$ 0.184  \\ 
4959\ [O\ {\sc iii}]\                          &  0.157$\pm$ 0.042 &  0.153$\pm$ 0.042      &  0.684$\pm$ 0.037 &  0.501$\pm$ 0.037     &  0.703$\pm$ 0.084 &  0.469$\pm$ 0.084  \\ 
5007\ [O\ {\sc iii}]\                          &  0.463$\pm$ 0.044 &  0.450$\pm$ 0.043      &  2.037$\pm$ 0.079 &  1.490$\pm$ 0.079     &  1.818$\pm$ 0.159 &  1.214$\pm$ 0.159  \\ 
6548\ [N\ {\sc ii}]\                           &       ...         &        ...             &  0.023$\pm$ 0.048 &  0.016$\pm$ 0.046     &  0.017$\pm$ 0.116 &  0.011$\pm$ 0.115  \\
6563\ H$\alpha$\                               &  2.847$\pm$ 0.097 &  2.701$\pm$ 0.102      &  3.616$\pm$ 0.164 &  2.750$\pm$ 0.185     &  3.793$\pm$ 0.355 &  2.762$\pm$ 0.424  \\      
6584\ [N\ {\sc ii}]\                           &       ...         &        ...             &  0.072$\pm$ 0.058 &  0.051$\pm$ 0.056     &  0.083$\pm$ 0.149 &  0.055$\pm$ 0.148  \\
 & & & &   & & \\
$C(H\beta)$\ dex                            & \MC {2}{c}{0.04$\pm$0.04}                  & \MC {2}{c}{0.04$\pm$0.06}                 & \MC {2}{c}{0.01$\pm$0.12}                 \\
$EW(abs)$\ \AA\                               & \MC {2}{c}{0.60$\pm$0.57}                  & \MC {2}{c}{3.95$\pm$0.22}                 & \MC {2}{c}{2.20$\pm$0.39}                  \\
$EW(H\beta)$\ \AA\                          & \MC {2}{c}{  24$\pm$ 1}                    & \MC {2}{c}{  11$\pm$ 1}                   & \MC {2}{c}{   4$\pm$ 1}                   \\
\hline
$T_{\rm e}$(OIII)(K)\                          & \MC {2}{c}{23542$\pm$1271}                 & \MC {2}{c}{18538$\pm$1076}                & \MC {2}{c}{19079$\pm$1412}     \\
$T_{\rm e}$(OII)(K)\                           & \MC {2}{c}{16499$\pm$208 }                 & \MC {2}{c}{14939$\pm$82  }                & \MC {2}{c}{14962$\pm$15  }     \\
O$^{+}$/H$^{+}$($\times$10$^5$)\               & \MC {2}{c}{0.666$\pm$0.087}                & \MC {2}{c}{1.536$\pm$0.120}               & \MC {2}{c}{1.519$\pm$0.291}     \\
O$^{++}$/H$^{+}$($\times$10$^5$)\              & \MC {2}{c}{0.188$\pm$0.027}                & \MC {2}{c}{1.002$\pm$0.133}               & \MC {2}{c}{0.796$\pm$0.151}     \\
O/H($\times$10$^5$)\                           & \MC {2}{c}{0.854$\pm$0.091}                & \MC {2}{c}{2.538$\pm$0.179}               & \MC {2}{c}{2.315$\pm$0.328}     \\
12+log(O/H)$_{c}$(se)  \                       & \MC {2}{c}{6.93$\pm$0.09}                  & \MC {2}{c}{7.39$\pm$0.07}                 & \MC {2}{c}{7.35$\pm$0.09}      \\
12+log(O/H)$_{c}$(s)         \                 & \MC {2}{c}{6.95$\pm$0.06}                  & \MC {2}{c}{7.30$\pm$0.06}                 & \MC {2}{c}{7.26$\pm$0.07}    \\
\MC{3}{l}{~~} \
\end{tabular}
 }
\end{table*}

\begin{table*}
\centering{
\caption{Line intensities and derived parameters  of PGC385975 and AGC188955=J0821+0419 }
\label{t:Intens}
\begin{tabular}{lcccccc} \hline
\rule{0pt}{10pt}
& \MC{2}{c}{PGC385975=J0616--5745}   
& \MC{2}{c}{AGC188955--bright component}   
& \MC{2}{c}{AGC188955--faint component}     \\ \hline
\rule{0pt}{10pt}
$\lambda_{0}$(\AA) Ion                  & $F(\lambda)/F(H\beta)$ & $I(\lambda)/I(H\beta)$ & $F(\lambda)/F(H\beta)$ & $I(\lambda)/I(H\beta)$ & $F(\lambda)/F(H\beta)$ & $I(\lambda)/I(H\beta)$  \\ \hline
%
3727\ [O\ {\sc ii}]\                    &  2.121$\pm$ 0.282 &  2.353$\pm$ 0.389             &  0.899$\pm$0.014  &  0.923$\pm$0.017              &   2.959$\pm$0.068  &   2.912$\pm$0.076  \\ 
3868\ [Ne\ {\sc iii}]\                  &       ...         &        ...                    &  0.331$\pm$0.008  &  0.336$\pm$0.009              &   0.222$\pm$0.041  &   0.218$\pm$0.041  \\
3889\ [He\ {\sc i}] +\ H8\              &       ...         &        ...                    &  0.162$\pm$0.016  &  0.241$\pm$0.028              &       ...          &       ...          \\
3967\ [Ne\ {\sc iii}]\ +\ H7\           &       ...         &        ...                    &  0.230$\pm$0.005  &  0.304$\pm$0.008              &       ...          &       ...          \\
4101\ H$\delta$\                        &  0.092$\pm$ 0.026 &  0.265$\pm$ 0.098             &  0.219$\pm$0.004  &  0.293$\pm$0.007              &   0.161$\pm$0.011  &   0.178$\pm$0.016  \\ 
4340\ H$\gamma$\                        &  0.338$\pm$ 0.050 &  0.481$\pm$ 0.084             &  0.402$\pm$0.011  &  0.457$\pm$0.014              &   0.369$\pm$0.013  &   0.383$\pm$0.017  \\
4363\ [O\ {\sc iii}]\                   &       ...         &        ...                    &  0.086$\pm$0.008  &  0.084$\pm$0.009              &       ...          &       ...          \\
4471\ [He\ {\sc i}]                     &       ...         &        ...                    &  0.035$\pm$0.001  &  0.034$\pm$0.001              &       ...          &       ...          \\
4861\ H$\beta$\                         &  1.000$\pm$ 0.176 &  1.000$\pm$ 0.204             &  1.000$\pm$0.020  &  1.000$\pm$0.021              &   1.000$\pm$0.028  &   1.000$\pm$0.029  \\ 
4959\ [O\ {\sc iii}]\                   &  0.433$\pm$ 0.061 &  0.370$\pm$ 0.061             &  1.633$\pm$0.031  &  1.541$\pm$0.031              &   0.581$\pm$0.025  &   0.572$\pm$0.025  \\ 
5007\ [O\ {\sc iii}]\                   &  1.216$\pm$ 0.155 &  1.032$\pm$ 0.152             &  4.866$\pm$0.084  &  4.580$\pm$0.083              &   1.658$\pm$0.044  &   1.632$\pm$0.044  \\
6300\ [O\ {\sc i}]\                     &       ...         &        ...                    &  0.014$\pm$0.002  &  0.013$\pm$0.002              &       ...          &       ...          \\
6312\ [S\ {\sc iii}]\                   &       ...         &        ...                    &  0.021$\pm$0.002  &  0.019$\pm$0.002              &       ...          &       ...          \\
6548\ [N\ {\sc ii}]\                    &  0.010$\pm$ 0.059 &  0.007$\pm$ 0.046             &  0.015$\pm$0.003  &  0.013$\pm$0.003              &   0.039$\pm$0.021  &   0.039$\pm$0.021  \\ 
6563\ H$\alpha$\                        &  4.003$\pm$ 0.520 &  2.759$\pm$ 0.449             &  3.176$\pm$0.054  &  2.800$\pm$0.054              &   2.807$\pm$0.078  &   2.773$\pm$0.085  \\ 
6584\ [N\ {\sc ii}]\                    &  0.043$\pm$ 0.079 &  0.029$\pm$ 0.061             &  0.047$\pm$0.007  &  0.041$\pm$0.007              &   0.124$\pm$0.027  &   0.122$\pm$0.027  \\ 
6678\ [He\ {\sc i}]                     &       ...         &        ...                    &  0.030$\pm$0.001  &  0.026$\pm$0.001              &       ...          &       ...          \\
  & & & & & & \\
$C(H\beta)$\ dex                      & \MC {2}{c}{0.34$\pm$0.17}                         & \MC {2}{c}{0.11$\pm$0.02}                         & \MC {2}{c}{0.00$\pm$0.03}  \\
$EW(abs)$\ \AA\                         & \MC {2}{c}{1.10$\pm$0.03}                         & \MC {2}{c}{6.30$\pm$0.25}                         & \MC {2}{c}{0.40$\pm$0.20}  \\
$EW(H\beta)$\ \AA\                    & \MC {2}{c}{   7$\pm$ 1}                           & \MC {2}{c}{118.2$\pm$1.7}                         & \MC {2}{c}{25$\pm$1 }  \\
\hline                                                                                                                                                       
$T_{\rm e}$(OIII)(K)\                   & \MC {2}{c}{18393$\pm$1380}                        & \MC {2}{c}{14692$\pm$701}                         & \MC {2}{c}{16554$\pm$1014}  \\
$T_{\rm e}$(OII)(K)\                    & \MC {2}{c}{14926$\pm$130 }                        & \MC {2}{c}{13743$\pm$382}                         & \MC {2}{c}{14547$\pm$323}  \\
O$^{+}$/H$^{+}$($\times$10$^5$)\        & \MC {2}{c}{2.165$\pm$0.362}                       & \MC {2}{c}{1.108$\pm$0.104}                       & \MC {2}{c}{2.906$\pm$0.218}  \\ 
O$^{++}$/H$^{+}$($\times$10$^5$)\       & \MC {2}{c}{0.718$\pm$0.144}                       & \MC {2}{c}{5.320$\pm$0.653}                       & \MC {2}{c}{1.430$\pm$0.208}  \\ 
O/H($\times$10$^5$)\                    & \MC {2}{c}{2.883$\pm$0.390}                       & \MC {2}{c}{6.429$\pm$0.662}                       & \MC {2}{c}{4.336$\pm$0.302}  \\ 
12+log(O/H)(T$_{\rm e}$)\               & \MC {2}{c}{...}                                   & \MC {2}{c}{7.81$\pm$0.08}                         & \MC {2}{c}{...} \\
12+log(O/H)$_{c}$(se)  \                & \MC {2}{c}{7.45$\pm$0.10}                         & \MC {2}{c}{7.81$\pm$0.10}                         & \MC {2}{c}{7.64$\pm$0.08}  \\
12+log(O/H)$_{c}$(s)         \          & \MC {2}{c}{7.31$\pm$0.07}                         & \MC {2}{c}{...}                                   & \MC {2}{c}{...}  \\
\MC{3}{l}{~~} \
\end{tabular}
 }
\end{table*}

\begin{table*}
\centering{
\caption{Line intensities and derived parameters of AGC198454, PGC1314481 and PGC1230703 }
\label{t:Intens}
\begin{tabular}{lcccccc} \hline
\rule{0pt}{10pt}
& \MC{2}{c}{AGC198454=J0928+0732}     
& \MC{2}{c}{PGC1314481=J0948+0707}     
& \MC{2}{c}{PGC1230703=J1004+0233}                 \\ \hline
\rule{0pt}{10pt}
$\lambda_{0}$(\AA) Ion                  & $F(\lambda)/F(H\beta)$ & $I(\lambda)/I(H\beta)$ & $F(\lambda)/F(H\beta)$ & $I(\lambda)/I(H\beta)$ & $F(\lambda)/F(H\beta)$ & $I(\lambda)/I(H\beta)$  \\ \hline
%
3727\ [O\ {\sc ii}]\                    &  1.338$\pm$ 0.059 &  1.462$\pm$ 0.072     &  4.964$\pm$ 0.853 &  3.705$\pm$ 1.174       &  3.199$\pm$ 0.136 &  2.908$\pm$ 0.149              \\ 
4101\ H$\delta$\                        &  0.203$\pm$ 0.033 &  0.258$\pm$ 0.054     &       ...         &        ...              &       ...         &        ...                     \\
4340\ H$\gamma$\                        &  0.426$\pm$ 0.061 &  0.472$\pm$ 0.075     &       ...         &        ...              &  0.260$\pm$ 0.019 &  0.340$\pm$ 0.032              \\ 
4861\ H$\beta$\                         &  1.000$\pm$ 0.038 &  1.000$\pm$ 0.047     &  1.000$\pm$ 0.219 &  1.000$\pm$ 0.420       &  1.000$\pm$ 0.044 &  1.000$\pm$ 0.050              \\ 
4959\ [O\ {\sc iii}]\                   &  0.924$\pm$ 0.038 &  0.880$\pm$ 0.038     &  1.237$\pm$ 0.275 &  0.710$\pm$ 0.271       &  0.696$\pm$ 0.037 &  0.632$\pm$ 0.037              \\ 
5007\ [O\ {\sc iii}]\                   &  2.605$\pm$ 0.084 &  2.470$\pm$ 0.083     &  3.209$\pm$ 0.547 &  1.826$\pm$ 0.536       &  2.133$\pm$ 0.082 &  1.939$\pm$ 0.082              \\ 
6548\ [N\ {\sc ii}]\                    &  0.033$\pm$ 0.051 &  0.028$\pm$ 0.045     &  0.037$\pm$ 0.160 &  0.017$\pm$ 0.123       &  0.036$\pm$ 0.041 &  0.033$\pm$ 0.041              \\ 
6563\ H$\alpha$\                        &  3.290$\pm$ 0.134 &  2.772$\pm$ 0.128     &  5.773$\pm$ 0.940 &  2.798$\pm$ 0.854       &  2.984$\pm$ 0.140 &  2.774$\pm$ 0.155              \\ 
6584\ [N\ {\sc ii}]\                    &  0.106$\pm$ 0.060 &  0.088$\pm$ 0.052     &  0.115$\pm$ 0.185 &  0.051$\pm$ 0.143       &  0.118$\pm$ 0.052 &  0.107$\pm$ 0.052              \\ 
6716\ [S\ {\sc ii}]\                    &       ...         &        ...            &  0.693$\pm$ 0.221 &  0.305$\pm$ 0.172       &       ...         &        ...                     \\
6730\ [S\ {\sc ii}]\                    &       ...         &        ...            &  0.413$\pm$ 0.201 &  0.181$\pm$ 0.154       &       ...         &        ...                     \\
 & & & &  & & \\                          
$C(H\beta)$\ dex                      & \MC {2}{c}{0.18$\pm$0.05}                 & \MC {2}{c}{0.34$\pm$0.21}                   & \MC {2}{c}{0.00$\pm$0.06}                           \\
$EW(abs)$\ \AA\                         & \MC {2}{c}{0.75$\pm$0.45}                 & \MC {2}{c}{1.55$\pm$0.32}                   & \MC {2}{c}{1.20$\pm$0.14}                           \\
$EW(H\beta)$\ \AA\                    & \MC {2}{c}{  19$\pm$ 1}                   & \MC {2}{c}{   3$\pm$ 1}                     & \MC {2}{c}{  12$\pm$ 1}                             \\
\hline                                                                                                                                 
$T_{\rm e}$(OIII)(K)\                   & \MC {2}{c}{16920$\pm$1028}                & \MC {2}{c}{15374$\pm$2136}                  & \MC {2}{c}{16148$\pm$1041}                          \\
$T_{\rm e}$(OII)(K)\                    & \MC {2}{c}{14656$\pm$281 }                & \MC {2}{c}{14087$\pm$ 988}                  & \MC {2}{c}{14408$\pm$383 }                          \\
O$^{+}$/H$^{+}$($\times$10$^5$)\        & \MC {2}{c}{1.425$\pm$0.111}               & \MC {2}{c}{4.098$\pm$1.600}                 & \MC {2}{c}{2.991$\pm$0.297}                         \\
O$^{++}$/H$^{+}$($\times$10$^5$)\       & \MC {2}{c}{2.066$\pm$0.295}               & \MC {2}{c}{1.966$\pm$0.820}                 & \MC {2}{c}{1.769$\pm$0.279}                         \\
O/H($\times$10$^5$)\                    & \MC {2}{c}{3.490$\pm$0.315}               & \MC {2}{c}{6.064$\pm$1.798}                 & \MC {2}{c}{4.760$\pm$0.407}                         \\
12+log(O/H)$_{c}$(se)  \                & \MC {2}{c}{7.52$\pm$0.09}                 & \MC {2}{c}{7.75$\pm$0.15}                   & \MC {2}{c}{7.66$\pm$0.08}                           \\
\MC{3}{l}{~~} \
\end{tabular}
 }
\end{table*}

\begin{table*}
\centering{
\caption{Line intensities and derived parameters  of J1001+0846, PGC1178576 and AGC208397 }
\label{t:Intens}
\begin{tabular}{lcccccc} \hline
\rule{0pt}{10pt}
& \MC{2}{c}{J1001+0846}  
& \MC{2}{c}{PGC1178576=J1021+0054}   
& \MC{2}{c}{AGC208397=J1038+0352}      \\ \hline
\rule{0pt}{10pt}
$\lambda_{0}$(\AA) Ion                  & $F(\lambda)/F(H\beta)$ & $I(\lambda)/I(H\beta)$ & $F(\lambda)/F(H\beta)$ & $I(\lambda)/I(H\beta)$ & $F(\lambda)/F(H\beta)$ & $I(\lambda)/I(H\beta)$  \\ \hline
%
3727\ [O\ {\sc ii}]\                    &  3.402$\pm$ 0.189 &  2.467$\pm$ 0.215    &  2.202$\pm$ 0.137 &  1.774$\pm$ 0.161       &  1.237$\pm$ 0.081 &  0.977$\pm$ 0.083             \\ 
4340\ H$\gamma$\                        &  0.212$\pm$ 0.031 &  0.473$\pm$ 0.123    &  0.227$\pm$ 0.024 &  0.484$\pm$ 0.083       &  0.268$\pm$ 0.023 &  0.472$\pm$ 0.060             \\ 
4861\ H$\beta$\                         &  1.000$\pm$ 0.059 &  1.000$\pm$ 0.091    &  1.000$\pm$ 0.054 &  1.000$\pm$ 0.075       &  1.000$\pm$ 0.040 &  1.000$\pm$ 0.052             \\ 
4959\ [O\ {\sc iii}]\                   &  0.906$\pm$ 0.057 &  0.616$\pm$ 0.057    &  0.488$\pm$ 0.033 &  0.357$\pm$ 0.033       &  0.464$\pm$ 0.026 &  0.366$\pm$ 0.026             \\ 
5007\ [O\ {\sc iii}]\                   &  2.952$\pm$ 0.143 &  2.002$\pm$ 0.143    &  1.488$\pm$ 0.074 &  1.086$\pm$ 0.073       &  1.382$\pm$ 0.045 &  1.091$\pm$ 0.045             \\ 
6548\ [N\ {\sc ii}]\                    &  0.064$\pm$ 0.038 &  0.041$\pm$ 0.036    &  0.027$\pm$ 0.054 &  0.018$\pm$ 0.049       &  0.022$\pm$ 0.032 &  0.017$\pm$ 0.032             \\ 
6563\ H$\alpha$\                        &  3.992$\pm$ 0.204 &  2.776$\pm$ 0.226    &  3.865$\pm$ 0.234 &  2.744$\pm$ 0.246       &  3.306$\pm$ 0.117 &  2.725$\pm$ 0.132             \\ 
6584\ [N\ {\sc ii}]\                    &  0.170$\pm$ 0.051 &  0.109$\pm$ 0.048    &  0.083$\pm$ 0.076 &  0.056$\pm$ 0.070       &  0.076$\pm$ 0.041 &  0.060$\pm$ 0.041             \\ 
   & & & &  & & \\                          
$C(H\beta)$\ dex                      & \MC {2}{c}{0.09$\pm$0.07}                & \MC {2}{c}{0.12$\pm$0.08}                  & \MC {2}{c}{0.00$\pm$0.05}                          \\
$EW(abs)$\ \AA\                         & \MC {2}{c}{1.90$\pm$0.07}                & \MC {2}{c}{2.30$\pm$0.07}                  & \MC {2}{c}{4.65$\pm$0.15}                          \\
$EW(H\beta)$\ \AA\                    & \MC {2}{c}{   4$\pm$ 1}                  & \MC {2}{c}{   6$\pm$ 1}                    & \MC {2}{c}{  17$\pm$ 1}                            \\
\hline                                                                                           
$T_{\rm e}$(OIII)(K)\                   & \MC {2}{c}{16591$\pm$1111}               & \MC {2}{c}{19311$\pm$1115}                 & \MC {2}{c}{20970$\pm$1068}                         \\
$T_{\rm e}$(OII)(K)\                    & \MC {2}{c}{14559$\pm$349 }               & \MC {2}{c}{14961$\pm$20  }                 & \MC {2}{c}{15634$\pm$25  }                         \\
O$^{+}$/H$^{+}$($\times$10$^5$)\        & \MC {2}{c}{2.455$\pm$0.283}              & \MC {2}{c}{1.622$\pm$0.148}                & \MC {2}{c}{0.781$\pm$0.066}                        \\
O$^{++}$/H$^{+}$($\times$10$^5$)\       & \MC {2}{c}{1.690$\pm$0.283}              & \MC {2}{c}{0.665$\pm$0.089}                & \MC {2}{c}{0.568$\pm$0.061}                        \\
O/H($\times$10$^5$)\                    & \MC {2}{c}{4.145$\pm$0.400}              & \MC {2}{c}{2.287$\pm$0.172}                & \MC {2}{c}{1.348$\pm$0.090}                        \\
12+log(O/H)$_{c}$(se)  \                & \MC {2}{c}{7.60$\pm$0.08}                & \MC {2}{c}{7.35$\pm$0.07}                  & \MC {2}{c}{7.13$\pm$0.08}                          \\
12+log(O/H)$_{c}$(s)         \          & \MC {2}{c}{...}                          & \MC {2}{c}{7.25$\pm$0.06}                  & \MC {2}{c}{7.13$\pm$0.05}                          \\
\MC{3}{l}{~~} \
\end{tabular}
 }
\end{table*}

\begin{table*}
\centering{
\caption{Line intensities and derived parameters  of PGC044681, PGC135827 and AGC258574 }
\label{t:Intens}
\begin{tabular}{lcccccc} \hline
\rule{0pt}{10pt}
& \MC{2}{c}{PGC044681=J1259--1924}   
& \MC{2}{c}{PGC135827=J1328+0216}    
& \MC{2}{c}{AGC258574=J1545+0148}       \\ \hline
\rule{0pt}{10pt}
$\lambda_{0}$(\AA) Ion                  & $F(\lambda)/F(H\beta)$ & $I(\lambda)/I(H\beta)$ & $F(\lambda)/F(H\beta)$ & $I(\lambda)/I(H\beta)$ & $F(\lambda)/F(H\beta)$ & $I(\lambda)/I(H\beta)$  \\ \hline
%
3727\ [O\ {\sc ii}]\                    &  2.114$\pm$ 0.301 &  2.404$\pm$ 0.365   &  1.211$\pm$ 0.023 &  1.254$\pm$ 0.027        &  1.982$\pm$ 0.190 &  1.539$\pm$ 0.259                  \\ 
3798\ H10\                              &       ...         &        ...          &  0.040$\pm$ 0.007 &  0.115$\pm$ 0.028        &       ...         &        ...                         \\
3835\ H9\                               &       ...         &        ...          &  0.063$\pm$ 0.009 &  0.137$\pm$ 0.024        &       ...         &        ...                         \\
3967\ [Ne\ {\sc iii}]\ +\ H7\           &       ...         &        ...          &  0.215$\pm$ 0.009 &  0.291$\pm$ 0.016        &       ...         &        ...                         \\
4101\ H$\delta$\                        &       ...         &        ...          &  0.220$\pm$ 0.009 &  0.288$\pm$ 0.015        &       ...         &        ...                         \\
4340\ H$\gamma$\                        &       ...         &        ...          &  0.394$\pm$ 0.026 &  0.448$\pm$ 0.032        &       ...         &        ...                         \\
4363\ [O\ {\sc iii}]\                   &       ...         &        ...          &  0.077$\pm$ 0.021 &  0.076$\pm$ 0.022        &       ...         &        ...                         \\
4861\ H$\beta$\                         &  1.000$\pm$ 0.113 &  1.000$\pm$ 0.173   &  1.000$\pm$ 0.018 &  1.000$\pm$ 0.020        &  1.000$\pm$ 0.078 &  1.000$\pm$ 0.145                  \\ 
4959\ [O\ {\sc iii}]\                   &  0.099$\pm$ 0.024 &  0.095$\pm$ 0.024   &  1.348$\pm$ 0.024 &  1.284$\pm$ 0.024        &  0.665$\pm$ 0.082 &  0.390$\pm$ 0.081                  \\ 
5007\ [O\ {\sc iii}]\                   &  0.295$\pm$ 0.072 &  0.285$\pm$ 0.071   &  4.005$\pm$ 0.069 &  3.804$\pm$ 0.068        &  1.982$\pm$ 0.129 &  1.150$\pm$ 0.126                  \\ 
6548\ [N\ {\sc ii}]\                    &  0.034$\pm$ 0.114 &  0.029$\pm$ 0.097   &  0.020$\pm$ 0.010 &  0.018$\pm$ 0.009        &  0.105$\pm$ 0.101 &  0.048$\pm$ 0.077                  \\ 
6563\ H$\alpha$\                        &  3.250$\pm$ 0.330 &  2.724$\pm$ 0.316   &  3.147$\pm$ 0.059 &  2.793$\pm$ 0.059        &  5.578$\pm$ 0.375 &  2.746$\pm$ 0.339                  \\ 
6584\ [N\ {\sc ii}]\                    &  0.102$\pm$ 0.125 &  0.085$\pm$ 0.107   &  0.062$\pm$ 0.013 &  0.054$\pm$ 0.012        &  0.330$\pm$ 0.125 &  0.148$\pm$ 0.095                  \\ 
   & & & &  & & \\                        
$C(H\beta)$\ dex                       & \MC {2}{c}{0.21$\pm$0.13}               & \MC {2}{c}{0.11$\pm$0.02}                    & \MC {2}{c}{0.37$\pm$0.09}                              \\
$EW(abs)$\ \AA\                          & \MC {2}{c}{0.35$\pm$2.19}               & \MC {2}{c}{6.80$\pm$0.68}                    & \MC {2}{c}{4.75$\pm$0.33}                               \\
$EW(H\beta)$\ \AA\                     & \MC {2}{c}{  17$\pm$ 1}                 & \MC {2}{c}{ 155$\pm$ 2}                      & \MC {2}{c}{   7$\pm$ 1}                                \\
\hline                                                                            
$T_{\rm e}$(OIII)(K)\                   & \MC {2}{c}{20165$\pm$1469}              & \MC {2}{c}{15229$\pm$2022}                   & \MC {2}{c}{19574$\pm$1345}                              \\
$T_{\rm e}$(OII)(K)\                    & \MC {2}{c}{14902$\pm$179 }              & \MC {2}{c}{14018$\pm$971 }                   & \MC {2}{c}{14952$\pm$67  }                              \\
O$^{+}$/H$^{+}$($\times$10$^5$)\        & \MC {2}{c}{2.224$\pm$0.348}             & \MC {2}{c}{1.409$\pm$0.320}                  & \MC {2}{c}{1.409$\pm$0.238}                              \\
O$^{++}$/H$^{+}$($\times$10$^5$)\       & \MC {2}{c}{0.162$\pm$0.048}             & \MC {2}{c}{4.038$\pm$1.337}                  & \MC {2}{c}{0.690$\pm$0.119}                              \\
O/H($\times$10$^5$)\                    & \MC {2}{c}{2.386$\pm$0.351}             & \MC {2}{c}{5.448$\pm$1.374}                  & \MC {2}{c}{2.099$\pm$0.266}                              \\
12+log(O/H)(T$_{\rm e}$)\               & \MC {2}{c}{...}                         & \MC {2}{c}{7.74$\pm$0.11}                    & \MC {2}{c}{...}                                         \\
12+log(O/H)$_{c}$(se)  \                & \MC {2}{c}{7.37$\pm$0.09}               & \MC {2}{c}{7.71$\pm$0.08}                    & \MC {2}{c}{7.31$\pm$0.09}                               \\
12+log(O/H)$_{c}$(s)         \          & \MC {2}{c}{7.20$\pm$0.08}               & \MC {2}{c}{...}                              & \MC {2}{c}{7.23$\pm$0.07}                               \\
\MC{3}{l}{~~} \                                        
\end{tabular}
 }
\end{table*}

\begin{table*}
\centering{
\caption{Line intensities and derived parameters of KK246 and AGC335193 }
\label{t:Intens}
\begin{tabular}{lcccc} \hline
\rule{0pt}{10pt}
& \MC{2}{c}{KK246=J2003--3140}      
& \MC{2}{c}{AGC335193=J2303+0431}     \\ \hline
\rule{0pt}{10pt}
$\lambda_{0}$(\AA) Ion                  & $F(\lambda)/F(H\beta)$ & $I(\lambda)/I(H\beta)$ & $F(\lambda)/F(H\beta)$ & $I(\lambda)/I(H\beta)$  \\ \hline
%
3727\ [O\ {\sc ii}]\                    &  1.960$\pm$ 0.089 &  2.830$\pm$ 0.147       &  3.816$\pm$ 0.150 &  2.476$\pm$ 0.162            \\ 
4340\ H$\gamma$\                        &  0.357$\pm$ 0.030 &  0.472$\pm$ 0.050       &  0.098$\pm$ 0.014 &  0.449$\pm$ 0.137            \\
4861\ H$\beta$\                         &  1.000$\pm$ 0.040 &  1.000$\pm$ 0.048       &  1.000$\pm$ 0.036 &  1.000$\pm$ 0.058            \\ 
4959\ [O\ {\sc iii}]\                   &  0.765$\pm$ 0.041 &  0.700$\pm$ 0.040       &  0.929$\pm$ 0.035 &  0.598$\pm$ 0.035            \\ 
5007\ [O\ {\sc iii}]\                   &  2.291$\pm$ 0.092 &  2.063$\pm$ 0.089       &  2.769$\pm$ 0.089 &  1.783$\pm$ 0.089            \\ 
6548\ [N\ {\sc ii}]\                    &  0.059$\pm$ 0.042 &  0.035$\pm$ 0.027       &  0.049$\pm$ 0.041 &  0.031$\pm$ 0.040            \\ 
6563\ H$\alpha$\                        &  4.631$\pm$ 0.189 &  2.793$\pm$ 0.133       &  3.913$\pm$ 0.143 &  2.769$\pm$ 0.171            \\ 
6584\ [N\ {\sc ii}]\                    &  0.178$\pm$ 0.052 &  0.106$\pm$ 0.033       &  0.153$\pm$ 0.053 &  0.098$\pm$ 0.052            \\ 
   & & & &\\
$C(H\beta)$\ dex                       & \MC {2}{c}{0.60$\pm$0.05}                   & \MC {2}{c}{0.01$\pm$0.05}                        \\
$EW(abs)$\ \AA\                          & \MC {2}{c}{1.75$\pm$0.62}                   & \MC {2}{c}{2.80$\pm$0.04}                         \\
$EW(H\beta)$\ \AA\                     & \MC {2}{c}{  29$\pm$ 1}                     & \MC {2}{c}{   5$\pm$ 1}                          \\
\hline
$T_{\rm e}$(OIII)(K)\                   & \MC {2}{c}{16026$\pm$1042}                  & \MC {2}{c}{16865$\pm$1059}                       \\
$T_{\rm e}$(OII)(K)\                    & \MC {2}{c}{14362$\pm$399 }                  & \MC {2}{c}{14640$\pm$297 }                       \\
O$^{+}$/H$^{+}$($\times$10$^5$)\        & \MC {2}{c}{2.941$\pm$0.303}                 & \MC {2}{c}{2.421$\pm$0.222}                       \\
O$^{++}$/H$^{+}$($\times$10$^5$)\       & \MC {2}{c}{1.936$\pm$0.310}                 & \MC {2}{c}{1.479$\pm$0.223}                       \\
O/H($\times$10$^5$)\                    & \MC {2}{c}{4.877$\pm$0.433}                 & \MC {2}{c}{3.901$\pm$0.314}                       \\
12+log(O/H)$_{c}$(se)  \                & \MC {2}{c}{7.67$\pm$0.08}                   & \MC {2}{c}{7.57$\pm$0.08}                        \\
\MC{2}{l}{~~} \                                                                                                                         
\end{tabular}
 }
\end{table*}

\bsp

\label{lastpage}


\begin{thebibliography}{99}

\bibitem[\protect\citeauthoryear{Abazajian et al.}{2009}]{DR7}
   Abazajian K.N., Adelman-McCarthy J.K., Ag\"ueros M.A. et al.,
   2009, ApJS, 182, 543


\bibitem[\protect\citeauthoryear{Abolfathi et al.}{2018}]{DR14}
Abolfathi B., Aguado D.S., Aguilar G., et al.,  2018, ApJS, 335, 42

\bibitem[\protect\citeauthoryear{Annibali et al.}{2013}]{Annibali2013}
   Annibali F., Cignoni M., Tosi M., et al., 2013, AJ, 146, 144

\bibitem[\protect\citeauthoryear{Asplund et al.}{2009}]{Asplund09}
  Asplund M., Grevesse N., Sauval A. J., Scott P., 2009, ARA\&A, 47, 481

\bibitem[\protect\citeauthoryear{Berg et al.}{2012}]{Berg12}
    Berg D.A., Skillman E.D., Marble A., et al.  2012, ApJ, 754, 98

\bibitem[\protect\citeauthoryear{Buckley, Swart \& Meiring}{2006}]{Buck06}
	Buckley, D.A.H., Swart, G.P., Meiring, J.G., 2006, SPIE, 6267

\bibitem[\protect\citeauthoryear{Burgh et al.}{2003}]{Burgh03}
	Burgh, E.B., Nordsieck, K.H., Kobulnicky, H.A., Williams, T.B.,
	O'Donoghue, D., Smith, M.P., Percival, J.W., 2003, SPIE, 4841, 1463

\bibitem[\protect\citeauthoryear{Chengalur, Pustilnik}{2013}]{Triplet}
    Chengalur J.N., Pustilnik S.A.,  2013, \mnras, 428, 1579

\bibitem[\protect\citeauthoryear{Chengalur, Pustilnik, Egorova}
   {Chengalur et al.}{2016}]{U3672}
    Chengalur J.N., Pustilnik S.A., Egorova E.S., 2017, MNRAS, 465, 2342

\bibitem[\protect\citeauthoryear{Colless et al.}{2001}]{2dF_2001}
     Colless M., et al., 2001, MNRAS, 328, 1039

\bibitem[\protect\citeauthoryear{Crawford et al.}{2010}]{Cra2010}
    Crawford S. M. et al., 2010, in Silva D. R., Peck A. B., Soifer B. T.,
    Proc. SPIE Conf. Ser. Vol. 7737, Observatory Operations: Strategies,
    Processes, and Systems III. SPIE, Bellingham, p. 773725

\bibitem[\protect\citeauthoryear{Dey et al.}{2019}]{DECaLS}
   Dey A., Schlegel D., Lang D., et al. 2019, AJ, 57, id. 168


\bibitem[\protect\citeauthoryear{Dopita et al.}{2013}]{Dopita2013}
   Dopita M.A., Sutherland R.S., Nicholls D.C., Kewley L.J., Vogt F.P.A.
   2013, ApJS, 208, 10

\bibitem[\protect\citeauthoryear{Flewelling et al.}{2016}]{PS1-database}
Flewelling H.A., Magnier E.A., Chambers K.C., et al. 2016, arXiv:1612.05243v3

\bibitem[\protect\citeauthoryear{Guseva et al.}{2017}]{Guseva17}
    Guseva N.G., Izotov Y.I., Frieke K.J., Henkel C., 2017,
    A\&A,  599, A65


\bibitem[\protect\citeauthoryear{Haynes et al.}{2018}]{ALFALFA}
   Haynes M.P., Giovanelli R., Kent B., et al. 2018, ApJ, 861, 49

\bibitem[\protect\citeauthoryear{Hirschauer et al.}{2016}]{Hirschauer16}
   Hirschauer A.S., Salzer J.J., Skillman E.D., et al. 2016, ApJ, 822, 108

\bibitem[\protect\citeauthoryear{Hsyu et al.}{2017}]{LittleCub}
   Hsyu T., Cooke R.J., Prochaska J.X., Bolte M., 2017, ApJ Lett., 845, L22

\bibitem[\protect\citeauthoryear{Izotov \& Thuan}{2007}]{IT07}
    Izotov Y.I., Thuan T.X., 2007,   \apj,  665, 1115

\bibitem[\protect\citeauthoryear{Izotov et al.}{1990}]{Izotov1990}
   Izotov Y.I., Guseva N.G., Lipovetsky V.A., Kniazev A.Y., Stepanian J.A.,
   1990, Nature, 343, 238

\bibitem[\protect\citeauthoryear{Izotov et al.}{1994}]{ITL94}
   Izotov Y.I., Thuan T.X., Lipovetsky V.A., 1994, ApJ, 435, 647


\bibitem[\protect\citeauthoryear{Izotov, Thuan, Guseva}{2007}]{ITG07}
	Izotov Y.I., Thuan T.X., Guseva N.G., 2007, ApJ, 671, 1297

\bibitem[\protect\citeauthoryear{Izotov et al.}{2009}]{Izotov2009}
    Izotov Y.I., Guseva N.G., Frieke K.J., Papaderos P. 2009, A\&A, 503, 61

\bibitem[\protect\citeauthoryear{Izotov et al.}{2018}]{Izotov18}
  Izotov Y.I., Thuan T.X., Guseva N.G., Liss S.E.  2018,  MNRAS,  473, 1956

\bibitem[\protect\citeauthoryear{Izotov, Thuan, Guseva}{Izotov et al.}{2019a}]{Izotov19}
  Izotov Y.I., Thuan T.X., Guseva N.G., 2019a,  MNRAS, 483, 5491

\bibitem[\protect\citeauthoryear{Izotov et al.}{2019b}]{Izotov19DR14}
 Izotov Y.I., Guseva N.G., Frieke K.J., Henkel C., 2019b, A\&A, 523, A40



\bibitem[\protect\citeauthoryear{Kniazev et al.}{2004}]{Kniazev04}
  Kniazev A.Y., Pustilnik S.A., Grebel E.K., Lee H., Pramskij A.G.,
  2004, ApJS, 153, 429

\bibitem[\protect\citeauthoryear{Kniazev, Egorova, Pustilnik}{Kniazev et al.}{2018}]{Eridanus}
   Kniazev A.Y., Egorova E.S., Pustilnik S.A., 2018, MNRAS, 479, 3842

\bibitem[\protect\citeauthoryear{Kobulnicky et al.}{2003}]{Kobul03}
	Kobulnicky, H.A., Nordsieck, K.H., Burgh, E.B., Smith, M.P.,
	Percival, J.W., Williams, T.B., O'Donoghue, D., 2003, SPIE, 4841, 1634

\bibitem[\protect\citeauthoryear{Koleva et al.}{2009}]{Koleva2009}
   Koleva M., Prugniel P., Bouchard A., Wu Y., 2009, A\&A, 501, 1269

\bibitem[\protect\citeauthoryear{Kreckel et al.}{2011}]{Kreckel11}
   Kreckel K., Peebles P.J.E., van Gorkom J.H., van de Weygaert R.,
  van der Hulst J.M., 2011, AJ, 141, 204

\bibitem[\protect\citeauthoryear{Kunth \& \"Ostlin}{2000}]{Kunth2000}
   Kunth D., \& \"Ostlin G., 2000, A\&ARev, 10, 1

\bibitem[\protect\citeauthoryear{Leitherer et al.}{1999}]{Starburst99}
  Leitherer C., Schaerer D.,  Goldader J.D. et al. 1999, ApJS, 123, 1


\bibitem[\protect\citeauthoryear{Nicholls et al.}{2014}]{KK246_OH}
    Nicholls D.C., Jerjen H., Dopita M.A., Basurah H., 2014, ApJ, 780, 88

\bibitem[\protect\citeauthoryear{O'Donoghue et al.}{2006}]{Dono06}
	O'Donoghue, D., et al.\, 2006, \mnras, 372, 151

\bibitem[\protect\citeauthoryear{Papaderos, \"Ostlin}{2012}]{PO12}
  Papaderos P., \"Ostlin G., 2012, A\&A, 537, A126

\bibitem[\protect\citeauthoryear{Perepelitsyna, Pustilnik, Kniazev}{Perepelitsyna et al.}{2014}]
    {PaperIV}
 Perepelitsyna Y.A., Pustilnik S.A., Kniazev A.Y. 2014, Astrophys.Bull.,
   69, 247  (arXiv:1408.0613)

\bibitem[\protect\citeauthoryear{Pilyugin, Thuan}{2005}]{PT05}
      Pilyugin L.S., Thuan T.X., 2005, ApJ, 631, 231

\bibitem[\protect\citeauthoryear{Pustilnik, Pramskij, Kniazev}{2004}]{Pustilnik04}
    Pustilnik S.A., Pramskij A.G., Kniazev  A.Y.,
   2004, A\&A, 425, 51-65

\bibitem[\protect\citeauthoryear{Pustilnik, Tepliakova}{2011}]{PaperI}
      Pustilnik S.A., Tepliakova A.L., 2011, MNRAS, 415, 1188

\bibitem[\protect\citeauthoryear{Pustilnik, Kniazev, Pramskij}{2005}]{DDO68}
      Pustilnik S.A., Kniazev A.Y., Pramskij A.G., 2005, A\&A, 443, 91

\bibitem[\protect\citeauthoryear{Pustilnik et al.}{2010}]{J0926}
   Pustilnik S.A., Tepliakova A.L., Kniazev A.Y., Martin J.-M., Burenkov A.N.,
	 2010, MNRAS, 401, 333

\bibitem[\protect\citeauthoryear{Pustilnik, Perepelitsyna, Kniazev}{Pustilnik et al.}{2016}]
  {PaperVII}
 Pustilnik S.A., Perepelitsyna Y.A., Kniazev A.Y.,  2016, MNRAS, 463, 670

\bibitem[\protect\citeauthoryear{Pustilnik, Tepliakova \& Makarov}{2019}]{PTM19}
  Pustilnik S.A., Tepliakova A.L., Makarov D.I., 2019, MNRAS, 482, 4329 (PTM19)

\bibitem[\protect\citeauthoryear{Pustilnik et al.}{2019}]{PEPK19}
  Pustilnik S.A., Egorova E., Perepelitsyna Y.A., Kniazev A.Y.,
2019, MNRAS, submitted (PEPK19)


\bibitem[\protect\citeauthoryear{Sanchez Almeida et al.}{2016}]{Sanchez16a}
    Sanchez Almeida J., Perez-Montero E., Morales-Luis A.B., Munoz-Tunon C.,
Garcia-Benito R., Nuza S.E., Kitaura F.S.,   2016, ApJ, 819, 110


\bibitem[\protect\citeauthoryear{Schlafly \& Finkbeiner}{2011}]{Schlafly11}
Schlafly E.F., Finkbeiner D.P., 2011, ApJ, 737, 103, 13pp.

\bibitem[\protect\citeauthoryear{Searle \& Sargent}{1972}]{Searle72}
    Searle L., \& Sargent W.L.W., 1972,  ApJ, 173, 25


\bibitem[\protect\citeauthoryear{Skillman et al.}{2013}]{Skillman13}
Skillman E., Salzer J., Berg D.A. et al., 2013, AJ, 146, 3

\bibitem[\protect\citeauthoryear{Stasinska \& Izotov}{2003}]{SI2003}
  Stasinska G., \& Izotov Y.I., 2003, A\&A, 397, 71

\bibitem[\protect\citeauthoryear{Takashi et al.}{2019}]{Takashi19}
  Takashi K., Ouchi M., Rauch M., et al., 2019, arXiv:1910.08559

\bibitem[\protect\citeauthoryear{Tully et al.}{2008}]{Tully08}
  Tully B., Shaya E.J., Karachentsev I.D., Courtois H.M., Kocevski D.D.,
   Rizzi L., Peel A.,   2008, \apj, 676, 184

\bibitem[\protect\citeauthoryear{Tweed et al.}{2018}]{Tweed18}
   Tweed D.P., Mamon G.A., Thuan T.X., Cattaneo A., Dekel A., Menci N.,
   Calura F., Silk J.,  2018,   MNRAS, 477, 1427

\bibitem[\protect\citeauthoryear{Stasinska \& Izotov}{2003}]{SI2003}
  Stasinska G., \& Izotov Y.I., 2003, A\&A, 397, 71

\bibitem[\protect\citeauthoryear{Whitford}{1958}]{W1958}
  Whitford A.E., 1968, AJ, 63, 201

\end{thebibliography}
\end{document}